\documentclass[a4paper,twoside]{article}
\newcommand{\affil}[1]{$^{\rm #1}$}
\usepackage[authoryear]{natbib}
\bibpunct{(}{)}{;}{a}{}{,}

\usepackage{graphicx}
\usepackage{amssymb}
\date{} 
\newcommand{\aj}{AJ}
\newcommand{\apjs}{ApJS}
\newcommand{\apj}{ApJ}
\newcommand{\apjl}{ApJ}
\newcommand{\nat}{Nature}

\newcommand{\mnras}{MNRAS}

\newcommand{\aap}{A\&A}

\newcommand{\aaps}{A\&AS}

\newcommand{\apss}{Ap\&SS}
\newcommand{\araa}{ARA\&A}
\newcommand{\pasa}{PASA}
\newcommand{\pasp}{PASP}

\newcommand{\physrep}{Phys.\ Rep.}

\newcommand{\prd}{Phys.~Rev.~D}

\newcommand{\solphys}{Solar Physics}
\newcommand{\jgr}{Journal of Geophysical Research}

\title{\large\bf\flushleft Science with the Murchison Widefield Array}
\author{\parbox{\textwidth}{\flushleft
\vspace{-0.5cm}
{\it Judd D. Bowman\affil{A,W}, 
Iver Cairns\affil{P},
David L. Kaplan\affil{T},
Tara Murphy\affil{P,V},
Divya Oberoi\affil{L},
Lister Staveley-Smith\affil{S,V},
Wayne Arcus\affil{E}, 
David G. Barnes\affil{K},
Gianni Bernardi\affil{G},
Frank H. Briggs\affil{B,V},
Shea Brown\affil{N},
John D. Bunton\affil{D},
Adam J. Burgasser\affil{M},
Roger J. Cappallo\affil{H},
Shami Chatterjee\affil{C},
Brian E. Corey\affil{H},
Anthea Coster\affil{H},
Avinash Deshpande\affil{J},
Ludi deSouza\affil{B,P},
David Emrich\affil{E},
Philip Erickson\affil{H},
Robert F. Goeke\affil{I},
B. M. Gaensler\affil{P,V},
Lincoln J. Greenhill\affil{G},
Lisa Harvey-Smith\affil{D},
Bryna J. Hazelton\affil{R},
David Herne\affil{E},
Jacqueline N. Hewitt\affil{I},
Melanie Johnston-Hollitt\affil{U},
Justin C. Kasper\affil{G},
Barton B. Kincaid\affil{H},
Ronald Koenig\affil{D},
Eric Kratzenberg\affil{H},
Colin J. Lonsdale\affil{H},
Mervyn J. Lynch\affil{E},
Lynn D. Matthews\affil{H},
S. Russell McWhirter\affil{H},
Daniel A. Mitchell\affil{G,O,V},
Miguel F. Morales\affil{R},
Edward H. Morgan\affil{I},
Stephen M. Ord\affil{E},
Joseph Pathikulangara\affil{D},
Thiagaraj Prabu\affil{J},
Ronald A. Remillard\affil{I},
Timothy Robishaw\affil{P,F},
Alan E. E. Rogers\affil{H},
Anish A. Roshi\affil{J},
Joseph E. Salah\affil{H},
Robert J. Sault\affil{O},
N. Udaya Shankar\affil{J},
K. S. Srivani\affil{J},
Jamie B. Stevens\affil{D,Q},
Ravi Subrahmanyan\affil{J,V},
Steven J. Tingay\affil{E,V},
Randall B. Wayth\affil{E,G,V},
Mark Waterson\affil{B,E},
Rachel L. Webster\affil{O,V},
Alan R. Whitney\affil{H},
Andrew J. Williams\affil{S},
Christopher L. Williams\affil{I},
and J. Stuart B. Wyithe\affil{O,V}}\\
\vspace{0.4cm}
{\small \affil{A}\,Arizona State University, Tempe, AZ 85287, USA}\\
{\small \affil{B}\,Australian National University}\\
{\small \affil{C}\,Cornell University} \\
{\small \affil{D}\,CSIRO Astronomy and Space Science}\\
{\small \affil{E}\,Curtin University}\\
{\small \affil{F}\,Dominion Radio Astrophysical Observatory}\\
{\small \affil{G}\,Harvard-Smithsonian Center for Astrophysics}\\
{\small \affil{H}\,MIT Haystack Observatory}\\
{\small \affil{I}\,MIT Kavli Institute}\\
{\small \affil{J}\,Raman Research Institute}\\
{\small \affil{K}\,Swinburne University of Technology}\\
{\small \affil{L}\,Tata Institute for Fundamental Research}\\
{\small \affil{M}\,University of California, San Diego}\\
{\small \affil{N}\,University of Iowa}\\
{\small \affil{O}\,University of Melbourne}\\
{\small \affil{P}\,University of Sydney}\\
{\small \affil{Q}\,University of Tasmania}\\
{\small \affil{R}\,University of Washington-Seattle}\\
{\small \affil{S}\,University of Western Australia}\\
{\small \affil{T}\,University of Wisconsin-Milwaukee}\\
{\small \affil{U}\,Victoria University of Wellington, New Zealand}\\
{\small \affil{V}\,ARC Centre of Excellence for All-sky Astrophysics (CAASTRO)}\\
{\small \affil{W}\,Email: judd.bowman@asu.edu}}
}
{
\begin{document}
\twocolumn[
\begin{changemargin}{.8cm}{.5cm}
\begin{minipage}{.9\textwidth}
\vspace{-1cm}
\maketitle
%
%
\small{\bf Abstract:}
Significant new opportunities for astrophysics and cosmology have been identified at low radio frequencies.  The Murchison Widefield Array is the first telescope in the Southern Hemisphere designed specifically to explore the low-frequency astronomical sky between 80 and 300 MHz with arcminute angular resolution and high survey efficiency.   The telescope will enable new advances along four key science themes, including searching for redshifted 21 cm emission from the epoch of reionisation in the early Universe; Galactic and extragalactic all-sky southern hemisphere surveys;  time-domain astrophysics; and solar, heliospheric, and ionospheric science and space weather.  The Murchison Widefield Array is located in Western Australia at the site of the planned Square Kilometre Array (SKA) low-band telescope and is the only low-frequency SKA precursor facility.  In this paper, we review the performance properties of the Murchison Widefield Array and describe its primary scientific objectives.


\medskip
\medskip
\end{minipage}
\end{changemargin}
]
\small

\def\kperp{k_{\bot}}
\def\kpar{k_{\|}}


\section{Introduction}


The Murchison Widefield Array (MWA\footnote{http://mwatelescope.org}) is a low-frequency radio telescope operating between 80 and 300 MHz.  It is located at the Murchison Radio-astronomy Observatory (MRO) in Western Australia, the planned site of the future Square Kilometre Array (SKA\footnote{http://skatelescope.org}) low-band telescope, and is one of three telescopes designated as a Precursor for the SKA.    The MWA has been developed by an international collaboration, including partners from Australia, India, New Zealand, and the United States.   It has begun operations in 2013.  In this paper, we review the science capabilities of the telescope and the primary science objectives motivating its design and operation.   Details of the technical design and specifications of the MWA as built are presented in a companion article \citep{mwa_tech}, which supercedes the earlier \citet{2009IEEEP..97.1497L} description of plans for the instrument.

The four key science themes driving the MWA are:  1) searching for redshifted 21 cm emission from the epoch of reionisation (EoR) in the early Universe; 2) Galactic and extragalactic surveys; 3) time-domain astrophysics; and 4) solar, heliospheric, and ionospheric science and space weather.  We describe the plans and objectives for each of the key science themes in Sections 2 through 5.

The frequency range targeted by the MWA is one of the more difficult portions of the spectrum for ground-based radio telescopes due to strong distortions of incoming signals by the Earth's ionosphere and also due to congested terrestrial communications, including FM radio, TV stations, aircraft navigation, and satellite communications.  Hence, the low-frequency radio sky remains one of the least explored domains in astronomy.  The MRO site mitigates many issues associated with radio interference for the telescope, but processing of MWA observations requires particular care to correct for the ionospheric distortions. 

The unifying design theme for the MWA is that of high survey efficiency.  Many of the science goals require large areas of the sky to be surveyed and the EoR objective further requires very high surface brightness sensitivity.   It was recognized early in the development of the MWA that arcminute angular resolution is a convenient balance between the opposing tensions of maximizing scientific returns and minimizing technical complexity.  Arcminute angular scales are sufficient to probe typical ionised bubbles during the EoR, study diffuse and polarized emission structures in the Galaxy, reduce extragalactic source confusion, image  bursts in the solar corona near the Sun's surface, and localize transient events with sufficient precision to allow follow-on searches at other wavelengths.  In order to achieve arcminute resolution, the longest baseline of the MWA is 3~km.  This distance is within the scale size of a typical isoplanatic patch in the ionosphere, allowing simplying assumptions to be employed for widefield ionospheric calibration and keeping real-time imaging computations within reach of modest computer clusters.


\begin{table}
\begin{center}
\caption{System parameters for MWA}\label{t_system}
\begin{tabular}{lc}
\hline 
Parameter & $^a$Value \\
\hline
Number of tiles 					& 	128  \\
Tile area (m$^2$) 				& 	21.5 \\
Total collecting area (m$^2$) 		& 	2752 \\
Receiver temperature (K)			& 	50 	 \\
Typical sky temperature (K)		&	350$^b$	 \\
Field of view (deg$^2$)			&	610	 \\
Angular resolution (arcmin)		&	2 \\
Frequency range (MHz)			&	80--300  \\
Instantaneous bandwidth (MHz)	&	30.72	\\
Spectral resolution (MHz)			&	0.04 \\
Temporal resolution (s)			&	0.5 \\
Minimum baseline (m)				&	7.7  \\
Maximum baseline (m)			&	2864 \\
Estimated confusion limit (mJy)	&	10 \\
\hline
\end{tabular}
\medskip\\
$^a$For frequency-dependent parameters, listed values are given at 150~MHz \\
$^b$Sky temperature varies considerably with Galactic latitude.  Here, we use typical values from \citet{2009Nijboer} and \citet{2008AJ....136..641R}  \\
\end{center}
\end{table}

The system properties of the MWA are described in detail by \citet{mwa_tech}.  We summarize them here in Table~\ref{t_system}.  The time needed to reach a point source sensitivity of $\sigma_{\rm s}$ for the MWA is given by:
\begin{equation}
t = \left ( \frac{2 k_{\rm B} T}{A_{\rm eff} N \epsilon_{\rm c}} \right )^2 \frac{1}{\sigma^2_{\rm s} B n_{\rm p}}
\end{equation}
where $k_{\rm B}$ is the Boltzmann constant, $T=T_{\rm sky} + T_{\rm rcv}$ is the system temperature, $A_{\rm eff}$ is the effective area of each antenna tile, $N$ is the number of antenna tiles, $\epsilon_{\rm c}$ is the correlator efficiency (assumed to be unity), $B$ is the instantaneous bandwidth, and $n_{\rm p}$ is the number of polarizations.   For continuum imaging at 150~MHz, this reduces to $t \approx 8 \times 10^4 / ( B \sigma_s^2)$ seconds for $B$ in units of MHz and $\sigma_s$ in units of mJy.  The intrinsic source confusion limit  for the MWA angular resolution of $2'$ is estimated to be $\sim10$~mJy at 150~MHz \citep{1974ApJ...188..279C, 1985ApJ...289..494W, 1998AJ....115.1693C, 2002astro.ph..9569S}.  Hence, the MWA will reach the confusion limit in $\sim25$~seconds for continuum imaging or within several hours in a single 40~kHz spectral channel.  

The surface brightness sensitivity of the MWA varies significantly with angular scale.  The MWA antenna layout emphasizes a dense, central core in order to achieve a high surface brightness sensitivity on degree-scales for EoR science.  For a typical EoR spectral resolution of 1~MHz, the brightness temperature uncertainty is less than 150~mK after 1~hour of integration on degree-scales, whereas it is approximately 3~K on arcminute-scales.

In the remainder of the paper, we discuss the four key science themes.  Section~2 describes the epoch of reionisation science.  In Section~3, broad categories of radio astronomy science are discussed within the context of sky surveys.   Section~4 describes the time-domain astronomy that will be possible with the MWA and Section~5 presents the primary solar, heliospheric, and ionospheric science goals.  We conclude in Section~\ref{s_conclusion}.


\section{Epoch of Reionisation}
\label{s_eor}

One of the most significant events in the history of the intergalactic medium (IGM) was the reionisation of hydrogen.  The epoch of reionisation marks the time when the small fraction of matter within galaxies significantly affected their surroundings (and hence future generations of galaxies), ionizing the intergalactic gas and heating it from $10$s of Kelvin to $\sim 10^4\,$K.

The MWA has been designed with the goal of detecting 21~cm emission from the reionisation epoch.  It is one of three new radio arrays specifically targeting this science objective, along with PAPER\footnote{http://eor.berkeley.edu} and LOFAR\footnote{http://www.lofar.org}.   The 21~cm hyperfine transition line of neutral hydrogen is anticipated to provide an ideal probe of reionisation. Its weak oscillator strength (in contrast to the Ly-$\alpha$ line) allows 21~cm photons to propagate from fully neutral environments at high redshifts without appreciable absorption.  This property makes it possible to directly observe the neutral hydrogen gas before and during reionisation.

\begin{figure*}
\begin{center}
\includegraphics[scale=0.75, angle=0, trim=0in 0 0 0, clip=true]{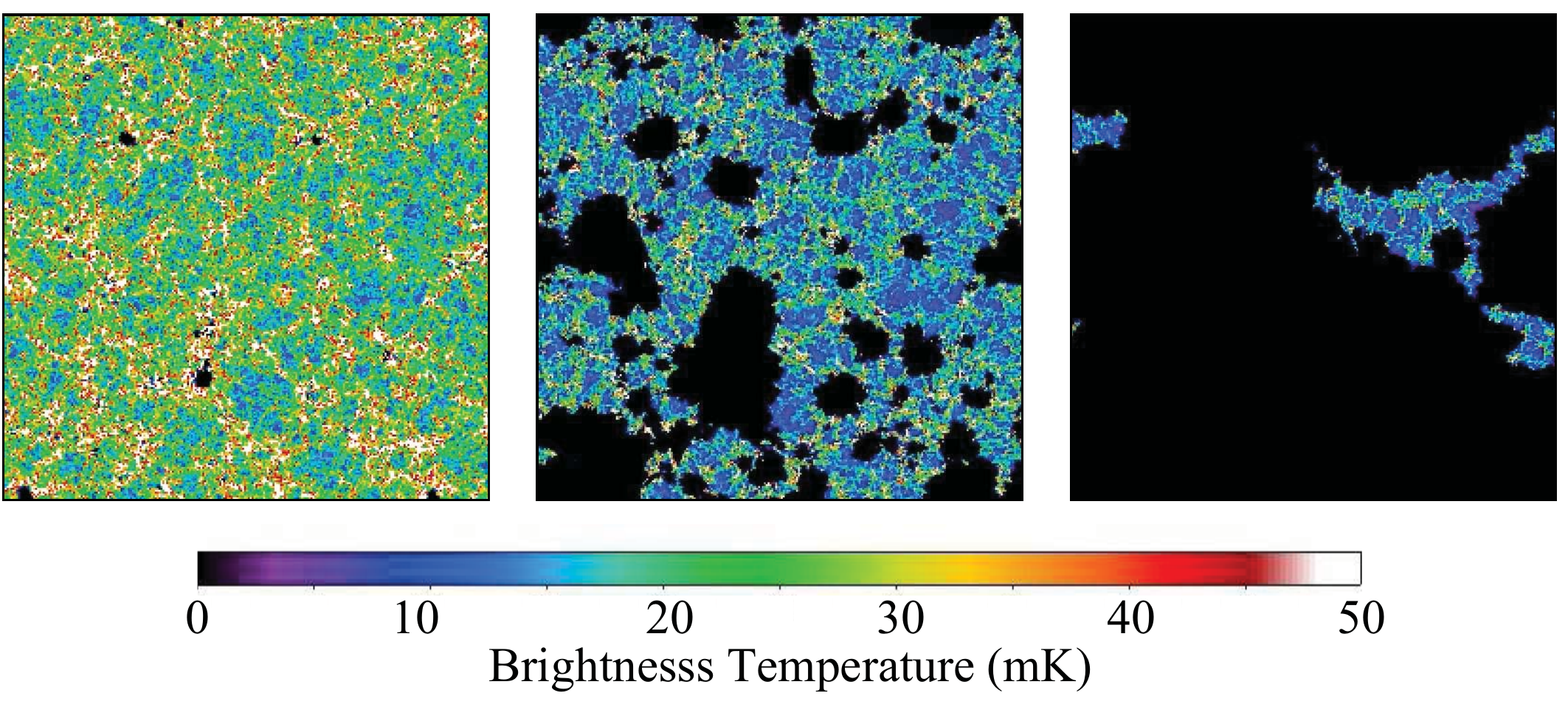}
\caption{Simulated 21~cm brightness temperature maps from simulation S1 by \citet{2007MNRAS.377.1043M} at three epochs calculated under the assumption that $T_S \gg T_\gamma$.  The left panel is an almost fully neutral IGM (before reionisation at $z=12$) such that the $21$cm emission traces the matter density. The middle panel is during reionisation ($x_{HI} \approx 0.5$, which occurs at $z=8$ in this simulation): The dark regions represent ionized regions around galaxies.  Lastly, the right panel shows the residual emission near the end of reionisation ($z= 6$).  Each panel is $100$~cMpc across and subtends $0.5^\circ$ (a small fraction of the MWA field of view).}\label{f_eor_sim}
\end{center}
\end{figure*}

Unlike the continuum emission of the cosmic microwave background (CMB), 21~cm emission contains information in the frequency (or redshift) dimension in addition to in angle, enabling the study of the full history of reionisation.  Redshifted 21~cm emission from neutral hydrogen gas in the IGM should appear as a faint, diffuse background in radio frequencies below $\nu \lesssim 200$~MHz for redshifts above $z>6$, where $\nu=1420/(1+z)$~MHz. At these redshifts, the expected unpolarized differential brightness temperature, $\delta T_{21}$, of the redshifted 21~cm line relative to the CMB is readily calculable from basic
principles \citep[see][for a complete review]{2006PhR...433..181F} and is given by
\begin{eqnarray}
\lefteqn{\delta T_{21}(\vec{\theta}, z) \approx~25~x_{HI} \left ( 1+\delta \right ) \left ( \frac{1+z}{10} \right )^{1/2}} \nonumber \\
& & \times \left [ 1 - \frac{T_\gamma(z)}{T_S} \right ] \left [ \frac{H(z) / (1+z)}{d\nu_{\|}/dr_{\|}} \right ] \mbox{mK}, \label{eqn_eor_temp}
\end{eqnarray}
where $x_{HI}$ is the neutral fraction of hydrogen, $\delta$ is the local over-density in baryons, $T_\gamma(z) = 2.73~(1+z)$~K is the temperature of CMB at the redshift of interest, $T_S$ is the spin temperature that describes the relative population of the ground and excited states of the 21~cm transition, $H(z)$ is the Hubble constant, and $d\nu_{\|}/dr_{\|}$ is the velocity gradient along the line of sight.

All of these factors contain unique astrophysical information. The dependence on $\delta$ traces the development of the cosmic web \citep{1990MNRAS.247..510S}, while the velocity factor creates line-of-sight distortions that can be used to separate different components of the signal \citep{2005ApJ...624L..65B}.  The other two factors depend on the ambient radiation fields in the early Universe. The local neutral fraction, $x_{HI}$, is determined by the ionizing
radiation emitted by galaxies and possibly quasars.  Theoretical studies suggest that the distribution of neutral gas during reionisation is very inhomogeneous (e.g. \citealt{2007MNRAS.377.1043M}).  Finally, the 21~cm spin temperature, $T_S$, is determined primarily by three mechanisms that affect the hyperfine level populations: the ambient radiation at 21~cm due to the CMB with $T_\gamma$, the ultraviolet background that tends to drive $T_S$ to the gas temperature \citep{1952AJ.....57R..31W, 4065250}, and heating from the cosmological X-ray background  \citep{2004ApJ...602....1C}.  By the time the Universe is a few percent ionized, most calculations predict that $T_S \gg T_\gamma$, such that fluctuations in $T_S$ are not important (e.g. \citealt{2006PhR...433..181F}).

Figure~\ref{f_eor_sim} illustrates how the factors in Equation~\ref{eqn_eor_temp} combine to create the observable 21~cm signal.  The panels in this figure are from the simulations of \citet{2007MNRAS.377.1043M} and are $100$~cMpc across or $\sim 0.5^\circ$ in angle (a small fraction of the $\sim 20^\circ$ MWA field of view). The left panel of Figure~\ref{f_eor_sim} shows the expected signal just before reionisation begins, when the fluctuations are dominated by local perturbations in the matter density.  In the middle panel, reionisation is underway ($x_{HI} \approx 0.5$), and the large-scale voids in the map signify ionized regions.  After reionisation has largely finished (right panel), only localized pockets of neutral hydrogen remain.

The MWA will attempt to probe structures in the redshifted 21~cm background in order to reveal the astrophysics of reionisation.  The MWA is expected to be sensitive to the power spectrum of the 21~cm signal over the redshift range $6<z\lesssim10$ (Section~\ref{s_eor_pow}).  Its observations have the potential to characterise the nature of the sources that are responsible for ionizing the IGM, chart the evolution of the global neutral fraction, and probe the nature of quasar emissions by constraining the properties of their ionized proximity zones.  
%

\subsection{21 cm Power Spectrum}
\label{s_eor_pow}

\begin{figure}
\begin{center}
\includegraphics[scale=0.5, angle=0, trim=1.01in 0.1in 0 0.7in, clip=true]{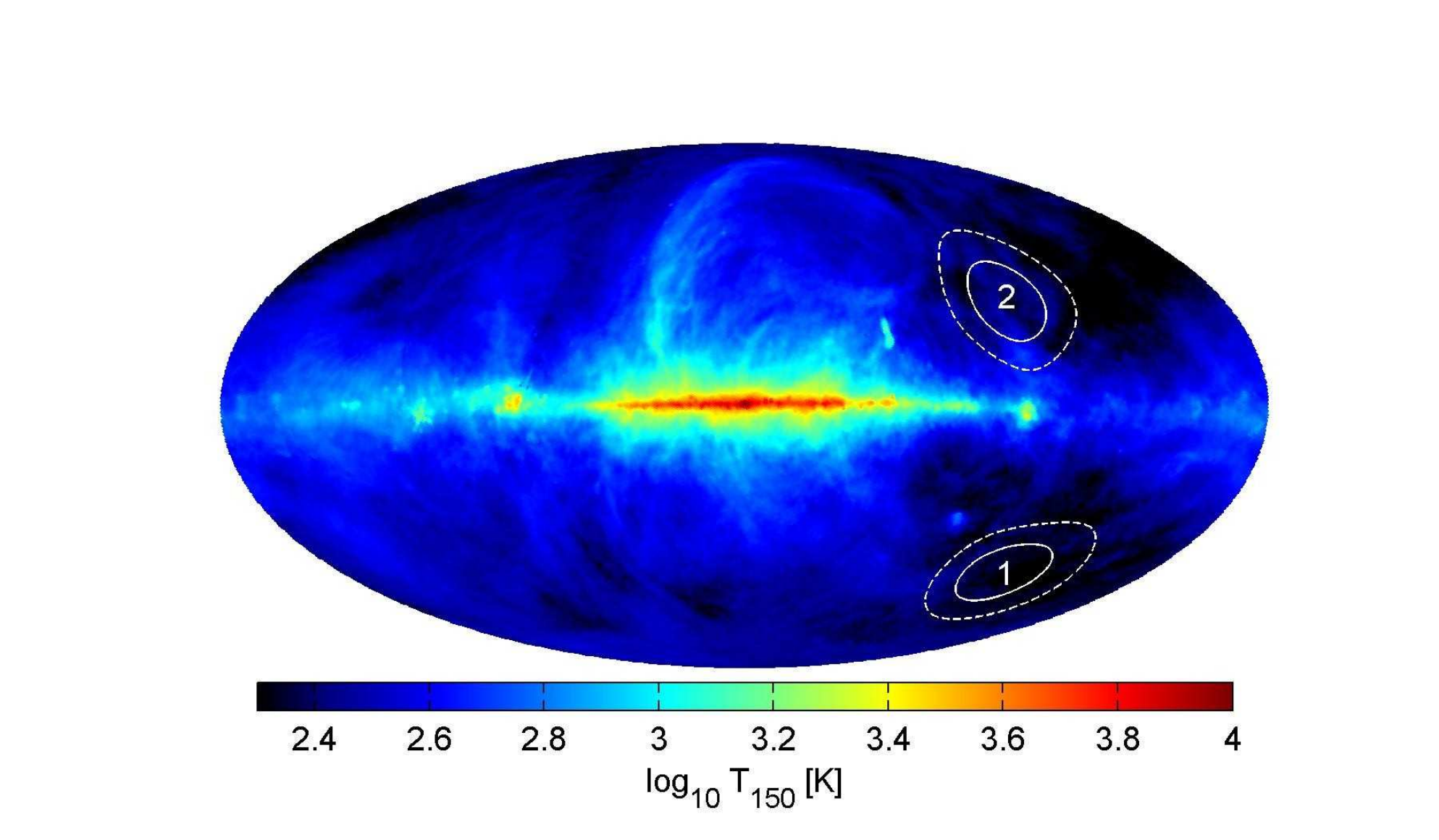}
\caption{Projected all-sky brightness temperature map of \citet{2008MNRAS.388..247D} at 150~MHz.  The primary (1) and secondary (2) target regions for EoR observations are centred at $\alpha,\delta=60^\circ, -30^\circ$ ($\ell,b=228^\circ, -49^\circ$) and $\alpha,\delta=155^\circ, -10^\circ$ ($\ell,b=253^\circ, +38^\circ$), respectively. The solid white curves indicate the 50\% response power contour of the primary antenna tile beams for reference and the dashed lines show the 10\% power contour.}\label{f_targetfields}
\end{center}
\end{figure}

Initial MWA observations aimed at detecting 21~cm emission will focus on two target regions at relatively high Galactic latitudes (see Figure~\ref{f_targetfields}).  Over the course of one year, approximately 1000~hours of optimal nighttime observations will be divided between these two regions.  The signal-to-noise ratio (S/N) in each pixel will be $\lesssim1$, with the highest S/N on the largest angular modes.  Hence, rather than imaging the 21~cm brightness temperature in the target fields, reionisation science with the MWA will focus on statistical measures of the 21~cm emission.  The number of independent measurements in a raw, multi-frequency MWA data cube will be $\sim10^8$.  These samples record largely Galactic synchrotron and free-free emission ($\gtrsim200$~K) and extragalactic continuum sources ($\sim50$~K), with only a small contribution from the redshifted 21~cm signal ($\sim25$~mK).  The primary approach planned for the MWA is to calculate the power spectrum of spatial fluctuations of the 21~cm signal.  

The three-dimensional power spectrum of redshifted 21~cm emission is the square of the Fourier transform of $\delta T_{21}$ (Eq. ~\ref{eqn_eor_temp}).  It has a large spherically-symmetric component due to the isotropy of space, but has slightly more power for modes that point in the line-of-sight direction that results from peculiar velocities.  The predicted components in the full three-dimensional power spectrum can be decomposed into terms with different powers of $\mu\equiv\hat{k}\cdot\hat{n}$, where $\mu$ is the cosine of the angle between the line of sight and the wavevector \citep{1987MNRAS.227....1K, 2005ApJ...624L..65B, 2006ApJ...653..815M}. Thus,
\begin{equation}
P_{21}\left(\vec{k}\right) = P_{\mu^0}(k) + \mu^2 P_{\mu^2}(k) + \mu^4 P_{\mu^4}(k). \label{eqn_eor_mu}
\end{equation}
During reionisation, the $P_{\mu^0}$ term typically dominates, and the most important contribution to $P_{\mu^0}$ is the fluctuations in the neutral fraction \citep{2006ApJ...653..815M}.

Averaging $P_{21}(\vec{k})$ over angle to yield $P_{21}({k})$ loses some statistical information, but at the sensitivity of MWA this loss is minor during reionisation.  With one year of the planned observing program, the MWA can constrain $P_{21}({k})$ at a given redshift and across a range of scales spanning $0.01\lesssim k \lesssim1$~cMpc$^{-1}$ with total $S/N \approx14$\citep{2004ApJ...615....7M, 2006ApJ...638...20B, 2006ApJ...653..815M, Beardsley2012}.  The sensitivity of the MWA to the spherically-averaged power spectrum is shown in Figure~\ref{f_eor_sim_pow}.

The properties of the power spectrum, $P_{21}({k})$, during reionisation are dominated by the characteristics of the ionized bubbles (Figure~\ref{f_eor_sim}, middle panel) that are determined by the ionizing radiation produced by the first galaxies and quasars.  Figure~\ref{f_eor_sim_pow} illustrates how $\Delta_{21} \equiv (k^3 P_{21}(k)/2 \pi^2)^{1/2}$ contains information about the reionisation process.  It plots $\Delta_{21}$ for the fiducial model of \citet{2008ApJ...680..962L} at several neutral fractions.  The shape of the power spectrum evolves considerably as the hydrogen in the IGM becomes ionized, with power shifting from small scales initially, to larger scales later when large ionized bubbles have formed.  Eventually, the amplitude decreases as hydrogen reionisation completes.

The sensitivity forecasts shown in Figure~\ref{f_eor_sim_pow} utilize both the primary and secondary target fields in combination, assuming that each field has been observed over an 8~MHz bandwidth (corresponding to $\Delta z\approx0.5$).  The full 32~MHz instantaneous bandwidth of the MWA will yield four such measurements spanning a contiguous redshift range of $\Delta z \approx 2$.  Initial observations will focus on the lower redshifts ($6<z<8$), where the sensitivity of the instrument is highest due to the lower sky temperature.  Observations at low redshifts will crucially demonstrate whether the foreground subtraction is successful, since the quality of the foreground subtraction will be readily apparent at $z<6$, where no 21~cm signal should be detectable by the MWA since $x_{HI}\lesssim2\%$ after reionisation \citep{2008MNRAS.383..606W}.

\subsubsection{Sources of reionisation}
\label{s_eor_sources}

The 21~cm sky contains a wealth of information about the properties of the ionizing sources because they strongly affect the topology of the EoR.  Reionisation scenarios dominated by stars in galaxies produce well-defined ionized regions, while ``miniquasars'' (small accreting black holes) yield more diffuse features \citep{2005MNRAS.360L..64Z} and reionisation processes involving decaying/annihilating dark matter particles would be very inhomogeneous \citep{2009arXiv0904.1210B}.  The 21~cm observations could even distinguish between different stellar reionisation scenarios \citep{2008ApJ...680..962L}. For example, more massive galaxies produce larger ionized regions \citep{2004ApJ...613....1F,2007MNRAS.377.1043M}.  If it is able to constrain the shape and evolution of the power spectrum, the MWA will reveal the nature of some of the first cosmological objects.   

\subsubsection{The Neutral Fraction}
\label{s_eor_neutral}

Existing observations have provided tantalizing hints about reionisation, but even more unanswered questions.  CMB observations find that the average redshift of reionisation is $z = 10 \pm 2$ \citep{2009ApJS..180..330K}, and quasar absorption spectra imply that the Universe was totally ionized by $z=6$, one billion years after the Big Bang.  In addition, several studies have interpreted the large saturated regions at $z>6$ in these absorption spectra as  evidence for the end of reionisation (e.g., \citealt{2006AJ....132..117F}), but this conclusion is disputed \citep{2007ApJ...662...72B, 2010MNRAS.407.1328M}.

\begin{figure}[t]
\begin{center}
\includegraphics[scale=0.8]{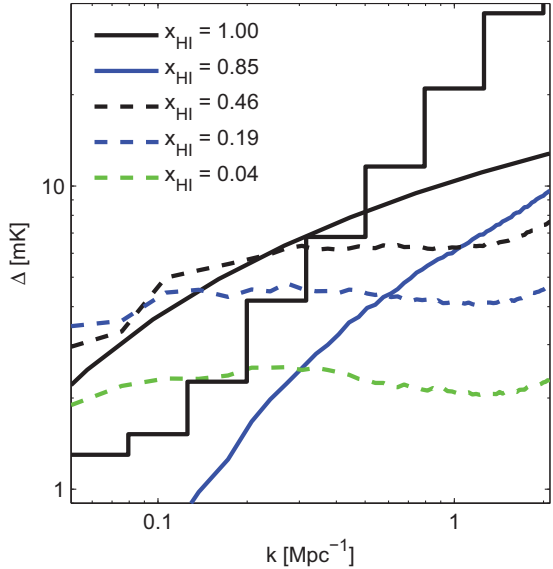}
\caption{MWA thermal uncertainty at $z=8$ with simulated 21~cm power spectra.  The black stepped line is the $1\sigma$ thermal uncertainty modeled for a combined observation of the primary and secondary EoR fields with a total of 1600~hours \citep{Beardsley2012}.  The 21~cm curves are calculated using the fiducial simulation described by \citet{2008ApJ...680..962L}.  The power spectrum is very steep at the beginning of reionisation ($x_{HI}\gtrsim0.8$) and falls rapidly at the scales probed by the MWA.  Large-scale power returns as large ionized regions form by $x_{HI}\approx0.5$ and then it falls again with decreasing neutral fraction.  The MWA should be able to detect the power spectrum at $z=8$ for any appreciable neutral fraction outside of the range $0.95\gtrsim x_{HI} \gtrsim0.8$.  }
\label{f_eor_sim_pow}
\end{center}
\end{figure}

Detection of the 21~cm power spectrum will yield measures of $\bar{x}_{HI}(z)$ through predictable features that depend on the neutral fraction.  As evident in Figure~\ref{f_eor_sim_pow}, the intensity of $10$~cMpc-scale fluctuations in 21~cm brightness temperature briefly fades when reionisation begins and the first galaxies ionize the overdense regions within a couple of cMpc of them.  Next, the ionized regions surrounding individual galaxies overlap, creating $\sim 10$~cMpc ionized regions and an increase in power on these scales.  Finally, all of the fluctuations fade as these bubbles overlap, swallowing the neutral hydrogen that remains.  Each of these trends can be used to disentangle $\bar{x}_{HI}(z)$.

For many possible sources that have been considered in the literature (such as galaxies, quasars, and mini-halos), the amplitude of the power spectrum peaks at $\bar{x}_{HI}\approx50$\%.  The MWA could probe the global ionization history by tracking the evolution of this amplitude and shape of the 21~cm power spectrum.  Using these features, the volume filling factor of neutral hydrogen (comparable to the mass-averaged neutral fraction) can be constrained after approximately two years of observations to between $0.4 < x_{HI} < 0.75$ at $2\sigma$ significance for the redshift where the power spectrum achieves its maximum amplitude \citep{2008ApJ...680..962L}.  While unlikely, it is possible that the IGM is highly ionized over the range of redshifts measured by the MWA. \citet{2008ApJ...680..962L} found that a null detection with the MWA at $z\leq8$ would be able to constrain the neutral fraction to $\lesssim4$\%.

Observations of individual quasar HII regions offer an additional probe of the global neutral hydrogen fraction since the contrast between the ionized HII region and neutral IGM can provide a measurement of the neutral fraction at the redshift of the quasar. \citet{2008MNRAS.390.1496G} have found that the shape of the quasar HII regions can be recovered from foreground subtracted MWA maps by using the entire data cube to help determine the contrast between the external IGM and ionized interior.  

\subsection{Quasar HII Regions}
\label{s_eor_quasars}

\begin{figure*}
\begin{center}
\includegraphics[scale=0.85, trim=0 0 0 0in, clip=true]{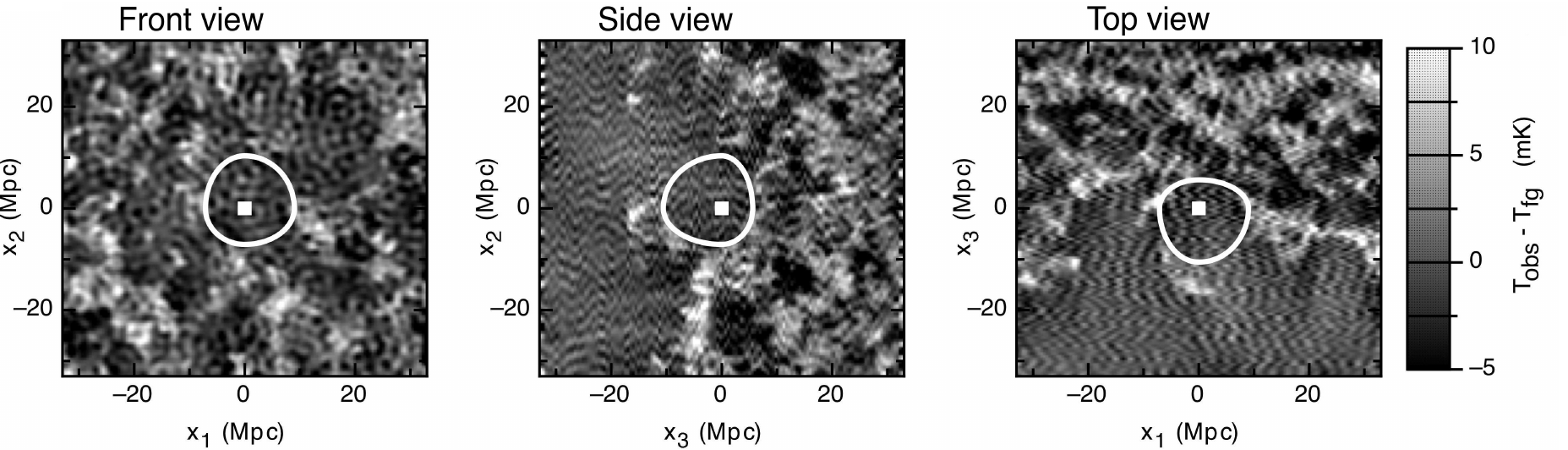}
\caption{Simulated recovery of a quasar HII region in an evolving IGM, as observed by an MWA-like instrument \citep{2008MNRAS.390.1496G}.  Foregrounds were simulated and removed for the maps shown.  The data cube is only a small part of a full MWA field and spans $\sim3.3$ degrees on a side and 33~MHz deep.  The three panels show different slices through the centre of the box when viewed from the front, side, and top.  Each slice is 6~cMpc thick, which corresponds to $\sim3$~MHz along the $x_3$-axis. The extracted shape of the HII region is shown, based on a 9~parameter model.  The quasar was assumed to have a lifetime of $4\times10^7$~years centred on $z=6.65$ and to have produced an HII region with a radius of $34$~cMpc at that redshift. The mass-averaged IGM neutral fraction at the redshift of the quasar is $\sim15\%$.
}\label{f_eor_quasar}
\end{center}
\end{figure*}

The detection of large quasar-generated HII regions in redshifted 21~cm emission late in the epoch of reionisation may be possible with the MWA \citep{2005ApJ...634..715W, 2008MNRAS.387..469W}.  \citet{2008MNRAS.390.1496G} have modeled the observation of quasar HII regions in a partially ionized IGM.  Figure~\ref{f_eor_quasar} shows a simulated example of a recovered quasar HII region with the MWA after approximately one observing season.  The recovery is possible only for quasars with positions known \textit{a priori}.  Although the EoR fields will not overlap the Sloan Digital Sky Survey (SDSS) fields, where the existing known $z\approx6$ quasars reside, the contemporaneous SkyMapper optical survey \citep{2007PASA...24....1K} should find around one quasar at $z > 6.25$, and around 5 quasars at $z > 6$ per MWA field.

Like all 21~cm signatures of the reionisation era, the detection of a quasar HII region will need to overcome the difficulties associated with the removal of bright Galactic and extragalactic foregrounds. \citet{2008MNRAS.390.1496G} find the primary effect of continuum foreground removal is to reduce contrast in the image. In particular, contributions to the 21~cm intensity fluctuations that have a scale length comparable to the frequency bandpass of the observation are removed by foreground subtraction.  This loss of contrast, however, does not severely affect the ability of 21~cm observations to measure the size and shape of the HII region. To demonstrate this quantitatively, \citet{2008MNRAS.390.1496G} modeled the HII region shape as a spheroid described by six parameters and showed that the shape recovered following foreground removal agrees well with the shape derived directly from fitting using the input HII region model.

Using the recovered best-fit shape of the HII region, the global neutral fraction of hydrogen in the IGM can be measured directly from the contrast in intensity between regions that are within and beyond the HII region. However, since foreground removal lowers the observed contrast between the HII region and the IGM, such a measurement of the neutral fraction requires a correction factor. The value of this correction factor depends on the reionisation history and the polynomial used for foreground subtraction and can be accounted for using astrophysical models.

\subsection{Non-Gaussian Statistics}
\label{s_eor_non}

During the reionisation epoch, fluctuations in the 21~cm emission are expected to be highly non-Gaussian due to galaxy bias and the patchy structure imparted by the ionized bubbles in the IGM.  The power spectrum (or two-point correlation function), therefore, is not a sufficient descriptor of the properties of the field, and additional statistical measures can provide complementary information.  Several extensions to the power spectrum have been proposed in this context, including the 21~cm-galaxy cross spectrum \citep{2006PhR...433..181F, 2007MNRAS.380.1087W, 2009ApJ...690..252L}, directly measuring the one-point probability distribution function (PDF) of the 21~cm brightness temperature distribution as a function of smoothing scale \citep{2006MNRAS.372..679M, 2010MNRAS.406.2521I, 2011MNRAS.413.2103P} or its higher moments \citep{2007MNRAS.379.1647W}, the three-point correlation function (bispectrum), matched filters \citep{2008MNRAS.391.1900D}, the threshold clustering function, and other derived metrics \citep{2008MNRAS.384.1069B}.  

\subsection{Foreground Subtraction}
\label{s_eor_sub}

The ability of the MWA to achieve its reionisation science is entirely contingent on the successful separation of the 21~cm signal from astrophysical foregrounds.  Hence, developing and demonstrating the techniques required to subtract foregrounds is a critical research objective.  

\subsubsection{Reference Pipeline}
\label{sec:pipeline}

The baseline foreground subtraction strategy is a multi-stage process consisting of the following principal stages:

\textit{1 - Bright source subtraction.}  The first step in foreground subtraction is the removal of bright points sources (e.g. extragalactic AGN). This is necessary because the sidelobes of these sources mask the 21~cm signal.  Recent theoretical studies have indicated that bright point source subtraction must be complete for sources with flux greater than $S\gtrsim100$~mJy in order to allow for successful subsequent Galactic continuum foreground subtraction \citep{2009MNRAS.394.1575L}.  The brightest sources in the field will be ``peeled'' (self-calibrate and subtract)  from the raw visibility measurements \citep{2008ISTSP...2..707M}.  Efficient techniques to remove point sources from gridded and integrated maps are also under investigation \citep{2011PASA...28...46P, 2011MNRAS.413..411B}.  

\textit{2 - Diffuse continuum subtraction.}  A variety of mechanisms contribute to diffuse continuum emission in the low-frequency radio sky including Galactic synchrotron emission, Galactic free-free emission, extragalactic free-free emission, and a sea of faint extragalactic point sources that are too numerous to identify individually in maps and are, thus, said to be ``confused''.  All of these foregrounds are significantly stronger than the redshifted 21~cm signal and have considerable angular structure on the sky \citep[see Figure~\ref{f_targetfields} and ][their Figure 5]{2005ApJ...625..575S}.  But the emission mechanisms of these foregrounds produce smooth spectra that follow power-law profiles, making them possible to separate from the redshifted 21~cm emission by their long spectral coherence scales compared to the small, $\sim1$~MHz spectral coherence of the reionisation signal.  A variety of techniques have been proposed to accomplish this separation including polynomial fits along the spectral axis of each pixel in an image map or its uv-map conjugate, Minkowsky-functionals, and non-parametric techniques.  The baseline approach planned for the MWA is to fit and subtract a third-order polynomial from each pixel in the uv-domain representations of the bright source cleaned maps.  This technique has been demonstrated for the MWA in the limit of perfect instrumental calibration \citep{2009MNRAS.398..401L, 2009ApJ...695..183B, 2009MNRAS.394.1575L}.  

\textit{3 - Polarized leakage subtraction.} In addition to the foreground issues discussed thus far, gain and phase calibration errors and non-ideal feeds conspire to contaminate all Stokes parameters further. This leakage leads to some portion of the complex linear polarization signal finding its way into the Stokes $I$ intensity maps that will be used for the redshifted 21~cm measurements.  By using  rotation measure (RM) synthesis (\citealt{2005A&A...441.1217B}, see also \S~\ref{sec:geg_faraday}) to exploit the Fourier relationship between the polarized signal and the Faraday dispersion function, it is possible to identify components of polarized signal with specific Faraday depth.  The three-dimensional Stokes $I$ data set is first re-sampled along the frequency axis into the natural $\lambda^2$-basis for Faraday rotated emission. A one-dimensional Fourier transform is then applied along each line of sight. Any feature observed with a non-zero Faraday depth in the three-dimensional RM synthesis map will, by definition, be polarized contamination. By applying an inversion process together with knowledge of the instrument's specific leakage behaviour, this contamination can be subtracted \citep{2011MNRAS.418..516G}. 

\textit{4 - Statistical template fitting and power spectrum estimation.} Many classes of errors and residuals remaining in the maps after the foreground cleaning stages described above can be modeled given sufficient knowledge of the instrumental response \citep{2006ApJ...648..767M, 2009ApJ...695..183B, 2010ApJ...724..526D, 2012arXiv1202.3830M}.  The statistical properties of these residuals can be exploited to provide an additional layer of foreground removal by fitting templates of the residual statistical structures during the final redshifted 21~cm power spectrum parameter estimation.  This technique is now commonly employed in CMB power spectrum analysis to account for the angular power of faint point sources and the effects of gravitational lensing.

\subsubsection{Unified Parameter Estimation}  

The stages of foreground subtraction outlined above can be performed discretely, resulting in a series of output products, the last of which is used as the input into power spectrum estimation and other statistical tests to extract information from the redshifted 21~cm signal.  Alternatively, they can be performed in a unified matrix framework that simultaneously fits models for the foreground contributions and the 21 cm statistical signal \citep{2011PhRvD..83j3006L, 1997PhRvD..55.5895T}.  The latter has the advantage of producing an optimal result that minimizes both bias (from thermal noise and foreground contamination) and correlations between parameters, and yields the smallest possible error bars, but at the expense of a large computational overhead.  Advances in processing power and new efforts at sparse matrix reduction have begun to enable the trial application of unified parameter estimation in simulations and preliminary data, and it is expected this approach will ultimately be the preferred technique for data analysis.

\section{Galactic and Extragalactic Science}
\label{s_geg}

The MWA offers several exciting science opportunities under the collective banner ``Galactic and Extragalactic'' science.  The major input into this science will come from a deep all-sky survey over the MWA frequency range with full polarimetry and spectral resolution.  This survey is a critical undertaking as it will also be used by the other science programs as a basis for an all-sky foreground model and the identification of calibrators. The data will also be combined with other new surveys, including GALFACTS and POSSUM (1.4 GHz), S-PASS (2.4 GHz), and Planck (27 GHz to 1 THz) surveys.  

The following sections describe a selection of Galactic and extragalactic experiments for which the MWA is well suited. These projects capitalize on three advantages of the MWA's low observing frequencies: (1) non-thermal sources will be very bright; (2) Faraday rotation and depolarization will be very strong; and (3) along some sightlines free-free absorption will be significant. 

\begin{figure*}
\begin{center}
\includegraphics[scale=0.9]{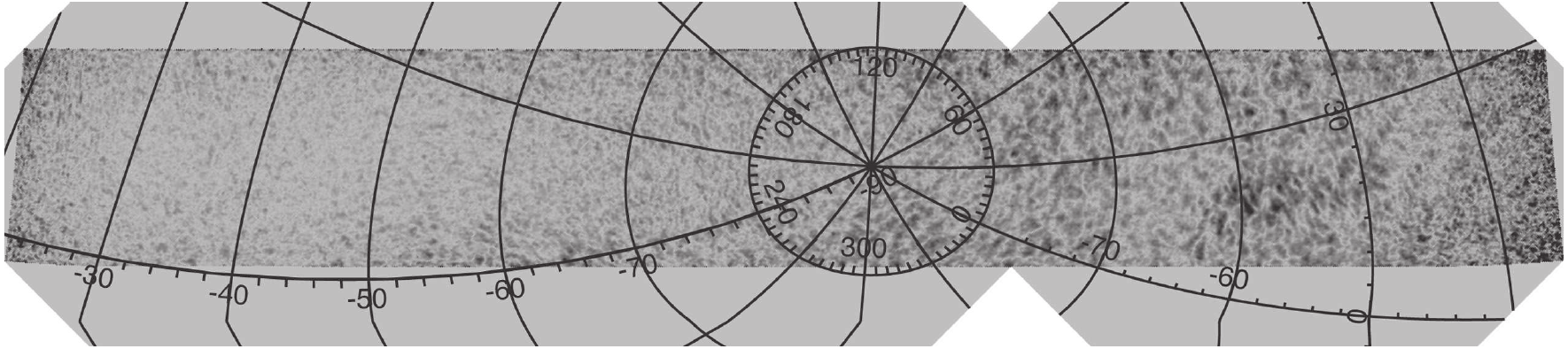}
\end{center}
\caption{Polarized intensity at Faraday depth zero at 188~MHz from the MWA development system \citep{Bernardi13}. The field spans 2400 square degrees with 15 arcminute angular resolution and is roughly centered on the southern Galactic pole, extending to within 20~degrees of the Galactic plane.   The radial lines are Galactic latitude and the circular arcs indicate Galactic longitude.  Flux density is plotted on a linear color scale from 0 to 0.2 Jy~beam$^{-1}$~RMSF$^{-1}$ (white to black), where RMSF is the rotation measure spread function.  Polarized maps from the MWA showing similar complex structures and discontinuties that often have no counterparts in total intensity will be used to determine the properties of both ordered and turbulent magnetic fields in the Milky Way's ISM (see Figure~\ref{fig:turbulence}).}
\label{fig:wsrt_pol}
\end{figure*}

\subsection{The Cosmic Web}
\label{sec:geg_igm}

Theory suggests that the infall of baryons into the cosmic web results in high Mach number intergalactic shocks (e.g., \citealt{1999ApJ...514....1C, 2003ApJ...593..599R}). When combined with seed magnetic fields, efficient cosmic ray acceleration is possible, resulting in steep-spectrum synchrotron emission \citep{2004ApJ...617..281K, 2006MNRAS.367..113P,  2008MNRAS.385.1211P,  2011ApJ...735...96S, 2012MNRAS.423.2325A}. This may be a ubiquitous component of the extragalactic radio sky \citep{2004ApJ...617..281K}, and recent simulations suggest that the majority of the luminous radio shocks can be found in the outskirts of massive galaxy clusters \citep{2012MNRAS.423.2325A}. The MWAs centrally condensed antenna configuration provides the theoretical sensitivity required to make the maps of this relativistic cosmic web. However, the accurate subtraction of discrete sources requires long baselines and high image fidelity \citep{2011JApA...32..577B}. This may limit direct to the densest portions of the web, though statistical detection may be possible \citep{2004ApJ...617..281K, 2010MNRAS.402....2B}.

\subsection{Radio Relics and Clusters}
\label{sec:relics}

The MWA is also well suited for studying the diffuse, steep-spectrum synchrotron emission from clusters of galaxies. These objects, which include the centrally located 'radio halos' and peripheral 'radio relics' \citep{2004rcfg.proc..335K}, are probes of relativistic electrons and large-scale magnetic fields in the intra-cluster medium (ICM; \citealt{2008SSRv..134...93F, 2009astro2010S.253R}). Possible mechanisms for accelerating the cosmic ray electrons responsible for this emission are turbulent re-acceleration following a cluster merger \citep{2001MNRAS.320..365B, 2001ApJ...557..560P} and hadronic collisions between cosmic-ray and thermal protons \citep{1980ApJ...239L..93D, 1999APh....12..169B}. Low frequency observations of clusters can probe these different models \citep{2010A&A...509A..68C, 2012_Cassano}.  The MWA's surface-brightness sensitivity will enable it to detect very steep-spectrum halos and relics. Predictions for the number of such halos and relics detectable with other low frequency instruments are in the range of several hundred out to $z < 0.6$ \citep{2012_Cassano}.

At 150~MHz, the MWA will resolve 1~Mpc structures out to a redshift of 0.2 and 500~kpc structures out to $z<0.07$. As with the detection of the cosmic web, confusion will be the major obstacle. Use of complementary higher resolution observations to model and subtract confusing sources or will be required \citep{2011JApA...32..577B}. The TIFR GMRT Sky Survey (TGSS) that covers 32,000 square degrees north of declination~$-30^\circ$ with angular resolution~$20''$ presents a good comparison survey to the MWA for this work.

\subsection{Faraday Tomography and Magnetic Fields}
\label{sec:geg_faraday}

The Milky Way and most other spiral galaxies all show well-organized, large-scale magnetic fields, but the processes that amplify and maintain these fields are still not understood. On smaller scales, fluctuations in these magnetic fields regulate the dissipation of large-scale gas flows via a turbulent cascade. However, it is as yet unclear what processes generate turbulence in the interstellar medium (ISM), and how the subsequent spectrum and outer scale of turbulence depend on the surrounding conditions.  

Polarized radio waves experience Faraday rotation when they propagate through a magnetized plasma. The amount of rotation depends on the square of the wavelength, and on the path integral of $n_e B_{||}$ along the line of sight (where $n_e$ is the electron density of the ISM, and $B_{||}$ is the magnetic field strength parallel to the line of sight).  By exploiting this phenomenon, the MWA will provide a new view of the large- and small-scale properties of otherwise unobservable magnetic fields in the Milky Way.

At the low observing frequencies of the MWA, small rotation measures have substantial effects on the distribution of polarized emission from the sky. This emission directly traces invisible foreground magnetic fields, which imprint their structure on the smooth polarized synchrotron emission from the Milky Way via foreground Faraday rotation. By imaging the polarized sky at a range of frequencies, the MWA will provide direct measurements of the turbulent cascade in the local interstellar medium, and will reveal the detailed magnetic geometries of discrete large-scale structures such as the Local Bubble, the North Polar Spur and the Magellanic Clouds.  Using the high spectral resolution of the MWA, we can also apply RM synthesis techniques (see also \S~\ref{sec:pipeline}) to isolate overlapping RM components.  An example is provided in Fig.~\ref{fig:wsrt_pol}, which shows the turbulent pattern seen at 188~MHz in a large region of the sky.  The MWA will provide a unique tomographic view of magneto-ionized gas, allowing the identification of multiple emitting and rotating regions along the same line of sight.

The MWA is expected to make the largest contributions to Galactic science by observing at mid- and high-latitude regions.  Depolarisation is already severe in the inner Galactic plane at 1.4~GHz \citep{1990A&AS...83..539R,wlr06,hgm06} and it is likely that much of the inner Galactic plane will appear completely depolarised below 300~MHz, although \citet{2009A&A...500..965B} have detected polarized emission at arcminute scales at 150~MHz. Regions at mid and high latitudes, on the other hand, show considerable diffuse Galactic polarised structure at 350~MHz \citep{hkb03,dkh06,2009A&A...494..611S}. The high
sensitivity of the MWA to extended structure will enable it to map the diffuse
polarised emission down to very low intensities, thus making visible
weak Galactic magnetic fields in the local interstellar medium and
Galactic extended disk and halo that are undetectable at higher
frequencies. RM synthesis across the entire MWA frequency band will
result in a RM resolution of 0.3~rad~m$^{-2}$, which gives
unprecedented detail in RM structure, allowing detailed studies of
e.g.\ interstellar magnetised turbulence, or magnetism in old
supernova remnants. Increasing depolarisation as a function of
decreasing frequency will give independent measurements of the amount
and scale of fluctuations in the magneto-ionised medium \citep{1998MNRAS.299..189S,gdm01}.

\begin{figure*}
\begin{center}
\includegraphics[trim=1.5in 0in 0in 0in, clip=true, scale=0.62, angle=90]{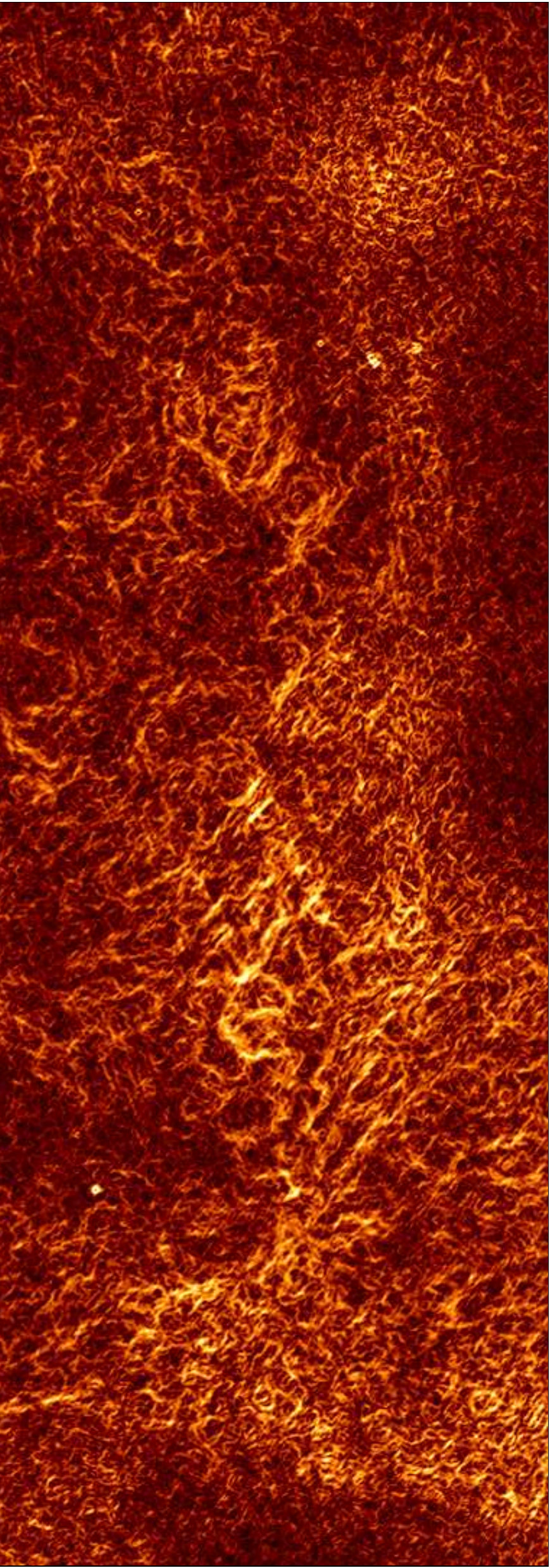}
\end{center}
\caption{Gradient image of linear polarization, $|\nabla \bf{P}|$, for an 18~deg$^2$ region of the Southern Galactic Plane Survey \citep{2011Natur.478..214G} that has been used to study magnetized turbulence in the interstellar medium.  Similar statistical techniques will be applied to MWA polarized maps to investigate magnetic field strength, Mach number, Reynolds number, and othe properties in the interstellar medium.}
\label{fig:turbulence}
\end{figure*}

A powerful new additional probe of the magnetised ISM comes from the polarisation gradient, in which the spatial derivatives of Stokes $Q$ and $U$ are combined to reveal locations on the sky at which the electron density or magnetic field strength show sharp fluctuations \citep{2011Natur.478..214G}. Application of the polarisation gradient at higher frequencies reveals a complex tangled network of narrow filaments (see Figure~\ref{fig:turbulence}). These structures are a superset of the ``polarisation canals'' seen in many radio polarisation images \citep{2004A&A...421.1011H, 2007EAS....23..109F, 2009A&A...494..611S}, and directly reveal the locations of shocks, cusps and shear in the turbulent ISM. Calculation of the polarisation gradient for magnetohydrodynamic simulations of interstellar turbulence shows that the intensity, alignment and morphology of these filaments depend sensitively on input parameters such as the sonic Mach number and magnetic field strength \citep{2012ApJ...749..145B}. By applying statistical metrics such as the genus and kurtosis, one can directly compare observations of the polarisation gradient with a grid of simulations, and thereby extract quantitative information on the otherwise unobservable turbulent parameters of the ISM.

Two effects distinguish the applicability of this technique for MWA data. First, the ``polarisation horizon'' \citep{2003ApJ...585..785U} at the MWA's low observing frequencies is very nearby ($\sim 100$~pc, depending on observing frequency and Galactic coordinates; \citealt{Bernardi13}), meaning that we can probe small-scale local features. Second, the change in electron density or magnetic field strength needed to produce a fluctuation in $Q$ or $U$ is very small (since the amount of Faraday rotation is proportional to $\lambda^2$). For these two reasons, application of the polarisation gradient to MWA data should provide a detailed view of turbulence in the solar neighbourhood, and will be able to extend studies of the global turbulent cascade to smaller amplitudes and smaller angular scale than has been possible from previous observations.

Rotation measures from extragalactic polarised point sources can be used to probe the intervening medium between the source and the observer, where generally the Galactic RM  component dominates. Little is known about the polarisation percentages of extragalactic point sources at low frequencies. At 1.4~GHz, the expected number of linearly polarised sources down to a $3\sigma$ sensitivity of 1~mJy is about $1.2\times10^4$ per steradian \citep{2004NewAR..48.1289B}. The number of polarised sources may decrease at low frequencies due to increased internal depolarisation.   At 1.4	~GHz, \citet{2002A&A...396..463M} found a median polarisation fraction of 2.2\% for sources above 80~mJy in the NVSS and \citet{2009ApJ...702.1230T} have compiled a catalog from the NVSS of over 30,000 polarised sources detected at better than $8\sigma$ and selected for polarisation fraction larger than 0.5\%.  The average polarisation fraction of sources in this catalog is $\sim7$\%.  At 350~MHz \citet{2009A&A...494..611S} found an average polarisation fraction of 3\% for polarized sources brighter than 3~mJy (see also \citealt{hav03b}).  \citet{Bernardi13}, have surveyed 2400 square degrees at 188~MHz using the 32-element MWA development system.  They achieved noise levels of 15~mJy in polarisation  and found only one polarized source, PMN J0351-2744, which had a polarization fraction of 1.2\% and an RM of~+34~rad~m$^{-2}$.  The source was detected at 20~$\sigma$.  They also placed an upper limit of $\sim1$\% on the polarisation fraction of over one hundred other bright sources above 4~Jy and detected diffuse polarised emission from the Galaxy with a peak level of 13~K.    

Beam depolarization is likely to be partially responsible for the lack of detected sources by \citet{Bernardi13}  since the MWA demonstrator had only 15~arcmin resolution.  Using 1\% polarization fraction as a lower limit, along with source flux counts from the 6C survey \citep{1988MNRAS.234..919H} and the ratio of polarised sources to unpolarized sources of 14\%  found by \citet{2011ApJ...733...69B}, we estimate the number of sources that will be detectable by the MWA assuming a fiducial sensitivity of 1~mJy to be approximately $\sim1$ source per square degree.   These extragalactic source RMs will provide a ``grid'' \citep{2004NewAR..48.1289B, 2009ApJ...702.1230T} that can be used to model large-scale magnetic fields in the Milky Way and, if deeper observations can provide a sufficiently dense grid, in nearby galaxies. Provided that polarisation calibration is accurate enough, the MWA will observe point source RM values with superb precision.

\subsection{The Magellanic Clouds}
\label{sec:magellanic}

The Magellanic System, consisting of the Large Magellanic Cloud (LMC), Small Magellanic Cloud (SMC), and other components, such as the Magellanic Stream and the Bridge and Leading Arm, is the Milky Way's most significant neighbour. The Clouds and the Milky Way are engaged in a three-body interaction allowing us to study at close hand the effect of dynamical interaction processes on the physical structure, star-formation, magnetic fields and chemistry of such galaxies.  The location of the Clouds in the deep southern sky makes them ideal candidates to observe with the MWA in preparation for much more detailed studies at similar frequencies with the SKA.

A comparison of radio and far-infared emission in the Magellanic Clouds and other nearby galaxies allows for a more detailed study of the correlation of cosmic rays, magnetic fields and star-formation than is possible for distant galaxies. At GHz frequencies, ATCA and IRAS studies allowed Hughes et al. (2006) to find that the radio/far-infrared (FIR) relation breaks down at $\sim 50$ pc (3 arcmin). Residual correlation at smaller scales remains due to the thermal component of the radio emission. Low-frequency observations of 6C galaxies have shown that the radio/FIR correlation remains good at 151 MHz (Fitt et al. 1988). Such observations are useful for further separating thermal and non-thermal sources of emission, for the study of synchrotron spectral aging and cosmic ray diffusion, and for allowing k-corrections in cosmological studies. 

The Magellanic Clouds are among only a handful of galaxies where the magnetic field can be studied via both Faraday rotation of the polarized emission from background sources (sensitive to the line of sight field) and the polarized component of the diffuse emission (sensitive to the field direction in the plane of the sky). Studies of the Clouds at GHz frequencies \citep{hmk86,ghs05,mgs08} have been useful in determining field strengths and morphology. Low-frequency observations will contribute by locating regions, perhaps associated with wind-blown superbubbles, of warm ionized gas unseen in X-rays, and by allowing more detailed studies of Faraday structures. The MWA will investigate the existence of a possible pan-Magellanic magnetic field via the old relativistic electron population which will have travelled far from its source.

\subsection{Cosmic Ray Mapping}
\label{sec:geg_cr}

One of the most exciting possibilities of the MWA survey at the lowest frequencies is the mapping of the three-dimensional cosmic ray emissivity of the Milky Way.  Within the MWA frequency range below approximately 200~MHz, Galactic HII regions turn optically thick, blocking emission from the synchrotron background of the Galaxy.  Typical optical depths of HII regions are expected to be $\tau\approx2$~to~5.  These HII regions, therefore, appear in absorption relative to the surroundings and the residual surface brightness at these positions allows a direct measurement of the integrated emissivity of cosmic-ray electrons out to the HII region.  

\citet{nhr+06} used 74 MHz VLA observations of HII regions to detect such absorption against the Galactic synchrotron background.  They found 92 regions with absorption features, 42 of which have measured distances. While such features have been seen since the 1950s, previous observations had poor resolution.  The 10~arcmin VLA beam allowed for better spatial resolution while still remaining sensitive to low surface brightness.  A number of surveys at higher frequencies have been conducted. \citet{2003A&A...397..213P} have compiled a catalog of 1442 diffuse and compact HII regions based on 24 prior surveys, although a number of regions are omitted in their catalog due to their emphasis on regions of order 10~arcmin (for an example of more recent identifications, see \citealt{2010ApJ...718L.106B}).  Approximately 800 of these regions have radio line velocity measurements providing distances.   \citet{2004MNRAS.347..237P} report on the distribution of the HII regions, most of which have Galactic latitude within $\pm2$~degrees of the plane and are distributed throughout the disk, with the spiral arm structure of the Galaxy evident in their distribution.

Because many Galactic HII regions have well-constrained distance estimates, MWA measurements of the synchrotron absorption features will allow for a three-dimensional map of the relativistic gas content in the Milky Way (e.g. \citealt{2008A&A...477..573S}).  For HII regions with kinematic distance degeneracies, MWA detections may be able to break the ambiguities.  Since the MWA will have a view of the inner and outer Galaxy, it can provide the first accounting of the overall distribution of Galactic cosmic rays and of the large-scale magnetic fields that cause them to emit.  In addition, due to its ability to measure the absorption in HII regions as function of frequency, the MWA should be able to yield the cosmic ray electron emissivity spectrum in the direction of the HII regions.  As noted by \citet{nhr+06},  multi-frequency observations should lead, therefore, to the mapping of the emissivity spectrum of cosmic rays.

Combined with the MWA's expected abilitity to detect and characterise supernova remnants (described below) these measurements will open new probes into the full energy budget of the Galaxy, including the role of supernovae in the production of energetic cosmic rays, as well as the diffusion and aging of the cosmic rays.

\subsection{Galactic Supernova Remnants}
\label{sec:geg_snr}

Statistical studies of supernova rates suggest that there should be $\sim 1000 - 2000$ supernova remnants (SNRs) in our Galaxy \citep{1991ApJ...378...93L, 1994ApJS...92..487T}, in stark contrast to the $\sim 300$ SNRs currently known \citep{2004BASI...32..335G, bgg+06}. This deficit is likely due to the observational selection effects, which discriminate against the identification both of old, faint, large SNRs, and young, small SNRs. Discovery of these `missing' SNRs, and hence a characterisation of the full SNR population, is crucial for  understanding the production and energy density of Galactic cosmic rays and the overall energy budget of the interstellar medium. 

\begin{figure*}
\begin{center}
\includegraphics[width=\textwidth]{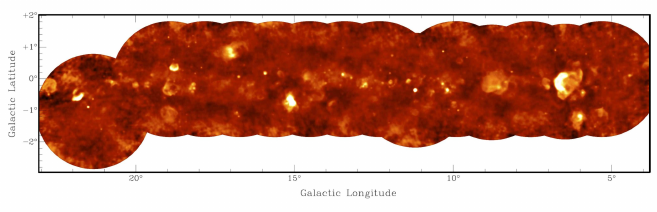}
\end{center}
\caption{A 330-MHz panorama of part of the Galactic first quadrant, derived by smoothing VLA data from \citet{bgg+06} to 3 arcmin resolution. At these low observing frequencies, most of the bright sources correspond to non-thermal emission from SNRs. The MWA will be able to carry out such observations over almost the entire inner Galaxy.}
\label{fig:bgg}
\end{figure*}

Recent studies have demonstrated that low-frequency interferometric maps are a very efficient way of identifying new SNRs, because of the comparatively steep radio spectra of these systems. Most notable has been the survey of \citet{bgg+06}, shown in Fig.\ref{fig:bgg}, which tripled the number of known SNRs in a 40 deg$^2$ patch of the Galactic plane. The MWA will have a unique view of virtually all the inner Galaxy. The high-quality images that we expect to derive, across a range of frequencies, will provide a superb data set from which to identify hundreds of previously unidentified SNRs. Additionally, attempts to uncover old, faint sources will be greatly enhanced by development of new detection algorithms which we plan to implement on MWA data (e.g., circle Hough transforms, see \citealt{2012PASA...29..309H}).  These sources will be of interest both as a statistical ensemble, and because among them will be individual objects with unique or extreme properties.

\subsection{Radio Recombination Lines}
\label{sec:geg_rrl}

Low-frequency radio recombination lines (RRLs) are a sensitive tool for probing interstellar plasma. Below 100~MHz, carbon RRLs have been detected, but only in absorption, indicating an association with cold ($<50$ K) neutral gas or molecular clouds. At frequencies above 100~MHz, both carbon and hydrogen recombination lines have been observed, with the hydrogen lines likely associated with normal HII regions as at higher frequencies.  Both carbon and hydrogen lines have been found in observations along the Galactic plane as far as $\pm4$~degrees off the plane in Galactic latitude. 

Measurements of RRLs in the MWA frequency range will provide an opportunity to study both emission and absorption processes.  They will enable new studies of diffuse photodissociation zones that lie between molecular clouds and the general ISM and yield estimates for electron temperature and density in these regions.  Carbon RRLs at 75~MHz exhibit a linewidth range from 5~to 50 km~s$^{-1}$ \citep{1995ApJ...454..125E}, corresponding to 1--12 kHz.  Although, the MWA spectral resolution is not ideal for such narrow lines, single-pixel RRL observations have previously been used with much success \citep{2012MNRAS.422.2429A}. Sensitivity can also be regained by stacking different transitions within the observing band.


\section{Time-domain Astrophysics}
\label{s_transients}

Systematic exploration of the time domain is becoming a dominant
frontier in astronomy, and significant resources have been deployed
across the electromagnetic spectrum in order to probe and understand
transient and time-variable phenomena.  Radio transients vary on time
scales ranging from less than one second to many days. The sources are compact objects or
energetic/explosive events, and each type is being investigated at the
forefront of astrophysics.

Radio observations can provide keys to understanding
the explosive and transient Universe by directly measuring the effects
of magnetic fields and non-thermal processes that drive the outbursts
\citep[e.g.,][]{clm04}.  Even more intriguingly, synergies with
multi-messenger astronomy (correlating electromagnetic transients with
events detected in gravitational waves or neutrinos) have the
potential to unlock substantial new physics.
The MWA spans the upper frequencies associated with coherent radio
sources and the lower frequencies of non-thermal magnetohydrodynamic processes.  This makes observations of transients with the MWA particularly useful for
understanding the non-equilibrium processes that drive dynamic
astrophysical systems.

The potential sources of radio transient emission fall into five broad
categories: coronal emission from nearby stars/substellar objects,
emission from compact objects such as neutron stars and accreting
black holes, explosive events such as gamma-ray bursts (GRBs) and radio supernovae,
planetary emission from nearby systems, and new phenomena discovered
through the opening of new regions of observational phase space.  Of
these, the first three are known sources of transient radio emission
(at least at some frequencies), while the last two categories
constitute more speculative targets.  However, there are firm
theoretical expectations that these latter objects should give rise to
transient radio emission, and our searches will be far deeper than
previous surveys.

First, we discuss in detail some of the expected physical constraints
to be gained from the MWA observations of known sources of transients
(\S~\ref{sec:sciguar}).  We then explore some of the more speculative
science targets that we hope to explore with the new phase space
opened up by the MWA (\S~\ref{sec:serendip}).  We describe the
new capabilities afforded by the MWA in \S~\ref{sec:mwatransients} and in
\S~\ref{sec:pulsars} we discuss observations of pulsars, which can be both steady and transient sources.

\subsection{Known Transient Targets}

\label{sec:sciguar}

\begin{figure}
\begin{center}
\includegraphics[width=\hsize, angle=0]{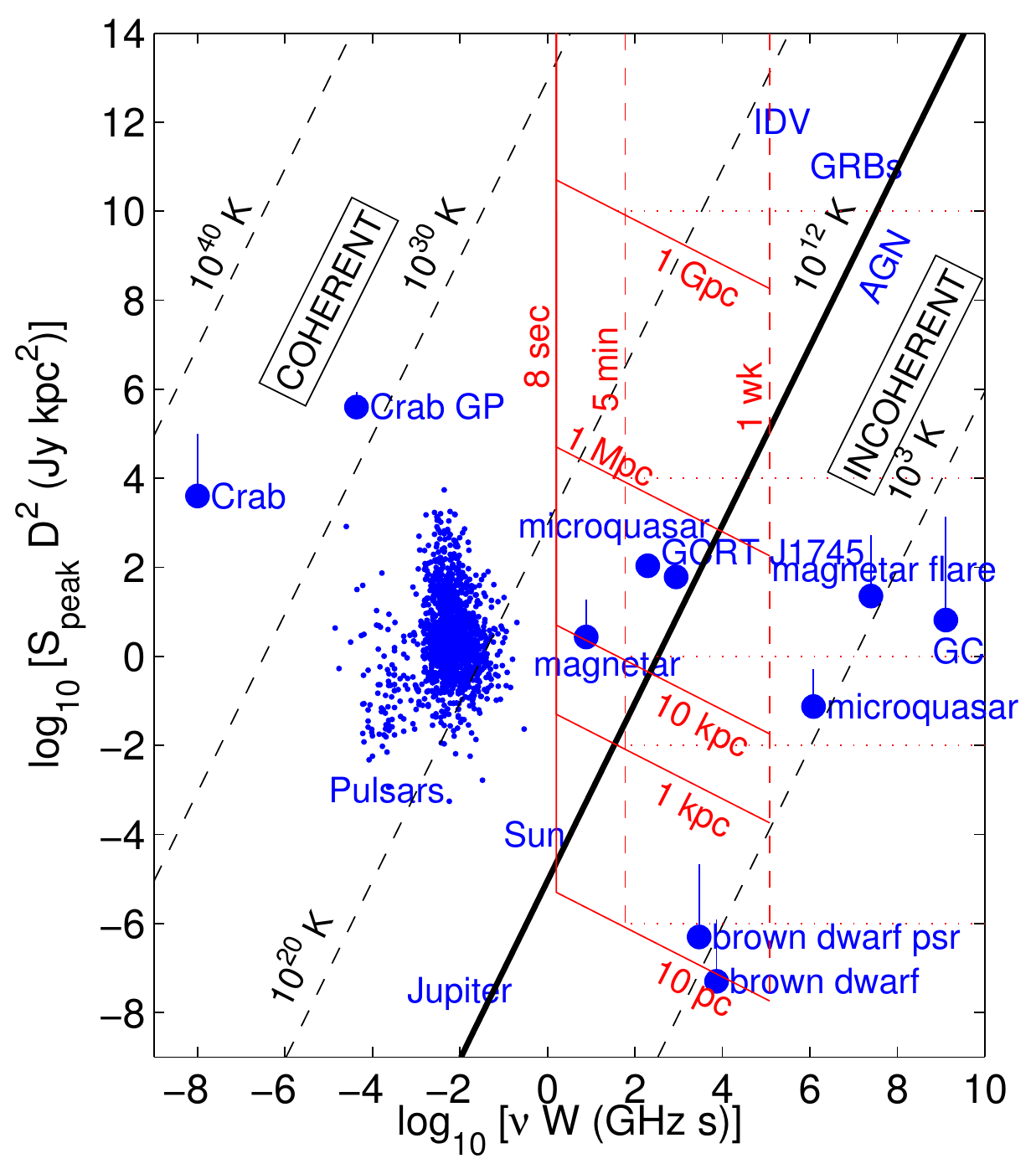}
\caption{Phase space for radio transients observable with the MWA at a
  nominal frequency of 200\,MHz, updated and adapted from
  \citet{clm04}.  We plot the product of the observed peak flux
  density $S_{\rm peak}$ and the square of the distance $D^2$ (like a
  luminosity) against the product of the emission frequency $\nu$ and
  the transient duration $W$.  In the Rayleigh-Jeans approximation,
  these quantities are directly proportional and related to the
  brightness temperature $T$ (given by the diagonal lines); we use a
  brightness temperature of $10^{12}\,$K (thick diagonal line) to
  divide coherent and incoherent processes \citep{readhead94}. The red
  lines show the predicted sensitivity of the MWA, assuming a 50\,mJy
  source can be detected in 8\,s.  The vertical red lines give the
  timescales for individual snapshot observations (8\,s), and short-
  and long-term surveys (5\,m and 1\,week).  The red diagonal lines
  give the $S_{\rm peak}D^2$ limits corresponding to distances of
  10\,pc (appropriate for sources such as low-mass stars, brown
  dwarfs, and planets), 1\,kpc (local Galactic sources), 10\,kpc (the
  Galactic Center), 1\,Mpc (local group) and 1\,Gpc (extragalactic
  sources); the dotted lines show an estimated confusion limit of
  10\,mJy, although we can search below this limit using image
  subtraction or for polarized emission }
\label{fig:cordes}
\end{center}
\end{figure}

\subsubsection{Low-Mass Stars and Brown Dwarfs} Magnetized plasmas
in stellar coronae (the outer layers of a star's magenetosphere)
produce strong radio emission through non-thermal, collective, and/or
coherent processes.  This emission reflects local properties such as
density and magnetic field strength, and can be used to understand
magnetic field generation, magnetospheric structure, angular momentum
evolution, magnetic flaring mechanisms and accretion processes.
Furthermore, magnetic fields provide one of the few indirect probes
of stellar interiors.


While radio studies of late-type stars and active binaries are
well-established (see \citealt{gudel02} for a  review), radio
observations of the lowest mass stars and brown dwarfs --- objects
incapable of sustained core hydrogen fusion \citep{kumar62} --- have
only recently been demonstrated, and comprise a relatively new tool in
this active field of research.  M dwarfs with masses
0.1--0.5$\,{\rm M}_{\odot}$ are a dominant stellar component of the
Galaxy, comprising $\sim$70\% of all stars.  They are also known to
exhibit high levels of magnetic activity, with strong surface magnetic
fields ($\sim$2--4$\,$kG, compared to $\sim$10$\,$G on the Sun;
\citealt{sl85}; \citealt{jv96}) and substantial filling factors
(50--80\% of the surfaces of M dwarfs are covered by magnetic fields;
\citealt{jv96}).
Magnetic radio emission from M dwarfs has been observed in many nearby
systems, both flaring and quiescent, predominantly in the
1--10$\,$GHz band.
These studies have confirmed the existence of large-scale,
organized magnetic fields around stars with fully convective interiors
\citep{dfc+06}, despite the loss of the $\alpha\Omega$ dynamo that powers
the Solar field \citep{parker55}.  

For even lower mass stars and brown dwarfs, the detection of both
quiescent and flaring radio emission was initially a surprise
\citep{bbb+01}, as the flux from these sources is orders of magnitude
greater than that expected from the G\"{u}del-Benz relation\footnote{A
correlation between radio flux and X-ray luminosity that extends over
several orders of magnitude; see \citealt{gb93}.}.  Several late-type
sources have now been detected at GHz frequencies
(Fig.~\ref{fig:cordes}), in many cases despite the absence of optical
or X-ray emission (see \citealt{berger06} for a review).  Indeed,
while H$\alpha$ and X-ray luminosities relative to bolometric
luminosities appear to decline with spectral type, relative radio
emission inexorably increases up to the substellar regime (see
\citealt{bp05,berger06,aob+07}).  Understanding why radio emission is
so prominent among these very low mass objects, and whether such
emission correlates with age, rotation or other physical parameters,
remain outstanding questions.

Radio flares in particular
demonstrate that this emission is associated with magnetic events,
occurring over timescales of microseconds to minutes (e.g.,
\citealt{ob06}).  The brightness temperatures of bursting emissions
frequently exceed $10^9\,$K, indicating coherent processes as
confirmed by high circular polarizations (30--100\%). Yet radio
variability from these sources is remarkably diverse, from strong
(30$\,$mJy peak flux), narrowband, fully polarized, bursting emission
with frequency drifts \citep{bp05}; to periodic bursts suggestive of
pulsar-like beaming \citep[][and Fig.~\ref{fig:cordes}]{hbl+07}; to
order-of-magnitude variations spanning years \citep{adh+07}.  It has
even been suggested that the observed ``quiescent'' emission from
these sources may be the result of low-level, sustained electron
cyclotron maser emission (e.g., \citealt{hbl+07}).  The physical
origin of transient radio bursts from low-mass stars, their incidence
rates (duty cycle estimates range from 1--30\%) and their spectral
characteristics are either poorly constrained or entirely unknown.
These questions have a direct impact on whether low mass stars comprise
a significant fraction of radio transients in general
\citep[e.g.,][]{kp05,khr+08}.

The MWA has the potential to contribute substantially to our understanding
of both quiescent and flaring magnetic emission from low mass stars
and brown dwarfs.  By sampling the 80--300$\,$MHz band, MWA
observations can be used to constrain the peak frequency ($\nu_{\rm
pk}$) and spectral indices about $\nu_{\rm pk}$, enabling
determination of coronal field strengths (currently estimated at
10--100$\,$G; \citealt{bp05,berger06}), electron densities and the
fundamental nature of the emission.  \citet{aob+07} argue that radio
emission from very low mass stars may in fact peak in the
$\sim$100$\,$MHz range if it arises from electron cyclotron emission
from weak fields ($\nu_{\rm pk} = \nu_{c} = 30-300\,$MHz for 10--100 G
fields; see also \citealt{ohbr06}). The sensitivity of the MWA is
essential for this work, given the intrinsic faintness, but variable
nature, of radio emission from low-mass stars; estimates range over
$10\,\mu$Jy to 100$\,$mJy at 100$\,$MHz at a distance of 10$\,$pc,
depending on the emission mechanism.  Similarly, the MWA's wide-field
capabilities are necessary to build up statistically robust samples,
particularly given the roughly uniform distribution of nearby low-mass
stars in the vicinity of the Sun.  High spectral resolution
observations enabled by the MWA will permit studies of field dynamics and
remote determinations of physical parameters such as the heights of
magnetic reconnection regions, electron speeds, and the density
profile of the stellar coronae.  

\begin{figure}
\includegraphics[width=\hsize]{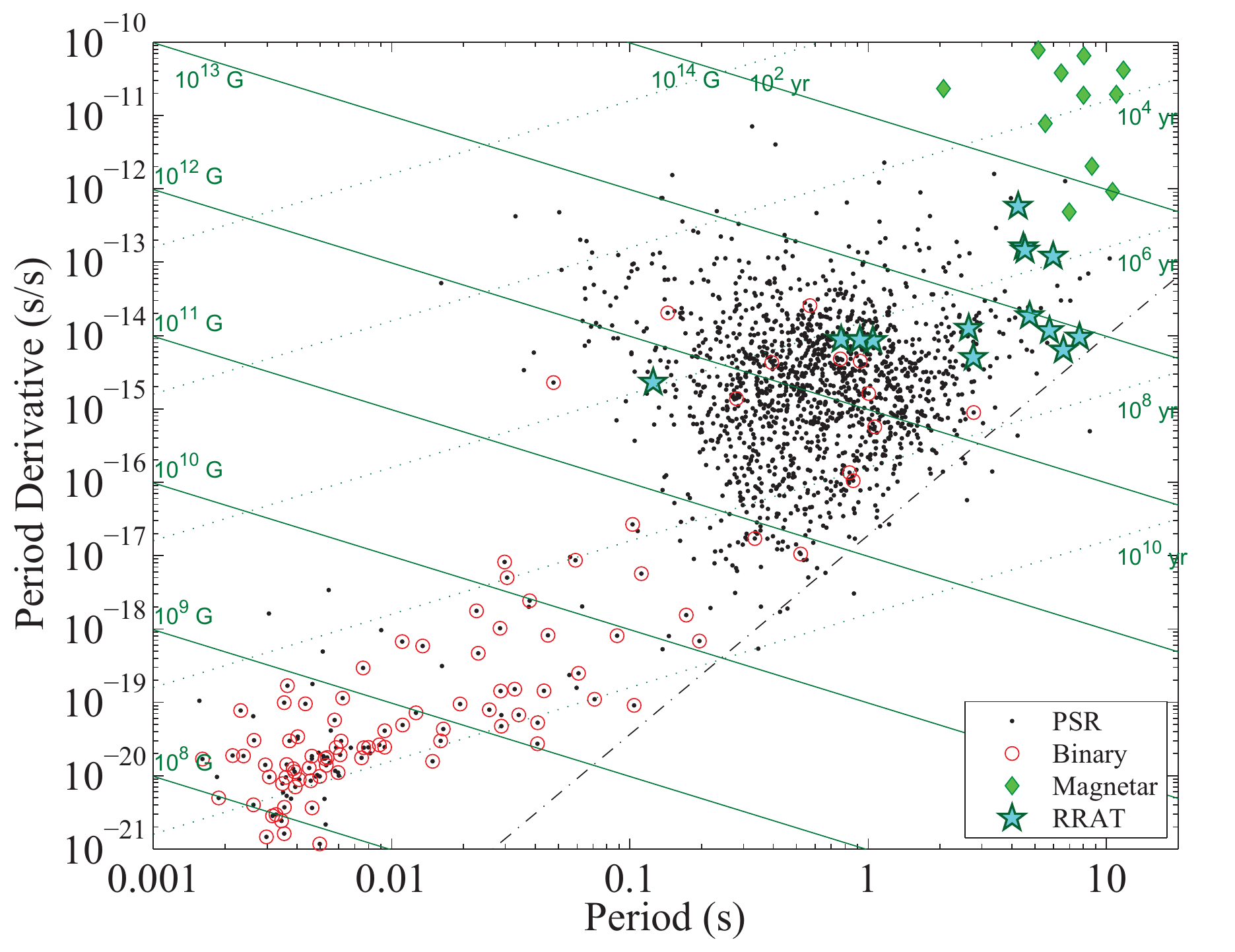}
\caption{$P$-$\dot P$ diagram showing the distribution of the known
  pulsar population. The diamonds show the magnetars (SGRs and AXPs)
  and the stars the RRATs; binary systems are circled.  The diagonal
  lines show loci of constant dipole magnetic field (solid) and
  spin-down age (dotted), while the dot-dashed line is an approximate
  ``death-line'' for pulsar activity.  The MWA will be uniquely
  sensitive to the slower, transient population while maintaining
  sensitivity to a considerable fraction of the millisecond pulsar
  population in the lower left.}
\label{fig:ppdot}
\vspace{-6mm}
\end{figure}

\subsubsection{Magnetars}
While a variety of neutron stars can produce transient or steady radio
emission (discussed in more detail in \S~\ref{sec:pulsars}), some
subclasses of neutron stars can give rise to extremely energetic,
explosive phenomena.  With magnetic fields of $\sim$$10^{15}\,$G,
magnetars \citep{dt92,td95} undergo catastrophic reconnection events known as
``giant flares'' that release $>10^{43}\,$erg.  With each flare the
magnetars release blobs of relativistic plasma
\citep{fkb99,gkg+05,ccr+05} with peak fluxes of hundreds of mJy at
frequencies from 200$\,$MHz up to 100$\,$GHz (Fig.~\ref{fig:cordes}).
Such events are detected through high-energy satellites, but the radio
monitoring provides vital diagnostics of the flare properties and
energetics.

Beyond this, some magnetars unexpectedly have steady radio emission
that is modulated at the X-ray pulse period (2--12$\,$s).  Prior
searches for steady-state radio emission from magnetars had not been successful
\citep[e.g.,][]{kkfvk02}, but \citet{hgb+05} found a radio source
coincident with the magnetar XTE~J1810--197 \citep[also][ and others]{crh+06}.
Monitoring the magnetars through both
individual pulses and phase-averaged emission, looking for both giant
flares and the onset of bright phases, will help understand the duty
cycles of the radio-bright phases and hence unravel the puzzles that
they present.

\subsubsection{X-ray Binaries} Relativistic outflows or jets are a
widespread yet transient consequence of accretion onto compact
objects.  Radio jets are seen for black-hole binaries (BHBs) and some
neutron-star binaries (NSBs). These sources are collectively called
``microquasars'' (Fig.~\ref{fig:cordes}), since they exhibit
scaled-down and rapidly evolving versions of the jets in active
galactic nuclei.  Microquasars exhibit several types of jets
with different ranges of radio power and ejection timescale. Their
study is crucially dependent on the correlated variations in radio and
X-ray properties.  




The most common type of jet in microquasars is the quasi-steady jets associated with the X-ray hard state \citep{fbg04,mr06} that can
persist for months in some sources. Steady jets are estimated to be
mildly relativistic (i.e. $\sim$0.7$c$; see \citealt{fender06}), but
there is uncertainty about the baryonic content and hence the scale of
the mass outflow and the fraction of accretion energy needed to power
such jets.

Impulsive ejections are also produced by BHBs and some NSBs and the
brightest of these are seen as bipolar radio jets that move with
velocities as high as $0.98 c$ \citep{mr99}.  Impulsive jets are
associated with X-ray state transitions \citep{fbg04}. In two BHBs
superluminal radio ejections, coinciding with major X-ray flares, led
to the appearance of X-ray jets (more than one year later), that were
imaged with {\it Chandra} as the jets impacted the local ISM
\citep{tcf+03,ckf+05}. 
Smaller and more frequent ejections in BHBs are seen as correlated
X-ray and radio flares with $\sim$10$\,$min timescales and radio-band
delays \citep{emm+00,wpp+07}.  The available X-ray data usually reveal
recurrent and abrupt changes in flux and spectral components that
associate the small jets with instability cycles in the inner
accretion disk.

Study of the connection between disks and jets is fundamentally based on correlated X-ray and
radio behaviour, as outlined above.  During the last decade, the
primary limitation for this enterprise has been the absence of
dedicated radio instruments that can partner with X-ray facilities,
but the MWA will radically change this landscape.  Additional partners
for jet investigations arise with \textit{Fermi} (10~keV to 300~GeV), and
the recent expansions of TeV Observatories (e.g., HESS and VERITAS;
50~GeV to 50~TeV).

Recent measurements with the GMRT (150--1450 MHz) in
India have shown variable radio emission in GRS~1915+105 \citep[][see Fig.~\ref{fig:cordes}]{irp+05}  and Sco X--1
\citep{prm+05}, with spectra that do not exhibit self-absorption cutoff at low frequencies.  This is interpreted as emission from the outer regions of a compact jet and from an optically-thin impulsive jet, respectively (also see \citealt{prp+06,pri+07}).

Overall, the MWA has the capability to monitor many known sources for which
X-ray activity is detected, establishing outburst statistics and
extending the spectra down to low frequencies where the total
energetics are constrained.  These observations can help to measure
the burst luminosity function and the X-ray/radio correlation, and
investigate their dependences on various intrinsic source parameters
such as mass and X-ray luminosity.  The MWA will provide data archives for the two microquasars that are
reliable sources of jets: SS~433 and GRS~1915+105, as well as the BHBs
and NSBs with the fastest outburst recurrence time (1--2 years:
4U~1630--47, and GX339--4, Aql X--1, 4U~1608--52, and the rapid
burster).  These monitoring observations will also find new outbursts
of known sources, acting as a trigger for observations across the electromagnetic
spectrum.  Finally, the MWA has the potential to discover new sources:
the large majority of the known BHBs and candidates are transient
sources that were discovered because of their X-ray outbursts
\citep{rm06}.  

\subsection{Serendipitous Transient Sources}
\label{sec:serendip}

Beyond the source categories discussed above, there are a number of
object classes which we might hope to see (see \citealt{clm04} and
\citealt{fws+06}).  These vary from the local to the cosmological, and
the population predictions also span a wide range.  Currently,
low-frequency transient radio emission has been proposed but not
observed for these objects, and there are only limited observational
constraints. Such sources are among the most exciting possibilities
for the MWA transient surveys (\S~\ref{sec:mwatransients}). Those
surveys cover  a large area of unexplored phase space --- orders
of magnitude in several different parameters --- and  provide a
real opportunity to discover unique sources.


\subsubsection{Extra-Solar Planets} 

In the Solar System, Jupiter is a
strong source of low-frequency radio emission produced by cyclotron
maser processes in its magnetosphere \citep{gurnett74,wl79}.
Observations have established that many of the Jovian planets 
discovered around other stars \citep[e.g.,][]{mbf+05} have magnetic fields
\citep{swb03}, although the field strengths are not known.
By analogy, many of those planets should have similar emission.  This
would then be a novel way of \textit{directly} detecting the planetary
companions, something that has only been possible in a few special
cases otherwise \citep{cbng02}.  This would allow an unprecedented
probe of the planetary systems, such as estimating properties of the
planetary magnetosphere, measuring its rotation rate, and observing
the interaction between the star and
planet \citep{lfd+04}.  \citet{gzs+07} discuss theoretical
expectations for radio emission from known extrasolar planets. Most of
the sources are predicted to emit at the microJansky level between
10--200\,MHz.  A handful of planets are expected to emit at the
mJy level in the MWA bandpass.  Their results are heavily
model-dependent and MWA observations of the currently known exoplanet
population will help constrain further models.

There have been a number of searches for emission from such planets
(e.g., \citealt{fdz99,bdl00,ztrr01,lfd+04,lf07}).  There are no
detections to date, although the results are not very constraining.
The emission is predicted to be very sporadic and bursty, and it is
difficult to search effectively for such emission with pointed
instruments like the VLA.  Instead, new instruments
like the MWA are likely to be the best way to detect extra-solar planets \citep{lfd+04}.

\subsubsection{Explosive Events}
\label{sec:explode}

Gamma Ray Bursts and radio supernovae may both produce short and long
duration transient radio signals.  The afterglow emission from each
type will provide signals at the upper frequencies accessible to the MWA that
are delayed from the initial explosion (due both to plasma in the ISM and to changing opacities) and slowly brighten over a few weeks to months \citep{mhk+05}.  These afterglows
can be important diagnostics for explosion energetics \citep{snc+06,scp+10}
and can help understand particle acceleration in the nearby
Universe \citep{crs+10}.  While the long timescales at low frequencies make the MWA
unfavorable for discovery of such events, the rarity of the more
extreme ones mean that the very large FOV  may compensate 

GRBs, supernovae, and other explosions may also produce prompt pulses
of coherent emission during the initial explosion \citep{cn71,np11}.
The theorized prompt signals are produced by coherent emission near
the explosive shock. Proposed mechanisms include current oscillations
\citep{uk00}, synchrotron maser activity behind the shock
\citep{sw02}, or collective plasma modes as seen in solar bursts
\citep{mk05}. Models by \citet{hl01} and \citet{bp98} also
predict detectable emission, depending on the plasma environment and
dynamics of the GRB.  Some of these signals will be accompanied by
bursts of radiation at other wavelengths, while some may offer the
potential for multi-messenger astronomy by correlating radio and
gravitational-wave signals \citep{np11,pll10}.


Detection of such prompt emission from GRBs offers a number of unique
insights into the properties of the bursts, the intergalactic medium,
and cosmology \citep{mhk+05}.  First, the details of the emission that
we see depend heavily on the specific model for the shock structure,
as discussed above.  Second, we can use the emission as a probe of the
media through which it propagates: we will see delays and dispersion
from the local plasma, the host galaxy's ISM, the intergalactic medium
(details of which can depend on cosmic reionisation), and the Milky
Way's ISM \citep{inoue04}. 

Past searches for prompt radio emission from GRBs
\citep{dgw+96,bp98,balsano99} were all unsuccessful.  Other projects
attempted to perform blind surveys for radio transients
\citep{alv89,bmj+76,khcm03}.  All of these attempts have suffered from
constraints inherent in narrowband receivers and simple transient
identification systems.  More modern projects like GASE and ETA
\citep{mhk+05,pem+08} were designed to correct many of these problems,
with digital baseband recording and interference rejection through
de-dispersion, but the collecting areas are still very small.

\subsubsection{New Opportunities} 

Above we discussed two classes of
sources that are expected to emit low-frequency radio transients, and
where detection of such transients would enable us to make significant
advances in our understanding of such objects, but where we do not
know the source properties enough to consider them ``guaranteed''
science.  Beyond even those classes are additional objects that may be found with blind searches (e.g.,  neutron
star binary inspiral events that may produce radio precursors; \citealt{hl01}).

To date, systematic transient searches with radio telescopes have
yielded some results, but they have largely been difficult to
interpret (e.g., \citealt{mdk+07,mnk+09}), due to poor statistics, limited localization, and a
long time before follow-up.  However, there have been successes that,
while enigmatic, hint at the potential offered by the MWA.  We do not
give a full review here, but point to particular detections that hold
promise for the MWA.

While searching for transient events at the
Galactic Centre (GC), \citet{hlkb02} found two sources, one of which
is an X-ray binary, but the other remains unclassified, although it
varies on long timescales ($\sim$months).  Later searches
\citep[][also see \citealt{hwl+08}]{hlk+05} found a much more rapidly
variable source, with flares on timescales of minutes that repeated
(sometimes) every 77$\,$min (GCRT~J1745 in Fig.~\ref{fig:cordes}).
The positional coincidence with the GC, along with the greatly
increasing source density near the Centre, suggests it is at a
distance of about 8~kpc, but this could be a selection effect because
all the fields searched for transients in the program were in that
direction.  \citet{bry+05} found another transient near the GC, but
they deduced that it was an X-ray binary with a jet outburst interacting with the dense medium in that
region.  

For searches of high-latitude fields, \citet{bsb+07} followed by
\citet{2012ApJ...747...70F} searched archival VLA data at cm
wavelengths and identified at least one very faint (sub-mJy) transient
source.  \citet{jhkl12} did repeated observations at a frequency of
325\,MHz (closer to the MWA band) and did identify one significant
transient of unknown origin, although they consider several
possibilities including a distant flare star.  Only recently have
surveys moved beyond examinations of archival data
\citep[e.g.,][]{bmg+11}, with a wide range of flux limits and
timescales, to systematic examinations of the variable sky
\citep[e.g.,][]{lgw+08,cbk+11}.  It is becoming clear that radio
transients, are rare but promissing targets that will lead to advances
in a number of areas of astrophysics.

\subsection{Planned Searches}
\label{sec:mwatransients}
We will conduct a continuous search for transient radio emission in
each image produced by the MWA's real-time system, looking for changes from the
previously observed sky on timescales ranging from 8\,s to months
(eventually using the high-quality integrated data from the sky
surveys). 
This search will be more than six orders of
magnitude more sensitive than previous transient surveys in the band
and will cover a much broader range of frequencies, transient
durations, and dispersion delays. 


Eventually the full transient capabilities of the MWA will: 1) cover
0.25~steradian instantaneously,---approximately $\sim$3\% of the visible sky---with
full polarization coverage; 2) search blindly over that area orders
of magnitude deeper than previous work; 3) simultaneously record
light curves and spectra for hundreds of sources and conduct routine observation
of tens of thousands of sources per week; 4) obtain typical exposures
of days/source, with hundreds--to--thousands of hours for sources in
selected areas (Galactic plane, EOR fields) over one year; and 5)
record full electric field time-series for a subset of observations.  
%

New sources detected by the MWA will have relatively poor (approximately arcminute) localization.  This will hinder
interpretation as many possible counterparts will be present at other
wavelengths \citep[e.g.,][]{khr+08}.  Prompt multi-wavelength follow-up
with wide-field imaging instruments will be crucial to help mitigate
this, as detection of contemporaneous variability at other wavelengths
will significantly improve the localization of sources and will pave the
way for more detailed follow-up observations such as spectroscopy.

Source confusion will limit the sensitivity of the MWA for source
detection.  However, time variability is a classic way to help beat
the confusion limit (i.e., searches for pulsars with single-dish
telescopes).  Subtracting images of the sky taken over successive time
periods will help beat down the confusion noise and allow transients
to emerge, especially when correlated with observations at other
wavelengths.  Finally, examination of polarised images that have much
lower background source levels will allow searches for bright and
highly polarized transients like GCRT~J1745$-$3009
\citep{hlk+05}.

\subsection{Pulsars}
\label{sec:pulsars}
With approximately 2000 now known (Fig.~\ref{fig:ppdot}), pulsars offer
opportunities to study a range of astrophysical phenomena, from the
microscopic to the Galactic.   The MWA's wide coverage in frequency and large collecting area, makes it a unique instrument
for low frequency detection and detailed studies of pulsars and
fast transients.  The areas of pulsar science that the MWA would probe at
low-frequencies include pulsar phenomenology and pulsar searches, in
general, and single-pulse and giant pulse studies, in particular.  This is highlighted by the successful detection of giant pulses from the Crab Nebula pulsar at 200~MHz with only three prototype MWA antenna tiles \citep{bwk+07}.


The MWA will be able to observe at rates faster than the 0.5--8~s
correlator integrations using a voltage capture system capable of
recording, for several hours, correlator input data.  This system will
provide $3072\times 10$\,kHz Nyquist sampled channels of undetected
voltages from each of the 128 antennas of the the MWA that will enable
high time-resolution studies of several types of objects, including
pulsars and fast transients over the $\sim 1000$\,deg$^2$ tile beam.

It is at MWA frequencies that pulsar emission peaks \citep{mkkw00,ks+12} and that
interstellar propagation effects are easiest to study, but this wavelength range has been
neglected for years because of difficulties in processing the data and
the increasing prevalence of radio frequency interference (RFI).  This
has left the low-frequency pulsar population largely unconstrained.
Since low frequencies select against distant objects (especially for
fast pulsars) the MWA will provide important new
constraints for the local, low-frequency population.  
As a baseline, in an 8\,hour pointing at 200\,MHz we should have a 8$\sigma$
sensitivity limit of 1.8\,mJy for long-period pulsars.  When corrected
to 1400~MHz using a spectral index of $-1.8$ \citep{mkkw00}, this
would be 0.05\,mJy, comparable to or better than ongoing surveys like
the High Time Resolution Universe survey \citep{kjvs+10} or the Green
Bank 350~MHz Driftscan Survey \citep{blr+12,lbr+12}.

Studies of individual pulsars help address a number of areas of
astrophysics, providing unique insight
into the as-yet poorly understood pulsar emission mechanism and
astrophysical plasmas \citep{melrose03}.  Assembling an average pulse
profile from thousands of individual pulses can help investigate the
emission further, probing the emission altitude, beam geometry, and
inclination of the magnetic field from the spin axis
\citep{jap01,gg03,ks+12,hsh+12}.  Statistics derived from average pulse profiles may be among the
best probes of complicated phenomena like core-collapse supernova
\citep[e.g.,][]{jhv+05}.  Longer timescale variations in the emission
can directly constrain the density of the magnetospheric plasma
\citep{klo+06}.  

%
Single-pulse studies should
be possible\footnote{The fractional magnitude of
  pulse-to-pulse variation in the conal profiles is expected to be
  close to unity. Therefore, although desirable, it is not crucial to
  have the ability to detect individual pulses from the sequences for
  assessing primary parameters of these fluctuations, since the
  detection of relatively stable fluctuation features/properties
  benefits from the total integration time in quite the same way as
  the average profile estimation does.} for a sample of pulsars in excess of 100, with the full
MWA system at frequencies $>100$\,MHz \citep{kondratiev+12}. In cases of bright pulsars, we
aim to probe the polar emission regions at different heights in
greater details to study single-pulse fluctuations simultaneously
across a wide range of radio frequencies, believed to come from
different heights from the star surface. Mapping of the emission
patterns corresponding to different heights simultaneously will enable
tomography of the pulsar emission cone
\citep{2013ApJS..204...12M, ad01, dr99}.  The wide bandwidth of
MWA can probe a large fraction of the emission cone, and at larger distances from the star than higher-frequency telescopes, revealing much needed clues about the mechanism of pulsar radio emission, which is still poorly understood even after 40~years of pulsar studies.  


\subsubsection{Rotating Radio Transients}
While observations of pulsars with regular steady pulses reveal much about those
sources, it is in sources where the pulsar emission becomes irregular that
we can hope to gain disproportionate insight about the pulsar emission
mechanism and the relation of pulsars to other more exotic classes of
neutron stars.  In particular, the rotating radio transients
(RRATs; \citealt{mll+06} and Figure~\ref{fig:ppdot}) likely serve as
links to other more extreme populations.  The RRATS emit only sporadic
dispersed radio bursts of between 2 and 30\,ms in duration that are
separated by 4\,min to 3\,hr, but nonetheless are identifiable as
rotating neutron stars by their long-term spin-down.  Initial
estimates suggest that there could be as many RRATs as traditional
pulsars \citep{mll+06,kle+10}, but their properties do not appear the
same, as they occupy different parts of the $P-\dot P$ diagram (see Figure~\ref{fig:ppdot}).
 
The study of RRATs and sources with and without radio emission may
help link the canonical rotation-powered pulsars---neutron stars that
emit primarily in the radio and that are powered by rotational
energy---to other classes of neutron stars that are powered via other
energy sources.  In particular, the X-ray and $\gamma$-ray sources
known as magnetars \citep{wt06}, which manifest observationally as the
currently\footnote{http://www.physics.mcgill.ca/~pulsar/magnetar/main.html} 11 confirmed anomalous X-ray pulsars (AXPs) and 9 soft gamma-ray repeaters (SGRs), are powered by decay of superstrong ($\gtrsim
10^{14}$\,G) magnetic fields.  In contrast, the soft X-ray sources
known as isolated neutron stars (INS; \citealt{kaplan08}) are powered
by residual heat, but may be the decayed remnants of old magnetars
\citep{kvk09b}.  There is currently a great deal of interest in
linking radio pulsars to these populations.  At least two magnetars
have been seen as transient radio pulsars \citep{crh+06,crhr07}, while
radio searches for the INS have so far been negative
\citep{johnston03,kkvk03,kml+09}.  The populations may be linked via
RRATs: the X-ray and timing characteristics of the RRATs offer
tantalizing clues  to their relation to both magnetars
\citep{lmk+09} and the INS \citep{ptp06,mrg+07}, but as yet nothing
firm has been established \citep{kk08}.

Identification of RRATs relies on detecting single, dispersed pulses
rather than steady repeating pulses \citep[as
  in][]{2011MNRAS.415.3065K, bsb10}.  They do appear as repeating
transients and so many of the techniques required will be similar to
those for other transients \citep{lbm+07,kskl12}.  We will also explore the bispectrum
techniques of \citet{lb12}, which enable computationally simple
searches for dispersed pulses at the expense of signal-to-noise,
although the small area of each antenna leads to a low signal-to-noise
ratio on each baseline, making such searches \citep{lbp+12}
significantly less sensitivity than traditional beamforming.

\subsubsection{Interstellar Medium}
\label{sec:probe}

Beyond the study of pulsars themselves, variations in individual pulses
with time and frequency can be used to probe the properties of the
ISM \citep{hsh+12} and the Galactic magnetic field \citep{njkk08}, as well as the solar corona \citep{ojs07}.  Signals of short intrinsic duration do not survive passage through interplanetary, interstellar, or intergalactic plasma unscathed \citep{rickett90,cwf+91,tc93} and are distorted via dispersion \citep{cl02}, scintillation \citep{rickett69}, scattering \citep{bcc+04}, and Faraday rotation.  

Dispersion makes lower frequencies arrive later.  Compared to a signal
at infinite frequency, a signal of frequency $\nu$ arrives with a time
delay $\propto \nu^{-2}$.  Dispersive delay can be both an advantage
and a disadvantage.  A radio counterpart to a transient
event, such as a gamma-ray burst, observed with a high-energy satellite, would be delayed by
several minutes to hours before it arrives in the MWA band.  This delay is beneficial since it allows time to prepare for the event at low frequencies \citep[e.g.,][]{cm03,macquart07}.  But the converse is also true.  If the MWA provides the first detection of a transient signal, it may be too late for other facilities at higher frequencies to observe it.

Scattering is the result of multi-path propagation in ionized media.
While it has other effects, such as spectral and angular broadening,
we are primarily concerned with temporal broadening here.  This leads
to information on time scales shorter than the scattering time being
lost, and this time scales as $\propto \nu^{-22/5}$.  Because of this
steep scaling this tends to set a minimum useful timescale for
low-frequency observations.  The scattering timescale becomes comparable to the standard MWA correlator integration time of 0.5~seconds in the Galactic plane for objects beyond $\sim5$~kpc \citep{cl02}.  Hence, the MWA will be most sensitive to scattering effects for sources within this distance.

\section{Solar, Heliospheric, and Ionospheric Science}
\label{s_shi}

The MWA will make significant contributions to the fields of solar, heliospheric, and ionospheric science through a diverse set of novel radio measurements, as described here and in earlier papers  \citep{bastian2004,cairns2004, OberoiandKasper2004, salahetal2005, OberoiandBenkevitch2010}.  These measurements directly address space weather, defined as the study of the impact of the Sun on the Earth and its neighborhood. Space weather effects are mostly driven by solar flares and coronal mass ejections (CMEs)  that produce electromagnetic radiation (from X-rays to radio wavelengths),  energetic particles, shock waves, changes in magnetic fields, and fast-moving plasma outflows that couple to the magnetosphere, ionosphere,  neutral atmosphere, and surface of the Earth.  These responses can directly affect humanity's environment, technology, and society \citep{schereretal2005, warmuthandmann2005}. The MWA will be able to observe and track CMEs, shocks, and other plasma structures from points close to their inception through their journey to the Earth and beyond, and it will be able to characterise their impact on the terrestrial ionosphere, making it a powerful instrument for space weather studies.

\subsection{Solar Science}
\label{sec:shi-solar}

Even the quiet radio Sun is a time-variable, spectrally and morphologically complex, extended source.  Emission features are especially remarkable during periods of high solar activity. The active Sun produces both broadband ($\Delta f / f \approx 1$) and narrowband ($\Delta f / f \ll 10^{-2}$) emissions at frequencies from below 1~MHz to approximately 30~GHz \citep[e.g.,][]{wildetal1963,bastianetal1998,caneetal2003,karlicky2003,cairns2011}. Variations in the emission can occur over time scales ranging from less than a millisecond to the eleven-year cycle of solar activity.  The complexity of solar radio emission presents a challenging radio imaging problem, but the unprecedented instantaneous dynamic range, wide-bandwidth, high spectral resolution, and full-Stokes capability of the MWA will enable spectroscopic imaging of the dynamic Sun.

The physical sources of solar radio emission in the MWA band fall into three distinct catagories:  1) thermal emission produced by hot electrons in all layers of the solar atmosphere, but especially the solar corona \citep{mcleanandlabrum1985}; 2) gyro-synchrotron and synchrotron emission with broad power-law spectra in frequency, created by electrons accelerated in magnetic loops and by shocks driven by coronal mass ejections \citep{bastianetal2001}; and 3) intense bursts of nearly-coherent narrow-band emissions that are produced as the so-called type I, II, III, IV, and V bursts \citep[e.g.,][]{wildetal1963, mcleanandlabrum1985, caneetal2003,cairns2011}. The presence of mulitple different emission mechanisms, along with propagation effects like refraction and scattering, endow radio observations with a rich and unique coronal diagnostics capability.   

\subsubsection{Electron Density and Temprature}

Large gradients in electron density and temperature from the solar chromosphere to the coronal regions have the consequence that the optical depth to radio waves tends to be a sensitive function of the ray path, which itself is determined by the refraction due to the density gradients. The optical depth of ray paths at different frequencies is dominated by contributions from different coronal heights.  Spectral imaging by the MWA, therefore, should make it possible to construct a three-dimensional distribution of electron density and temperature in the lower corona for the quiescent Sun.

\subsubsection{Electron Acceleration and Radio Bursts}

Solar radio bursts are one of the primary remote signatures of electron acceleration in the corona and inner heliosphere.  A wealth of information about bursts has been amassed, primarily from studies of their emission in frequency-time dynamic spectra. Their imaging studies remain rare. The standard models for these bursts involve generation of radiation near the electron plasma frequency, $\nu_{pe}$, as well as near $2\nu_{pe}$, where: 
\begin{equation}
\nu_{pe} \approx 9 n_e^{1/2}\ {\rm MHz}\ , 
\end{equation}
for electron number density, $n_{e}$, in m$^{-3}$.  This mechanism involves generation of electrostatic Langmuir waves and their conversion into radio emission via various linear and nonlinear processes \citep{ginzburgandzheleznyakov1958, melrose1980, robinsonandcairns2000, cairns2011}.   At MWA frequencies, beams of energetic electrons typically move away from the Sun, hence emission frequency is expected to decrease with time since the coronal electron density decreases with radius.

We will focus on type II and III bursts with the MWA.   Type II bursts are associated with suprathermal electrons accelerated at magnetohydrodynamic shocks to speeds of $400$--$2000$ km~s$^{-1}$, whereas type IIIs are  due to bursts of electrons moving at speeds of order one third  of the speed of light, produced in magnetic reconnection sites during solar flares \citep{wildetal1963, melrose1980, nelsonandmelrose1985, bastianetal1998, robinsonandcairns2000, cairns2011}.    
 
\begin{figure*}
\begin{center}
\includegraphics[angle=0, trim=0mm 0mm 0mm 0mm, clip=true, width=6in]{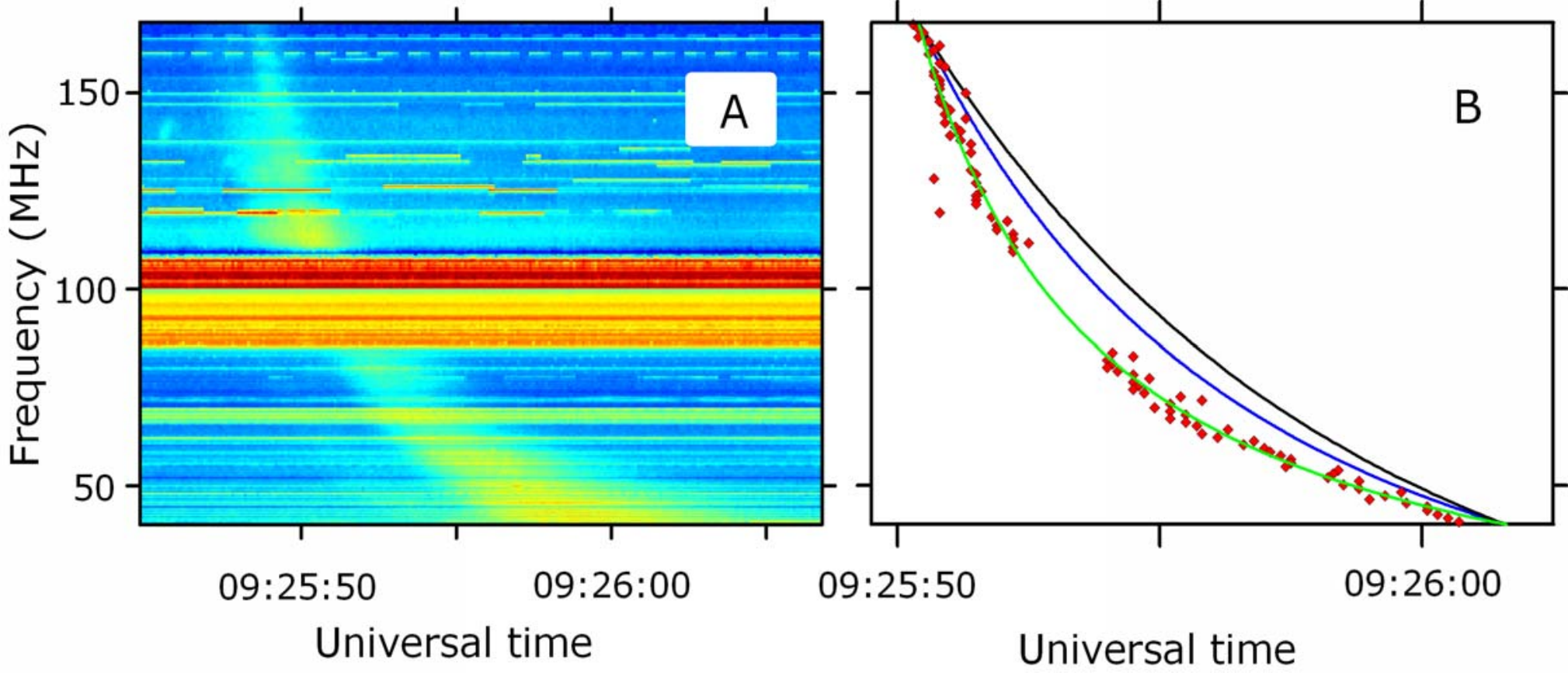}
\caption{(a) Dynamic spectrum of the type III burst of 09:26 UTC on 20 October 1999, observed by the Potsdam-Tremsdorf Radiospectrograph, with normalized intensity (color coded) as a function of frequency and time  \citep{cairnsetal2009}. Horizontal bands are due to RFI.  (b) Red diamonds show the $(f,t)$ locations of flux maxima for each time. The green line is the best fit to $\nu(t) = a (t - b)^{-\beta/2}$ and implies that $n_{e}(r) \sim (r - 1.0 R_{S})^{-\beta}$ with $\beta = 2.0 \pm 0.3$. This unexpected result is a solar wind-like profile for $n_{e}(r)$. Fits for the leading power-law term $\beta = 6$ of the standard density model based on coronagraph data (blue line) and an exponential gravitational-settling profile (black line) agree poorly with the type III data. }
\label{Fig:SHI1}
\end{center}
\end{figure*}

Several major unresolved questions exist.  The first is whether type II bursts are associated with a blast wave (i.e., a freely propagating shock) or the bow wave found upstream of a CME that is expanding into the heliosphere \citep{caneanderickson2005, vrsnakandcliver2008, cairns2011}.  The second problem is that estimates of density profiles that are derived by applying the plasma frequency relation to time-dependent spectra \citep{cairnsetal2009, lobzinetal2010}, have tended to be much shallower than expected, as shown in Figure~\ref{Fig:SHI1}.  The density profile is coupled to the formation mechanism of the corona, hence discrepancy with models casts doubt on the source of energetic electrons.  Lastly, unresolved theoretical issues exist due to the relatively small levels of circular polarization observed in type II and III bursts, despite existing theories predicting essentially 100\% circular polarization in the sense of the o~mode (lefthand) \citep{nelsonandmelrose1985, suzukianddulk1985, cairns2011}. 

\begin{figure*}
\begin{center}
{
\resizebox{0.30\hsize}{!}{\includegraphics[angle=0, trim=0mm 0mm 0mm 0mm, clip=true]{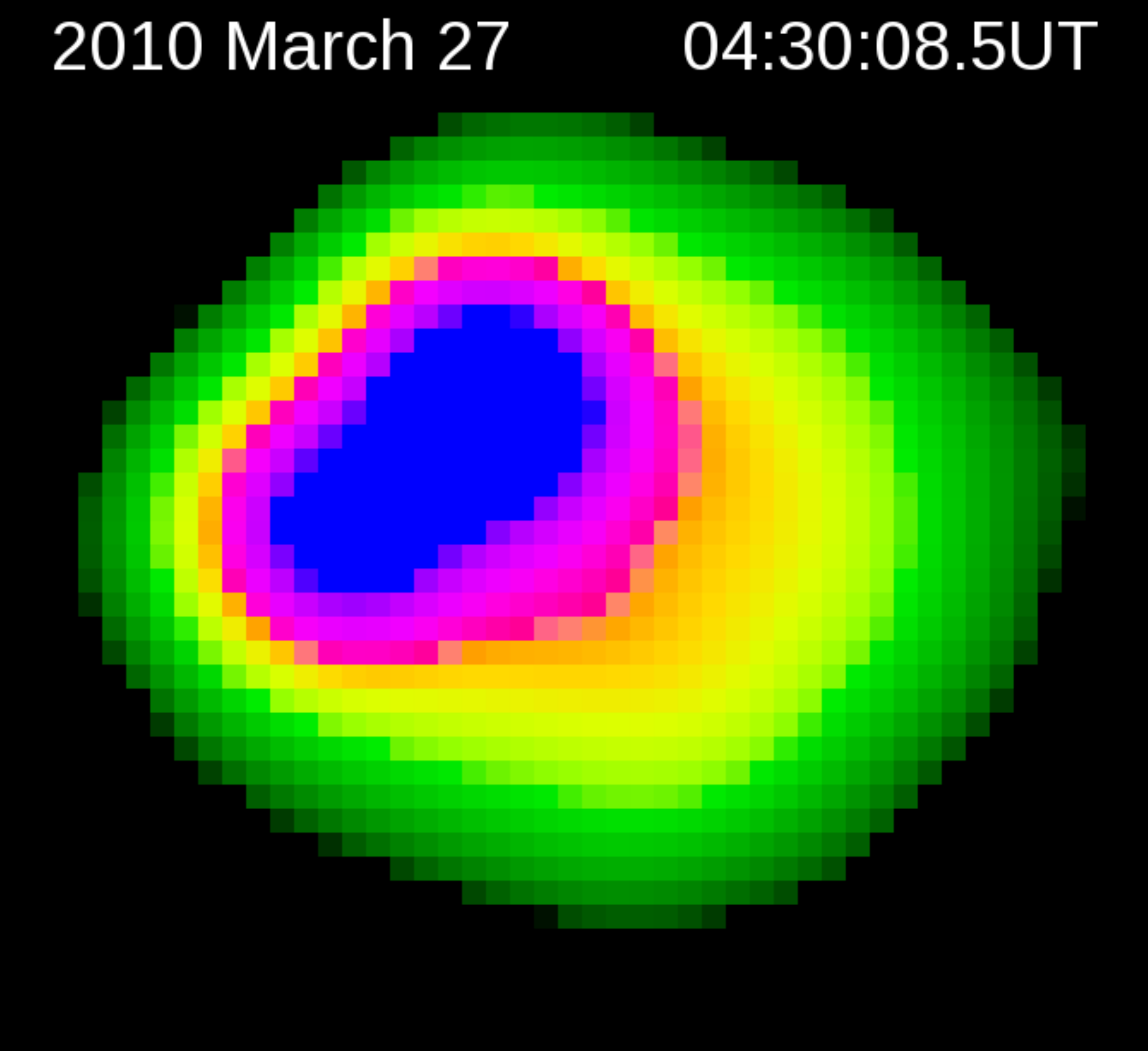}}
\resizebox{0.30\hsize}{!}{\includegraphics[angle=0, trim=0mm 0mm 0mm 0mm, clip=true]{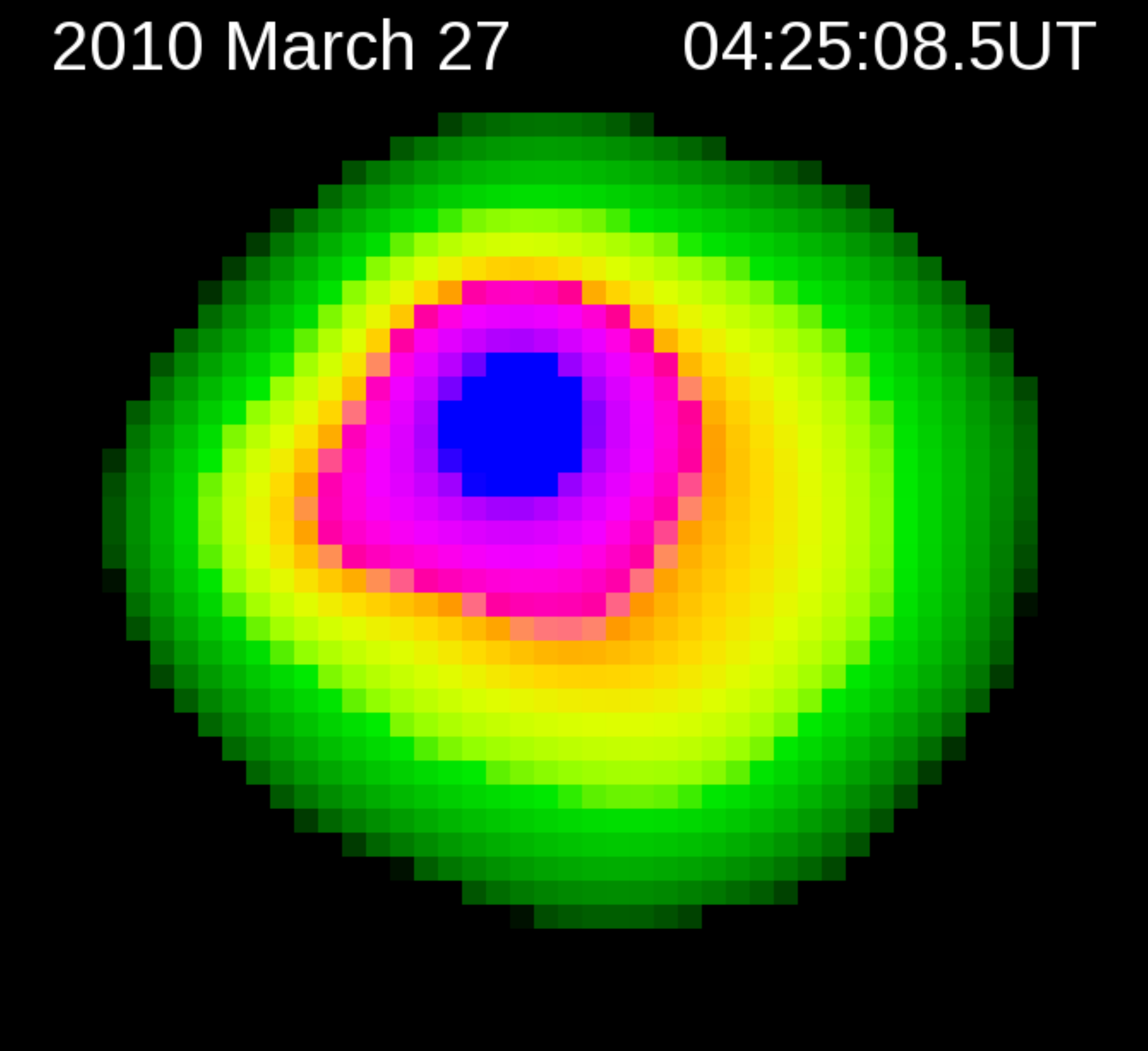}}
\resizebox{0.305\hsize}{!}{\includegraphics[angle=0, trim=0mm 0mm 0mm 0mm, clip=true]{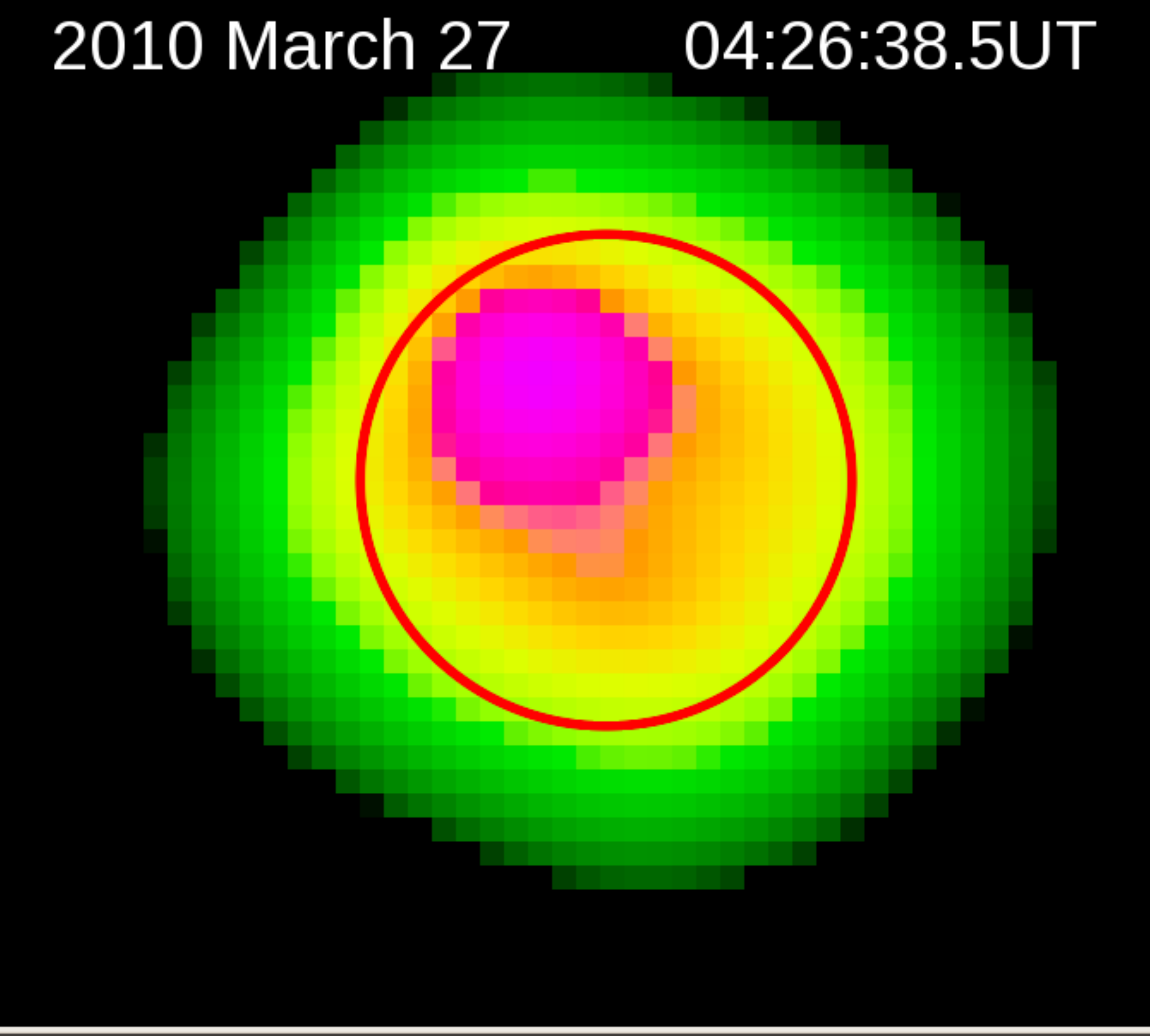}}
\resizebox{0.30\hsize}{!}{\includegraphics[angle=0, trim= 15mm 15mm 15mm 0mm, clip=true]{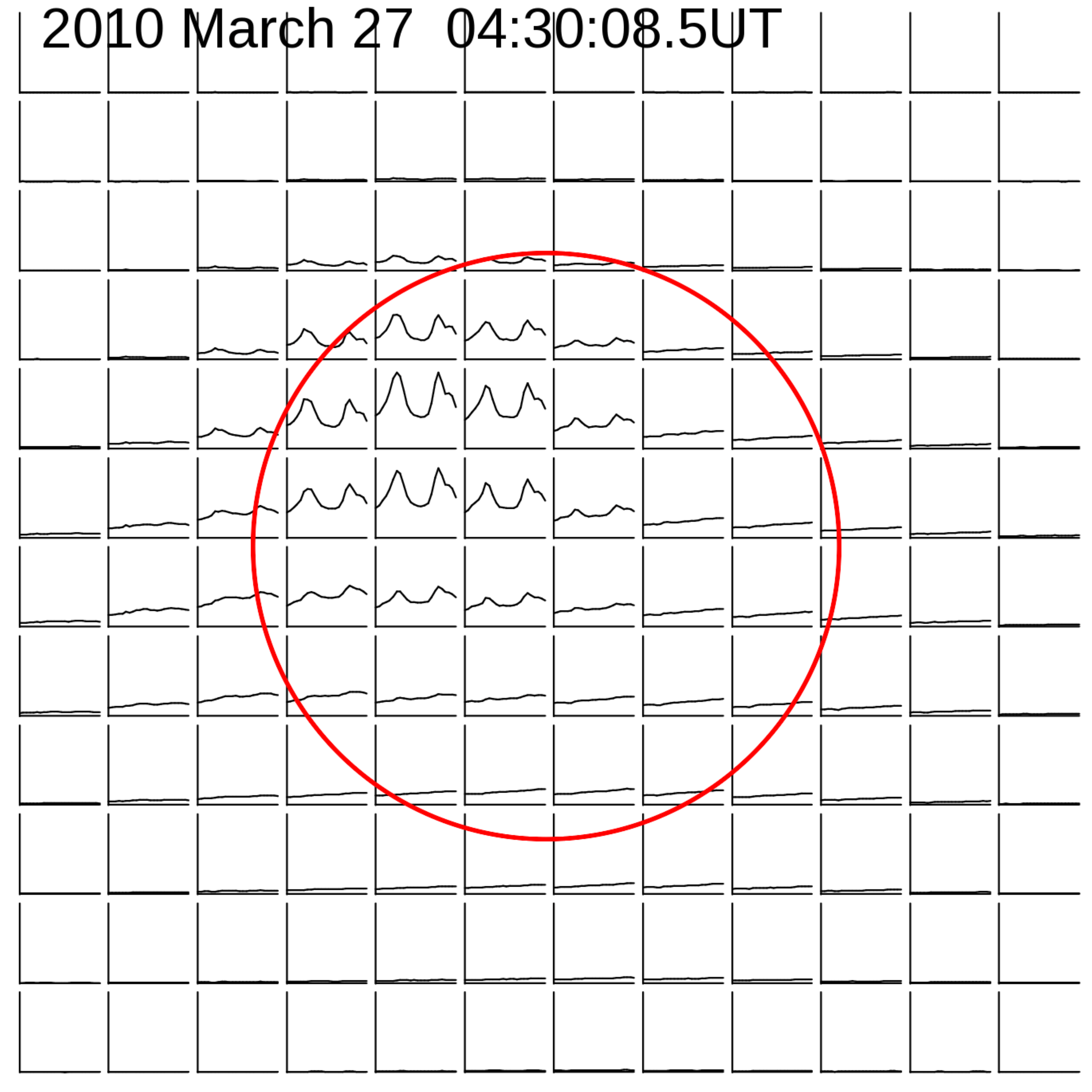}}
\resizebox{0.30\hsize}{!}{\includegraphics[angle=0, trim= 15mm 15mm 15mm 0mm, clip=true]{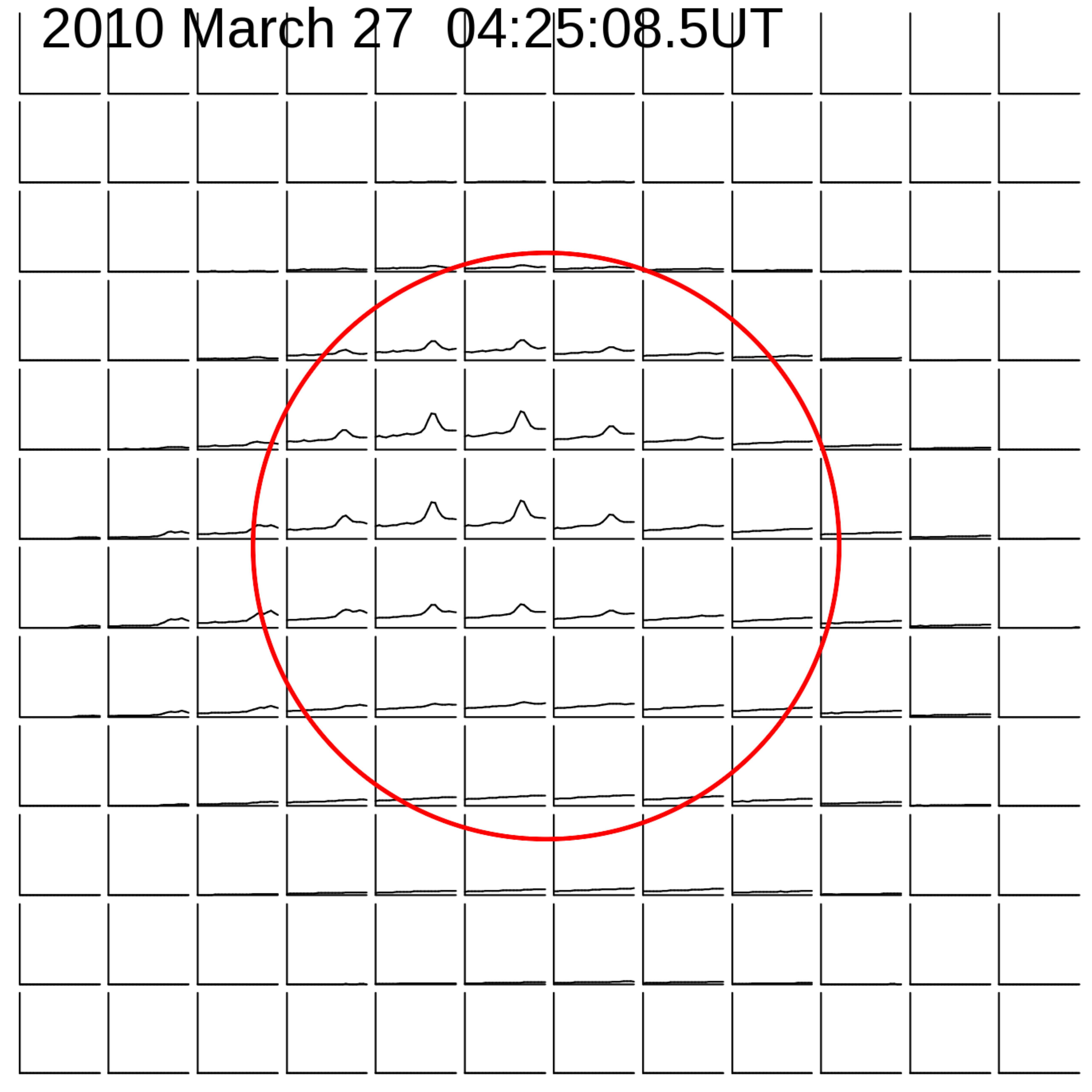}}
\resizebox{0.30\hsize}{!}{\includegraphics[angle=0, trim= 15mm 15mm 15mm 0mm, clip=true]{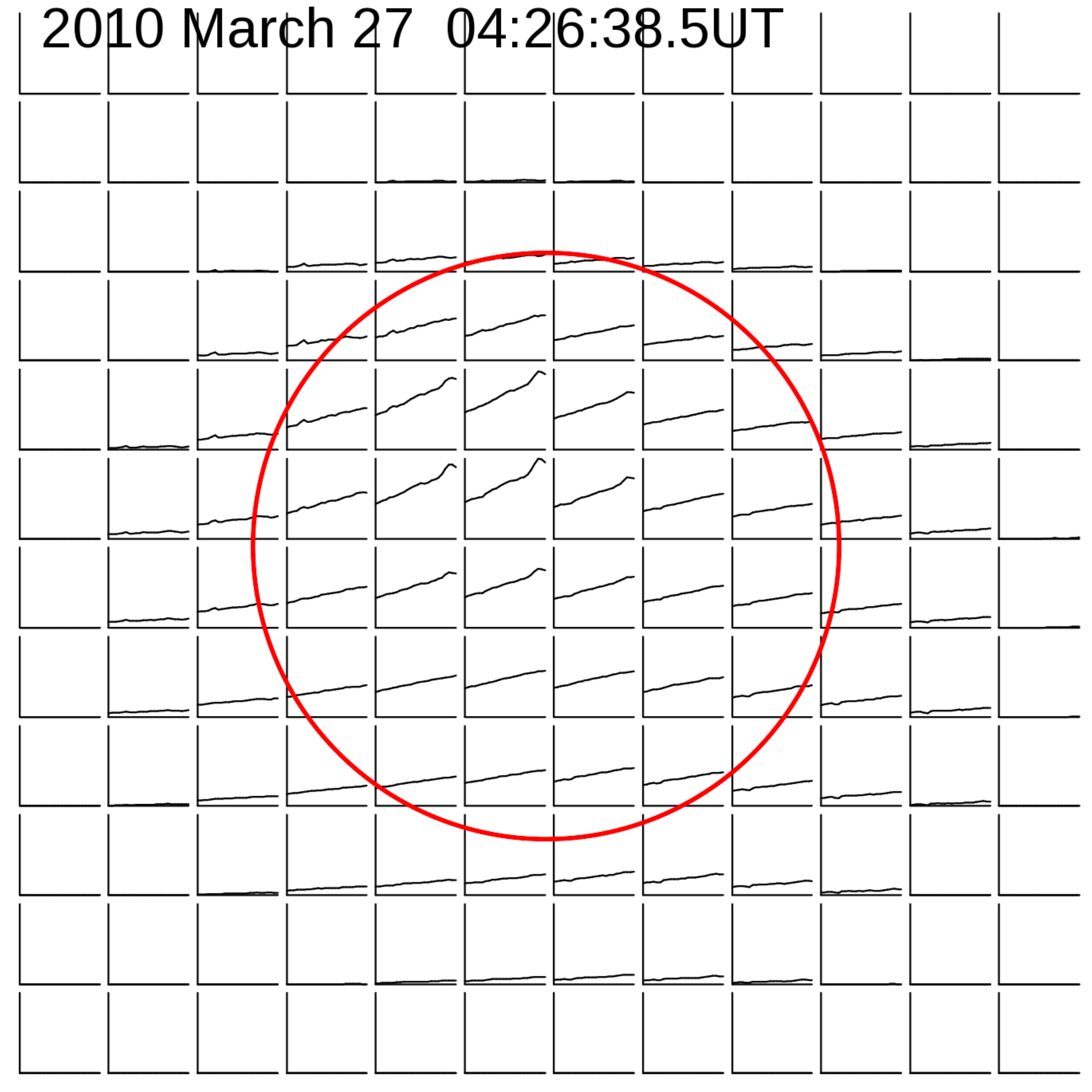}}
}
\caption{The top panels show a set of radio images at 193.3 MHz from the MWA development system on the same intensity scale \citep{oberoietal2011}.  From left to right, the first one comes from a time close to the peak of a weak non-thermal emission feature lasting $\sim$10~s and occupying the entire $\sim$30~MHz of observing bandwidth, the second is during one of numerous weaker events that spanned $\sim$5~s and $\sim$5~MHz and last one shows an image of steady solar emission for the nearly quiescent Sun, although with an active region still visible in the northern hemisphere. The dynamic range of these images is $\sim$2500 and exceeds that of earlier images by about an order of magnitude.  The bottom panels show spatially localized spectra of emission across the solar disc at the same times as the top panels.  The spectra are shown for every third pixel, the pixel size is 100$''\times$100$''$.  The spectra span the entire observing bandwidth, binned into 24 frequency bands spaced 1.28~MHz apart and averaged over 0.8~MHz.  The $y$-axis ranges in arbitrary units are (left to right) 50--4620, 100--6000 and 100--1800.  Celestial north is on top, and the red circles mark the size of the optical solar disc.
}
\label{Fig:SHI2}
\end{center}
\end{figure*}

The rapid spectroscopic imaging capability of MWA will address these issues by localizing burst emission.  An initial example of this capability is shown in Figure~\ref{Fig:SHI2}, in which the MWA development system imaged a solar burst at 15~arcminute angular resolution \citep{oberoietal2011}.  The 2~arcminute angular resolution of the full MWA will be nearly an order of magnitude better.   This will help to determine the shock mechanism by providing radio positions to compare with CME observations (e.g. from STEREO), and it will improve density profile estimates by eliminating the need to infer burst location based on plasma frequency.  In combination with the recent detailed models for the source regions and dynamics spectra of type II burst emission, as reviewed by \citet{cairns2011}, the MWA images and dynamic spectra should help to determine definitively the relationship between type II bursts, CMEs, and shocks and to provide strong tests of our understanding of the propagation of events between the source and Earth. 


\subsubsection{Direct Imaging of Emission from CME Plasma}

Though CMEs are routinely imaged in visible light using coronagraphs, and CME-driven shocks produce broadband radio emissions over several hundred~MHz to as narrow as 20~kHz, the CME plasma itself has been difficult to observe directly. Meter and decimeter frequencies are regarded as the most suitable in which to try to image the synchrotron, gyrosynchrotron, and/or thermal bremsstrahlung emission from the CME plasma \citep{bastian_and_gary1997}. Such measurements offer very useful and unique diagnostic capabilities, including the ability to constrain the CME magnetic field remotely. To date, however, only a few events have been confidently interpreted this way  \citep[e.g.][]{bastianetal2001, maiaetal2000}. Observing these events is particularly challenging for existing instruments due to the large contrasts between the inherent brightness of the non-thermal emission from the CME plasma (weak) and solar disc (strong), and other phenomenon which are often simultaneously present (e.g. noise storms), and the large angular scales of this relatively weak emission. With its significantly higher dynamic range imaging ability and higher sensitivity to extended emission scales, owing to its numerous short baselines, the MWA should be the most suitable instrument yet for direct imaging of emissions from the CME plasma.

\subsection{Heliospheric Science}
\label{shi-helio}

With an increasing reliance on the robust functioning of satellites, humankind is now more affected by space weather than ever before. Reliable predictions of space weather are needed to protect our technological assets in orbit as well as humans in space. The largest of the space weather events are caused by CMEs, and the geo-effectiveness of CMEs is closely tied to the presence of a southward directed magnetic field in the CME or behind the CME-driven shock.  Currently the only operational means of obtaining this information is from the {\em in-situ} observations by the near-Earth satellites. This approach suffers from two significant disadvantages: 1) These measurements provide a lead time of only a few hours, consequently the advance warning comes so close to the event itself that it leaves little opportunity to take any remedial measures.  2) At these crucial times, interactions between the CME's magnetic field and plasma significantly alter the coupling between Earth's ionosphere and magnetosphere and the solar wind, leading to major changes in energetic particle flux, plasma density, plasma motion, magnetic and electric fields, and ionospheric activity.  Degradation in satellite hardware and communication performance then occur, as well as many other space weather effects.

\subsubsection{Faraday Rotation}
\label{sec:shi-fr}

Observation of Faraday rotation of background polarized radio sources is the only known remote sensing technique sensitive to the orientation and magnitude of the interplanetary magnetic field.  Changes in FR associated with the passage of the solar corona or CMEs through the line of sight to spacecraft were convincingly detected using the telemetry signal from the Pioneer 6 and Helios satellites \citep{stelzried1970, bird1982, bird85}. More recently, coronal FR has been detected using VLA observations of radio galaxies as the Sun passes in front of them \citep{mancuso2000, Spangler09}.  The ability to routinely monitor and track CME magnetic fields would revolutionize the field of space weather. 

Faraday rotation measurements have been applied to many interesting studies of the nature of the solar corona and the solar wind, including to determine the mean magnetic field in the corona \citep{patzold1987}, to measure the power spectrum of coronal waves and turbulence as a function of distance \citep{efimov1996}, to suggest the existence of intermittent waves at single locations \citep{chashei1999}, to support theoretical models of wave dissipation as a coronal heating mechanism \citep{hollweg1982}, and to study the structure of the heliospheric current sheet \citep{mancuso2000}. 

Studies so far have been severely limited by the paucity of suitable linearly polarized sources and the narrow fields of view of existing instrumentation \citep{Spangler09}.  Observations using the WSRT at 340--370 MHz have lead to detection of polarized extra-galactic sources exhibiting typical polarized intensities of 20~mJy and readily measurable RMs \citep{wie93, hav03a, hav03b, hav03c}.  In the few areas of the sky investigated in detail, these efforts reveal about one suitable polarized source every $\sim$4~square degrees, as well as patches of diffuse polarized emission \citep{Bernardi13}, suggesting that there will be sufficiently numerous background sources to chart CMEs.  

The combination of the sensitivity of the MWA, its low frequency of operation, and its wide field of view, along with the availability of a large number of polarized sources with sufficient flux density, present the possibility of being able to measure the coronal and heliospheric FR simultaneously to several different lines of sight and out to much larger distances from the Sun ($\sim$ 50 solar radii) than previously accessible. No measurements of FR due to CME plasma are available yet at large solar offsets, hence we rely on simulations to build quantiative expectations. \citet{Jensen10} estimate the RM from CMEs to be $\sim$0.05~rad~m$^{-2}$ (12$^{\circ}$ at 150~MHz) at about 0.4~AU.  \citet{liu2007} also arrive at a similar estimate and also show that the FR time series from instruments like the MWA during the CME passage may be used to infer the helicity and size of the CME.

For multiple reasons, these are rather challenging measurements to make \citep{Oberoi13}.  Ionospheric RM, is 1--10~rad~m$^{-2}$, more than an order of magnitude larger than the expected heliospheric or CME RM signature.  This implies that we will need to achieve ionospheric RM calibration accuracies of a few percent over large fields of view in order to study heliospheric RM.
The quiet solar flux density is 10$^4$--10$^5$ Jy, hence detecting 20~mJy sources requires polarimetric imaging dynamic range of about 10$^6$, in the vicinity of a strong and variable Sun.

\subsubsection{Interplanetary Scintillation}

As radio waves from a distant compact radio source propagate through the turbulent and inhomogeneous solar wind plasma, the incident plane wavefronts pick up phase corrugations corresponding to density inhomogeneities.  These phase corrugations develop into intensity fluctuations by the time the wavefronts arrive at Earth-based telescope.
Due to the motion of the solar wind, the interference pattern produced by these intensity fluctuations is swept past the telescope giving rise to observed scintillations.

The evolution and dynamics of solar wind in the inner heliosphere dominates space weather effects. However, there are few techniques available to study the solar wind in this vast region.
Historically, interplanetary scintillation (IPS) was the first remote sensing technique used to investigate the solar wind in this region \citep{hewish64}.  More recently, routine availability of observations from heliospheric imagers  (e.g., STEREO) that image very large parts of the heliosphere using white light scattered from from the solar wind electrons, have emerged as very useful tools for investigations in the inner heliosphere.  These techniques, along with local {\em in-situ} sampling provided by most satellite based instrumentation, complement each other well since they cover different spatial scales and have lead to a more complete characterisation of the solar wind than would have been possible with either of the datasets \citep[e.g.][]{Jackson07, Dorrian08}.

Several new instruments are coming online to study IPS.  Apart from the MWA, the new facilities include the LOFAR \citep{Bisi11,Fallows12} and Mexico Array Telescope \citep{Mejia10}.  In addition, the Ooty Radio Telescope in India (Manoharan, P.~K., private communication) and Solar Wind Imaging Facility (SWIFT) in Japan \citep{tokumaru11} are both undegoing considerable upgrades which will substantially improve their scientific capabilities. 

 For the MWA, the primary IPS observable will be the power spectrum of observed intensity fluctuations, although we a will also employ a less commonly used observable, the normalized co-spectrum obtained from simultaneous observations at frequencies differing by a factor of $\sim$2 \citep{Scott83, Zhang07}.  Under the assumption of weak scattering, the interplanetary medium can be regarded as a collection of thin screens placed along the line-of-sight, and the IPS observable can be modeled as a weighted sum of the contributions from each of the thin screens.  The physical parmeters to which IPS is sensitive are: 1) the velocity of the solar wind; 2) the level of electron density fluctuations, $\delta n_e$, 3) the spectral index of the $\delta n_e$ power spectrum, 4) the inner scale for the power law governing $\delta n_e$; and 5) the axial ratio of $\delta n_e$.

Using an inversion process, it is possible to extract the global 3D distribution of IPM parameters \citep{Jackson98, Kojima98, Asai98, Oberoi00, Bisi08, Bisi10, Dorrian10, Xiong11}.  The most important bottleneck currently constraining the fidelity and resolution of inversion of IPS data, and hence its science output, are the sparseness (both in space and time) and the non-uniformity of the sampling of the heliosphere afforded by the present generation instrumentation \citep{OberoiandBenkevitch2010}.  The MWA will represent a significant advance in addressing the bottleneck of limited heliospheric sampling when its voltage capture system is completed.  The MWA will have a sensitivity of $\sim$70 mJy (1 second, 30.72 MHz), comparable to that of the SWIFT transit facility currently under commissioning in Japan.  The MWA will boost the number of independent IPS measurements by up to a factor of $\sim$30 over what is possible with most present day systems.  The southern hemisphere location of the MWA also complements the heliospheric coverage provided by the existing facilities, all of which happen to be in the northern hemisphere.  Combining these data with other MWA observations like solar imaging and FR observations, and the observations from other sources like STEREO, SDO and HINODE, will provide a very comprehensive dataset for charaterizing the inner heliosphere, monitoring and tracking CME evolution from close to the Sun to our terrestrial neighborhood.  Further, the software framework being developed for handling the MWA IPS data could provide a common data repository, interface, and analysis tools for IPS data from observatories across the world \citep{OberoiandBenkevitch2010}.

\subsection{Ionospheric Science}

Propagation through the dynamic and structured magnetoionic ionospheric plasma leaves an imprint on the incident low radio frequency radiation from astronomical sources.
In order to arrive at the true properties of the incident radiation, the ionospheric distortion must be identified and removed. The MWA can tackle this long standing challenge by measuring the refractive shift in the apparent position of dozens of bright and relatively compact sources at a cadence of 8 seconds.  The ionospheric calibration information thus comprises of a time series of vectors showing the refractive shift of calibrator sources from their fiducial positions.  To first order, the magnitude of the shift is directly proportional to the ionospheric gradient in the total electron content (TEC) towards the fiducial sky position of the calibrator source.   This information for all the observed calibrator sources leads to a series of differential TEC maps and an accuracy of $\sim$0.01 TECU (1~TECU$=10^{16}$~electrons~m$^{-2}$). 
These maps, produced routinely as a part of the MWA calibration will form the primary data products for ionospheric science.  The MWA beam will sample a region about $175$ km $\times175$ km at a  height of $\sim300$ km and the expected calibrator source density corresponds to spatial scales of $\sim10$ km at this ionospheric height.  

The high-precision and dense sampling of the ionosphere provided by the MWA offers new opportunities for ionospheric science.   The three areas of ionospheric research that the MWA will make significant contributions are in:  1) the measuring of waves in TEC associated with travelling ionospheric disturbances (TIDs);  2) the development of storm enhanced density (SED) and sharp TEC gradients associated with space weather events; and  3) the tracking of the onset and evolution of scintillation associated with the development of ionospheric irregularities.  

\subsubsection{Traveling Ionospheric Disturbances}

TIDs are manifestations of middle-scale ionospheric irregularities arising as response to acoustic gravity waves.   TIDs are frequently observed at high and middle latitudes \citep{bristowandgreenwald1996, nicollsandheinselman2007, oliveretal1994, oliveretal1997, djuthetal2004}, and their activity and amplitudes vary depending on latitude, longitude, local time, season, and solar cycle \citep[e.g.][]{kotakeetal2006, pajaresetal2006}.  They can be generated by auroral sources at the high latitude thermosphere, the passage of the solar terminator \citep{galushkoetal2003}, and  storms, hurricanes and tornados in the troposphere.  Detailed information about the sources, energetics, and scale sizes of energy coupling from the lower atmosphere to the upper atmosphere, along with resulting ionospheric effects, is critical to developing a deep understanding of overall upper atmospheric energy balance. Measurements of TIDs at the VLA have been made since the early 1990s \citep{1993AnGeo..11..869J, 1995JGR...100.1653J} and recent observations have been reported by \citet{2012RaSc...47.0L02H, 2012RaSc...47.5008H}.  The fine grid over which the MWA calibration solutions will provide differential TEC measurements will provide detailed measurements of the propagation of TIDs over the array and TEC gradients or irregularities. 

\subsubsection{Storm Enhanced Density}


The MWA is well-sited to contribute to coordinate international investigations of the SED phenomenon \citep{foster1993}.  This process is responsible for some of the largest space weather effects at mid-latitudes \citep{costerandfoster2007} since it produces severe, temporally dependent electron density gradients.  These gradients will be obvious in the differential TEC measurements made by the MWA.  The geographic/geomagnetic pole offset in the southern hemisphere is anticipated to influence the development of SED near the MWA 
\citep{costeretal2007, fosterandcoster2007, fosteranderickson2012}.  
SED associated gradients have a practical consequence, as they cause large errors 
in GPS range measurements, and significantly degrade the high precision 
positioning information provided by GPS systems. The MWA's longitude ($\sim116^{\circ}$E) is very close to the longitude 
being extensively instrumented in the Asian sector as part of the Chinese Meridian project ($\sim113^{\circ}$), and the conjugate longitude of $\sim64^{\circ}$W is within the field of view of the US sponsored chain of incoherent scatter radars, in particular Millstone Hill.  The combination of simultaneous measurements from these distributed sensors thus provides a unique way to determine mesoscale response to geomagnetic disturbances in the coupled geospace system.  In particular, it will be possible to investigate whether there is a seasonal and universal time dependence of geomagnetic storm response by comparing American and Asian/Australian sector response to solar wind and CME forcing.  Such effects have been predicted theoretically \citep{fosteranderickson2012} but have not yet been confirmed by observations.

\subsubsection{Ionospheric Scintillation}

Ionospheric scintillation can be defined as the rapid fluctuation of radio-frequency signal phase and/or amplitude, generated as a signal passes through the ionosphere.   MWA observations using voltage capture capabilities will be uniquely suited for the exploration of ionospheric irregularities and scintillation.  A voltage capture system that can record raw antenna voltages from the full array for 24~hours will be added to the MWA by the end of 2013.  This system should provide the ability to create array-beams in post-processing that use the entire bandwidth and collecting area of the MWA, and that can be pointed independently anywhere within the MWA field-of-view.  Such a system will provide scintillation measurements at multiple positions across the visible ionospheric region.  

MWA array-beam observations will be augmented by data from specially designed GPS receivers at and near the MWA site that can measure L-band scintillation statistics.  GPS TEC data from these receivers, when combined with data from a large 
number of newly deployed GPS receivers on the Australian continent, will 
provide important contextual information about the regional distribution 
of TEC around the MWA site.  Other satellite based remote sensing platforms, including 
COSMIC, DMSP, and JASON, will also be available during overflights to provide further information about background ionospheric plasma and electrodynamic conditions in the 
ionosphere.  A similar approach has been followed previously in the Combined Radio Interferometry and COSMIC Experiment in Tomography (CRICKET) campaign utilizing the VLA \citep{2008cosp...37..775D, 2009RaSc...44.0A11C}.  This mix of data types allows for ionospheric specification from the global scale down to the regional scale.



\section{Conclusion}
\label{s_conclusion}
The MWA is a unique low-frequency instrument with high survey efficiency and arcminute angular resolution.  It will provide deep integrations on EoR target regions that could enable the first detection of the redshifted 21~cm signal and detailed characterisation of diffuse Galactic emisison and faint extragalactic source populations.   The planned full-Stokes southern hemisphere sky survey will include the Galactic Centre, Galactic plane, and the Large and Small Magellanic Clouds, making the MWA particurly well-suited to identifying the missing population of SNRs, and thereby constraining the total energy budget of the Galaxy.  

Long integrations in the EoR fields and regular obserations along the Galactic plane will enable sensitive searches for pulsars and transient events on timescales from seconds to days.  Blind surveys for both rare and faint events are planned, as well as targeted obsrevations of known transient sources, including X-ray binaries, neutron stars, pulsars, and low-mass stellar objects. 

The MWA will open new windows into space weather and solar physics and provide new measurements of solar bursts throughout their journey from the surface of the Sun to the Earth and on to the outer Solar System.  High-dynamic range spectroscopic imaging will show the origin of bursts in the solar corona, while monitoring of interplanetary scintillation will probe the interstellar wind.  As events reach the Earth, the MWA's calibration system will yield detailed measurements of ionospheric disturbances traveling over the telescope.  

The telescope is complementary to other new low-frequency radio telescopes, falling between LOFAR and PAPER in both sensitivity and angular resolution.  It is the only low-frequency SKA precursor and is located on the very radio-quiet site planned for the SKA low-band telescope.  The combination of ASKAP and MWA at the MRO will yield a comprehensive program of precursor science for the SKA over the comining years.  The MWA posseses a versatile design with excess infrastructure capability that will allow the telescope and its users to embrace new scientific and technological opportunities over the coming decade, hopefully providing an enduring and fruitful future for the instrument.

\section*{Acknowledgments}

This scientific work makes use of the Murchison Radio-astronomy Observatory. We acknowledge the Wajarri Yamatji people as the traditional owners of the Observatory site. Support for the MWA comes from the U.S. National Science Foundation (grants AST-0457585, PHY-0835713, CAREER-0847753, and AST-0908884), the Australian Research Council (LIEF grants LE0775621 and LE0882938), the U.S. Air Force Office of Scientic Research (grant FA9550-0510247), and the Centre for All-sky Astrophysics (an Australian Research Council Centre of Excellence funded by grant CE110001020). Support is also provided by the Smithsonian Astrophysical Observatory, the MIT School of Science, the Raman Research Institute, the Australian National University, and the Victoria University of Wellington (via grant MED-E1799 from the New Zealand Ministry of Economic Development and an IBM Shared University Research Grant). The Australian Federal government provides additional support via the National Collaborative Research Infrastructure Strategy, Education Investment Fund, and the Australia India Strategic Research Fund, and Astronomy Australia Limited, under contract to Curtin University. We acknowledge the iVEC Petabyte Data Store, the Initiative in Innovative Computing and the CUDA Center for Excellence sponsored by NVIDIA at Harvard University, and the International Centre for Radio Astronomy Research (ICRAR), a Joint Venture of Curtin University and The University of Western Australia, funded by the Western Australian State government.


\begin{thebibliography}{}

\bibitem[\protect\citeauthoryear{{Alves} et~al.}{{Alves}
  et~al.}{2012}]{2012MNRAS.422.2429A}
{Alves}, M.~I.~R., {Davies}, R.~D., {Dickinson}, C., {Calabretta}, M., {Davis},
  R.,  \& {Staveley-Smith}, L. 2012, \mnras, 422, 2429

\bibitem[\protect\citeauthoryear{{Amy}, {Large}, \& {Vaughan}}{{Amy}
  et~al.}{1989}]{alv89}
{Amy}, S.~W., {Large}, M.~I.,  \& {Vaughan}, A.~E. 1989, Proceedings of the
  Astronomical Society of Australia, 8, 172

\bibitem[\protect\citeauthoryear{{Antonova} et~al.}{{Antonova}
  et~al.}{2007}]{adh+07}
{Antonova}, A., {Doyle}, J.~G., {Hallinan}, G., {Golden}, A.,  \& {Koen}, C.
  2007, \aap, 472, 257

\bibitem[\protect\citeauthoryear{{Araya-Melo} et~al.}{{Araya-Melo}
  et~al.}{2012}]{2012MNRAS.423.2325A}
{Araya-Melo}, P.~A., {Arag{\'o}n-Calvo}, M.~A., {Br{\"u}ggen}, M.,  \& {Hoeft},
  M. 2012, \mnras, 423, 2325

\bibitem[\protect\citeauthoryear{{Asai} et~al.}{{Asai} et~al.}{1998}]{Asai98}
{Asai}, K., {Kojima}, M., {Tokumaru}, M., {Yokobe}, A., {Jackson}, B.~V.,
  {Hick}, P.~L.,  \& {Manoharan}, P.~K. 1998, \jgr, 103, 1991

\bibitem[\protect\citeauthoryear{{Asgekar} \& {Deshpande}}{{Asgekar} \&
  {Deshpande}}{2001}]{ad01}
{Asgekar}, A.,  \& {Deshpande}, A.~A. 2001, \mnras, 326, 1249

\bibitem[\protect\citeauthoryear{{Audard} et~al.}{{Audard}
  et~al.}{2007}]{aob+07}
{Audard}, M., {Osten}, R.~A., {Brown}, A., {Briggs}, K.~R., {G{\"u}del}, M.,
  {Hodges-Kluck}, E.,  \& {Gizis}, J.~E. 2007, \aap, 471, L63

\bibitem[\protect\citeauthoryear{{Baird} et~al.}{{Baird} et~al.}{1976}]{bmj+76}
{Baird}, G.~A., {Meikle}, W.~P.~S., {Jelley}, J.~V., {Palumbo}, G.~G.~C.,  \&
  {Partridge}, R.~B. 1976, \apss, 42, 69

\bibitem[\protect\citeauthoryear{{Balsano}}{{Balsano}}{1999}]{balsano99}
{Balsano}, R.~J. 1999, Ph.D. thesis, Princeton University

\bibitem[\protect\citeauthoryear{{Banfield} et~al.}{{Banfield}
  et~al.}{2011}]{2011ApJ...733...69B}
{Banfield}, J.~K., {George}, S.~J., {Taylor}, A.~R., {Stil}, J.~M., {Kothes},
  R.,  \& {Scott}, D. 2011, \apj, 733, 69

\bibitem[\protect\citeauthoryear{{Bania} et~al.}{{Bania}
  et~al.}{2010}]{2010ApJ...718L.106B}
{Bania}, T.~M., {Anderson}, L.~D., {Balser}, D.~S.,  \& {Rood}, R.~T. 2010,
  \apjl, 718, L106

\bibitem[\protect\citeauthoryear{{Bannister} et~al.}{{Bannister}
  et~al.}{2011}]{bmg+11}
{Bannister}, K.~W., {Murphy}, T., {Gaensler}, B.~M., {Hunstead}, R.~W.,  \&
  {Chatterjee}, S. 2011, \mnras, 412, 634

\bibitem[\protect\citeauthoryear{{Barkana} \& {Loeb}}{{Barkana} \&
  {Loeb}}{2005}]{2005ApJ...624L..65B}
{Barkana}, R.,  \& {Loeb}, A. 2005, \apjl, 624, L65

\bibitem[\protect\citeauthoryear{{Barkana} \& {Loeb}}{{Barkana} \&
  {Loeb}}{2008}]{2008MNRAS.384.1069B}
{Barkana}, R.,  \& {Loeb}, A. 2008, \mnras, 384, 1069

\bibitem[\protect\citeauthoryear{{Bastian}}{{Bastian}}{2004}]{bastian2004}
{Bastian}, T.~S. 2004, Planetary and Space Science, 52, 1381

\bibitem[\protect\citeauthoryear{{Bastian}, {Benz}, \& {Gary}}{{Bastian}
  et~al.}{1998}]{bastianetal1998}
{Bastian}, T.~S., {Benz}, A.~O.,  \& {Gary}, D.~E. 1998, \araa, 36, 131

\bibitem[\protect\citeauthoryear{{Bastian}, {Dulk}, \& {Leblanc}}{{Bastian}
  et~al.}{2000}]{bdl00}
{Bastian}, T.~S., {Dulk}, G.~A.,  \& {Leblanc}, Y. 2000, \apj, 545, 1058

\bibitem[\protect\citeauthoryear{{Bastian} \& {Gary}}{{Bastian} \&
  {Gary}}{1997}]{bastian_and_gary1997}
{Bastian}, T.~S.,  \& {Gary}, D.~E. 1997, \jgr, 102, 14031

\bibitem[\protect\citeauthoryear{{Bastian} et~al.}{{Bastian}
  et~al.}{2001}]{bastianetal2001}
{Bastian}, T.~S., {Pick}, M., {Kerdraon}, A., {Maia}, D.,  \& {Vourlidas}, A.
  2001, \apjl, 558, L65

\bibitem[\protect\citeauthoryear{{Beardsley} et~al.}{{Beardsley}
  et~al.}{2013}]{Beardsley2012}
{Beardsley}, A.~P., et~al. 2013, \mnras, 429, L5

\bibitem[\protect\citeauthoryear{{Beck} \& {Gaensler}}{{Beck} \&
  {Gaensler}}{2004}]{2004NewAR..48.1289B}
{Beck}, R.,  \& {Gaensler}, B.~M. 2004, New Astronomy Reviews, 48, 1289

\bibitem[\protect\citeauthoryear{{Becker}, {Rauch}, \& {Sargent}}{{Becker}
  et~al.}{2007}]{2007ApJ...662...72B}
{Becker}, G.~D., {Rauch}, M.,  \& {Sargent}, W.~L.~W. 2007, \apj, 662, 72

\bibitem[\protect\citeauthoryear{{Belikov} \& {Hooper}}{{Belikov} \&
  {Hooper}}{2009}]{2009arXiv0904.1210B}
{Belikov}, A.~V.,  \& {Hooper}, D. 2009, \prd, 80, 035007

\bibitem[\protect\citeauthoryear{{Benz} \& {Paesold}}{{Benz} \&
  {Paesold}}{1998}]{bp98}
{Benz}, A.~O.,  \& {Paesold}, G. 1998, \aap, 329, 61

\bibitem[\protect\citeauthoryear{{Berger}}{{Berger}}{2006}]{berger06}
{Berger}, E. 2006, \apj, 648, 629

\bibitem[\protect\citeauthoryear{{Berger} et~al.}{{Berger}
  et~al.}{2001}]{bbb+01}
{Berger}, E., et~al. 2001, \nat, 410, 338

\bibitem[\protect\citeauthoryear{{Bernardi} et~al.}{{Bernardi}
  et~al.}{2009}]{2009A&A...500..965B}
{Bernardi}, G., et~al. 2009, \aap, 500, 965

\bibitem[\protect\citeauthoryear{{Bernardi} et~al.}{{Bernardi}
  et~al.}{2013}]{Bernardi13}
{Bernardi}, G., et~al. 2013, \apj, submitted

\bibitem[\protect\citeauthoryear{{Bernardi} et~al.}{{Bernardi}
  et~al.}{2011}]{2011MNRAS.413..411B}
{Bernardi}, G., {Mitchell}, D.~A., {Ord}, S.~M., {Greenhill}, L.~J., {Pindor},
  B., {Wayth}, R.~B.,  \& {Wyithe}, J.~S.~B. 2011, \mnras, 413, 411

\bibitem[\protect\citeauthoryear{{Bhat} et~al.}{{Bhat} et~al.}{2004}]{bcc+04}
{Bhat}, N.~D.~R., {Cordes}, J.~M., {Camilo}, F., {Nice}, D.~J.,  \& {Lorimer},
  D.~R. 2004, \apj, 605, 759

\bibitem[\protect\citeauthoryear{{Bhat} et~al.}{{Bhat} et~al.}{2007}]{bwk+07}
{Bhat}, N.~D.~R., et~al. 2007, \apj, 665, 618

\bibitem[\protect\citeauthoryear{{Bird}}{{Bird}}{1982}]{bird1982}
{Bird}, M.~K. 1982, Space Science Reviews, 33, 99

\bibitem[\protect\citeauthoryear{{Bird} et~al.}{{Bird} et~al.}{1985}]{bird85}
{Bird}, M.~K., et~al. 1985, \solphys, 98, 341

\bibitem[\protect\citeauthoryear{{Bisi} et~al.}{{Bisi} et~al.}{2011}]{Bisi11}
{Bisi}, M.~M., {Fallows}, R.~A., {Jensen}, E.~A., {Breen}, A., {Xiong}, M.,  \&
  {Jackson}, B.~V. 2011, AGU Fall Meeting Abstracts, C2020

\bibitem[\protect\citeauthoryear{{Bisi} et~al.}{{Bisi} et~al.}{2010}]{Bisi10}
{Bisi}, M.~M., {Jackson}, B.~V., {Breen}, A.~R., {Dorrian}, G.~D., {Fallows},
  R.~A., {Clover}, J.~M.,  \& {Hick}, P.~P. 2010, \solphys, 265, 233

\bibitem[\protect\citeauthoryear{{Bisi} et~al.}{{Bisi} et~al.}{2008}]{Bisi08}
{Bisi}, M.~M., {Jackson}, B.~V., {Hick}, P.~P., {Buffington}, A., {Odstrcil},
  D.,  \& {Clover}, J.~M. 2008, Journal of Geophysical Research (Space
  Physics), 113, 0

\bibitem[\protect\citeauthoryear{{Blasi} \& {Colafrancesco}}{{Blasi} \&
  {Colafrancesco}}{1999}]{1999APh....12..169B}
{Blasi}, P.,  \& {Colafrancesco}, S. 1999, Astroparticle Physics, 12, 169

\bibitem[\protect\citeauthoryear{{Bower} et~al.}{{Bower} et~al.}{2005}]{bry+05}
{Bower}, G.~C., {Roberts}, D.~A., {Yusef-Zadeh}, F., {Backer}, D.~C., {Cotton},
  W.~D., {Goss}, W.~M., {Lang}, C.~C.,  \& {Lithwick}, Y. 2005, \apj, 633, 218

\bibitem[\protect\citeauthoryear{{Bower} et~al.}{{Bower} et~al.}{2007}]{bsb+07}
{Bower}, G.~C., {Saul}, D., {Bloom}, J.~S., {Bolatto}, A., {Filippenko}, A.~V.,
  {Foley}, R.~J.,  \& {Perley}, D. 2007, \apj, 666, 346

\bibitem[\protect\citeauthoryear{{Bowman}, {Morales}, \& {Hewitt}}{{Bowman}
  et~al.}{2006}]{2006ApJ...638...20B}
{Bowman}, J.~D., {Morales}, M.~F.,  \& {Hewitt}, J.~N. 2006, \apj, 638, 20

\bibitem[\protect\citeauthoryear{{Bowman}, {Morales}, \& {Hewitt}}{{Bowman}
  et~al.}{2009}]{2009ApJ...695..183B}
{Bowman}, J.~D., {Morales}, M.~F.,  \& {Hewitt}, J.~N. 2009, \apj, 695, 183

\bibitem[\protect\citeauthoryear{{Boyles} et~al.}{{Boyles}
  et~al.}{2013}]{blr+12}
{Boyles}, J., et~al. 2013, \apj, 763, 80

\bibitem[\protect\citeauthoryear{{Brentjens} \& {de Bruyn}}{{Brentjens} \& {de
  Bruyn}}{2005}]{2005A&A...441.1217B}
{Brentjens}, M.~A.,  \& {de Bruyn}, A.~G. 2005, \aap, 441, 1217

\bibitem[\protect\citeauthoryear{{Bristow}, {Greenwald}, \&
  {Villain}}{{Bristow} et~al.}{1996}]{bristowandgreenwald1996}
{Bristow}, W.~A., {Greenwald}, R.~A.,  \& {Villain}, J.~P. 1996, \jgr, 101,
  15685

\bibitem[\protect\citeauthoryear{{Brogan} et~al.}{{Brogan}
  et~al.}{2006}]{bgg+06}
{Brogan}, C.~L., {Gelfand}, J.~D., {Gaensler}, B.~M., {Kassim}, N.~E.,  \&
  {Lazio}, T.~J.~W. 2006, \apjl, 639, L25

\bibitem[\protect\citeauthoryear{{Brown}, {Farnsworth}, \& {Rudnick}}{{Brown}
  et~al.}{2010}]{2010MNRAS.402....2B}
{Brown}, S., {Farnsworth}, D.,  \& {Rudnick}, L. 2010, \mnras, 402, 2

\bibitem[\protect\citeauthoryear{{Brown}}{{Brown}}{2011}]{2011JApA...32..577B}
{Brown}, S.~D. 2011, Journal of Astrophysics and Astronomy, 32, 577

\bibitem[\protect\citeauthoryear{{Brunetti} et~al.}{{Brunetti}
  et~al.}{2001}]{2001MNRAS.320..365B}
{Brunetti}, G., {Setti}, G., {Feretti}, L.,  \& {Giovannini}, G. 2001, \mnras,
  320, 365

\bibitem[\protect\citeauthoryear{{Burgasser} \& {Putman}}{{Burgasser} \&
  {Putman}}{2005}]{bp05}
{Burgasser}, A.~J.,  \& {Putman}, M.~E. 2005, \apj, 626, 486

\bibitem[\protect\citeauthoryear{{Burke-Spolaor} \& {Bailes}}{{Burke-Spolaor}
  \& {Bailes}}{2010}]{bsb10}
{Burke-Spolaor}, S.,  \& {Bailes}, M. 2010, \mnras, 402, 855

\bibitem[\protect\citeauthoryear{{Burkhart}, {Lazarian}, \&
  {Gaensler}}{{Burkhart} et~al.}{2012}]{2012ApJ...749..145B}
{Burkhart}, B., {Lazarian}, A.,  \& {Gaensler}, B.~M. 2012, \apj, 749, 145

\bibitem[\protect\citeauthoryear{{Cairns}}{{Cairns}}{2004}]{cairns2004}
{Cairns}, I.~H. 2004, Planetary and Space Science, 52, 1423

\bibitem[\protect\citeauthoryear{{Cairns}}{{Cairns}}{2011}]{cairns2011}
{Cairns}, I.~H. 2011, {Coherent Radio Emissions Associated with Star System
  Shocks}, ed. M.~P. {Miralles} \& J.~{S{\'a}nchez Almeida} 267

\bibitem[\protect\citeauthoryear{{Cairns} et~al.}{{Cairns}
  et~al.}{2009}]{cairnsetal2009}
{Cairns}, I.~H., {Lobzin}, V.~V., {Warmuth}, A., {Li}, B., {Robinson}, P.~A.,
  \& {Mann}, G. 2009, \apjl, 706, L265

\bibitem[\protect\citeauthoryear{{Cameron} et~al.}{{Cameron}
  et~al.}{2005}]{ccr+05}
{Cameron}, P.~B., et~al. 2005, \nat, 434, 1112

\bibitem[\protect\citeauthoryear{{Camilo} et~al.}{{Camilo}
  et~al.}{2007}]{crhr07}
{Camilo}, F., {Ransom}, S.~M., {Halpern}, J.~P.,  \& {Reynolds}, J. 2007,
  \apjl, 666, L93

\bibitem[\protect\citeauthoryear{{Camilo} et~al.}{{Camilo}
  et~al.}{2006}]{crh+06}
{Camilo}, F., {Ransom}, S.~M., {Halpern}, J.~P., {Reynolds}, J., {Helfand},
  D.~J., {Zimmerman}, N.,  \& {Sarkissian}, J. 2006, \nat, 442, 892

\bibitem[\protect\citeauthoryear{{Cane} \& {Erickson}}{{Cane} \&
  {Erickson}}{2005}]{caneanderickson2005}
{Cane}, H.~V.,  \& {Erickson}, W.~C. 2005, \apj, 623, 1180

\bibitem[\protect\citeauthoryear{{Cane}, {Erickson}, \& {Prestage}}{{Cane}
  et~al.}{2002}]{caneetal2003}
{Cane}, H.~V., {Erickson}, W.~C.,  \& {Prestage}, N.~P. 2002, Journal of
  Geophysical Research (Space Physics), 107, 1315

\bibitem[\protect\citeauthoryear{{Cassano} et~al.}{{Cassano}
  et~al.}{2012}]{2012_Cassano}
{Cassano}, R., {Brunetti}, G., {Norris}, R.~P., {R{\"o}ttgering}, H.~J.~A.,
  {Johnston-Hollitt}, M.,  \& {Trasatti}, M. 2012, \aap, 548, A100

\bibitem[\protect\citeauthoryear{{Cassano} et~al.}{{Cassano}
  et~al.}{2010}]{2010A&A...509A..68C}
{Cassano}, R., {Brunetti}, G., {R{\"o}ttgering}, H.~J.~A.,  \& {Br{\"u}ggen},
  M. 2010, \aap, 509, A68

\bibitem[\protect\citeauthoryear{{Cen} \& {Ostriker}}{{Cen} \&
  {Ostriker}}{1999}]{1999ApJ...514....1C}
{Cen}, R.,  \& {Ostriker}, J.~P. 1999, \apj, 514, 1

\bibitem[\protect\citeauthoryear{{Chakraborti} et~al.}{{Chakraborti}
  et~al.}{2011}]{crs+10}
{Chakraborti}, S., {Ray}, A., {Soderberg}, A.~M., {Loeb}, A.,  \& {Chandra}, P.
  2011, Nature Communications, 2

\bibitem[\protect\citeauthoryear{{Charbonneau} et~al.}{{Charbonneau}
  et~al.}{2002}]{cbng02}
{Charbonneau}, D., {Brown}, T.~M., {Noyes}, R.~W.,  \& {Gilliland}, R.~L. 2002,
  \apj, 568, 377

\bibitem[\protect\citeauthoryear{{Chashei} et~al.}{{Chashei}
  et~al.}{1999}]{chashei1999}
{Chashei}, I.~V., {Bird}, M.~K., {Efimov}, A.~I., {Andreev}, V.~E.,  \&
  {Samoznaev}, L.~N. 1999, \solphys, 189, 399

\bibitem[\protect\citeauthoryear{{Chen} \& {Miralda-Escud{\'e}}}{{Chen} \&
  {Miralda-Escud{\'e}}}{2004}]{2004ApJ...602....1C}
{Chen}, X.,  \& {Miralda-Escud{\'e}}, J. 2004, \apj, 602, 1

\bibitem[\protect\citeauthoryear{{Coker} et~al.}{{Coker}
  et~al.}{2009}]{2009RaSc...44.0A11C}
{Coker}, C., {Thonnard}, S.~E., {Dymond}, K.~F., {Lazio}, T.~J.~W., {Makela},
  J.~J.,  \& {Loughmiller}, P.~J. 2009, Radio Science, 44, 0

\bibitem[\protect\citeauthoryear{{Colgate} \& {Noerdlinger}}{{Colgate} \&
  {Noerdlinger}}{1971}]{cn71}
{Colgate}, S.~A.,  \& {Noerdlinger}, P.~D. 1971, \apj, 165, 509

\bibitem[\protect\citeauthoryear{{Condon}}{{Condon}}{1974}]{1974ApJ...188..279C}
{Condon}, J.~J. 1974, \apj, 188, 279

\bibitem[\protect\citeauthoryear{{Condon} et~al.}{{Condon}
  et~al.}{1998}]{1998AJ....115.1693C}
{Condon}, J.~J., {Cotton}, W.~D., {Greisen}, E.~W., {Yin}, Q.~F., {Perley},
  R.~A., {Taylor}, G.~B.,  \& {Broderick}, J.~J. 1998, \aj, 115, 1693

\bibitem[\protect\citeauthoryear{{Corbel} et~al.}{{Corbel}
  et~al.}{2005}]{ckf+05}
{Corbel}, S., {Kaaret}, P., {Fender}, R.~P., {Tzioumis}, A.~K., {Tomsick},
  J.~A.,  \& {Orosz}, J.~A. 2005, \apj, 632, 504

\bibitem[\protect\citeauthoryear{Cordes \& Lazio}{Cordes \& Lazio}{2002}]{cl02}
Cordes, J.~M.,  \& Lazio, T. J.~W. 2002, astro-ph/0207156

\bibitem[\protect\citeauthoryear{{Cordes}, {Lazio}, \& {McLaughlin}}{{Cordes}
  et~al.}{2004}]{clm04}
{Cordes}, J.~M., {Lazio}, T.~J.~W.,  \& {McLaughlin}, M.~A. 2004, New Astronomy
  Review, 48, 1459

\bibitem[\protect\citeauthoryear{{Cordes} \& {McLaughlin}}{{Cordes} \&
  {McLaughlin}}{2003}]{cm03}
{Cordes}, J.~M.,  \& {McLaughlin}, M.~A. 2003, \apj, 596, 1142

\bibitem[\protect\citeauthoryear{{Cordes} et~al.}{{Cordes}
  et~al.}{1991}]{cwf+91}
{Cordes}, J.~M., {Weisberg}, J.~M., {Frail}, D.~A., {Spangler}, S.~R.,  \&
  {Ryan}, M. 1991, \nat, 354, 121

\bibitem[\protect\citeauthoryear{{Coster} et~al.}{{Coster}
  et~al.}{2007}]{costeretal2007}
{Coster}, A.~J., {Colerico}, M.~J., {Foster}, J.~C., {Rideout}, W.,  \& {Rich},
  F. 2007, Geophysical Research Letters, 34, L18105

\bibitem[\protect\citeauthoryear{{Coster} \& {Foster}}{{Coster} \&
  {Foster}}{2007}]{costerandfoster2007}
{Coster}, A.~J.,  \& {Foster}, J.~C. 2007, {The Radio Science Bulletin}, 321

\bibitem[\protect\citeauthoryear{{Croft} et~al.}{{Croft} et~al.}{2011}]{cbk+11}
{Croft}, S., {Bower}, G.~C., {Keating}, G., {Law}, C., {Whysong}, D.,
  {Williams}, P.~K.~G.,  \& {Wright}, M. 2011, \apj, 731, 34

\bibitem[\protect\citeauthoryear{{Datta}, {Bowman}, \& {Carilli}}{{Datta}
  et~al.}{2010}]{2010ApJ...724..526D}
{Datta}, A., {Bowman}, J.~D.,  \& {Carilli}, C.~L. 2010, \apj, 724, 526

\bibitem[\protect\citeauthoryear{{Datta} et~al.}{{Datta}
  et~al.}{2008}]{2008MNRAS.391.1900D}
{Datta}, K.~K., {Majumdar}, S., {Bharadwaj}, S.,  \& {Choudhury}, T.~R. 2008,
  \mnras, 391, 1900

\bibitem[\protect\citeauthoryear{{de Bruyn} et~al.}{{de Bruyn}
  et~al.}{2006}]{dkh06}
{de Bruyn}, A.~G., {Katgert}, P., {Haverkorn}, M.,  \& {Schnitzeler},
  D.~H.~F.~M. 2006, Astronomische Nachrichten, 327, 487

\bibitem[\protect\citeauthoryear{{de Oliveira-Costa} et~al.}{{de
  Oliveira-Costa} et~al.}{2008}]{2008MNRAS.388..247D}
{de Oliveira-Costa}, A., {Tegmark}, M., {Gaensler}, B.~M., {Jonas}, J.,
  {Landecker}, T.~L.,  \& {Reich}, P. 2008, \mnras, 388, 247

\bibitem[\protect\citeauthoryear{{Dennison}}{{Dennison}}{1980}]{1980ApJ...239L..93D}
{Dennison}, B. 1980, \apjl, 239, L93

\bibitem[\protect\citeauthoryear{{Deshpande} \& {Rankin}}{{Deshpande} \&
  {Rankin}}{1999}]{dr99}
{Deshpande}, A.~A.,  \& {Rankin}, J.~M. 1999, \apj, 524, 1008

\bibitem[\protect\citeauthoryear{{Dessenne} et~al.}{{Dessenne}
  et~al.}{1996}]{dgw+96}
{Dessenne}, C.~A.-C., et~al. 1996, \mnras, 281, 977

\bibitem[\protect\citeauthoryear{{Djuth} et~al.}{{Djuth}
  et~al.}{2004}]{djuthetal2004}
{Djuth}, F.~T., {Sulzer}, M.~P., {Gonz{\'a}les}, S.~A., {Mathews}, J.~D.,
  {Elder}, J.~H.,  \& {Walterscheid}, R.~L. 2004, Geophysical Research Letters,
  31, L16801

\bibitem[\protect\citeauthoryear{{Donati} et~al.}{{Donati}
  et~al.}{2006}]{dfc+06}
{Donati}, J.-F., {Forveille}, T., {Cameron}, A.~C., {Barnes}, J.~R.,
  {Delfosse}, X., {Jardine}, M.~M.,  \& {Valenti}, J.~A. 2006, Science, 311,
  633

\bibitem[\protect\citeauthoryear{{Dorrian} et~al.}{{Dorrian}
  et~al.}{2008}]{Dorrian08}
{Dorrian}, G.~D., {Breen}, A.~R., {Brown}, D.~S., {Davies}, J.~A., {Fallows},
  R.~A.,  \& {Rouillard}, A.~P. 2008, grl, 35, 24104

\bibitem[\protect\citeauthoryear{{Dorrian} et~al.}{{Dorrian}
  et~al.}{2010}]{Dorrian10}
{Dorrian}, G.~D., et~al. 2010, \solphys, 265, 207

\bibitem[\protect\citeauthoryear{{Duncan} \& {Thompson}}{{Duncan} \&
  {Thompson}}{1992}]{dt92}
{Duncan}, R.~C.,  \& {Thompson}, C. 1992, \apjl, 392, L9

\bibitem[\protect\citeauthoryear{{Dymond} et~al.}{{Dymond}
  et~al.}{2008}]{2008cosp...37..775D}
{Dymond}, K., et~al. 2008, in COSPAR Meeting, Vol.~37, 37th COSPAR Scientific
  Assembly, 775

\bibitem[\protect\citeauthoryear{{Efimov} et~al.}{{Efimov}
  et~al.}{1996}]{efimov1996}
{Efimov}, A.~I., {Bird}, M.~K., {Andreev}, V.~E.,  \& {Samoznaev}, L.~N. 1996,
  Astronomy Letters, 22, 785

\bibitem[\protect\citeauthoryear{{Eikenberry} et~al.}{{Eikenberry}
  et~al.}{2000}]{emm+00}
{Eikenberry}, S.~S., {Matthews}, K., {Muno}, M., {Blanco}, P.~R., {Morgan},
  E.~H.,  \& {Remillard}, R.~A. 2000, \apjl, 532, L33

\bibitem[\protect\citeauthoryear{{Erickson}, {McConnell}, \&
  {Anantharamaiah}}{{Erickson} et~al.}{1995}]{1995ApJ...454..125E}
{Erickson}, W.~C., {McConnell}, D.,  \& {Anantharamaiah}, K.~R. 1995, \apj,
  454, 125

\bibitem[\protect\citeauthoryear{{Fallows} et~al.}{{Fallows}
  et~al.}{2012}]{Fallows12}
{Fallows}, R.~A., {Asgekar}, A., {Bisi}, M.~M., {Breen}, A.~R.,  \& {ter-Veen},
  S. 2012, \solphys

\bibitem[\protect\citeauthoryear{{Fan} et~al.}{{Fan}
  et~al.}{2006}]{2006AJ....132..117F}
{Fan}, X., et~al. 2006, \aj, 132, 117

\bibitem[\protect\citeauthoryear{{Farrell}, {Desch}, \& {Zarka}}{{Farrell}
  et~al.}{1999}]{fdz99}
{Farrell}, W.~M., {Desch}, M.~D.,  \& {Zarka}, P. 1999, \jgr, 104, 14025

\bibitem[\protect\citeauthoryear{{Fender}}{{Fender}}{2006}]{fender06}
{Fender}, R. 2006, in Compact stellar X-ray sources, ed. W.~Lewin \& M.~van~der
  Klis (Cambridge, UK: Cambridge University Press), 381

\bibitem[\protect\citeauthoryear{{Fender}, {Belloni}, \& {Gallo}}{{Fender}
  et~al.}{2004}]{fbg04}
{Fender}, R.~P., {Belloni}, T.~M.,  \& {Gallo}, E. 2004, \mnras, 355, 1105

\bibitem[\protect\citeauthoryear{{Fender} et~al.}{{Fender}
  et~al.}{2006}]{fws+06}
{Fender}, R.~P., et~al. 2006, in VI Microquasar Workshop: Microquasars and
  Beyond

\bibitem[\protect\citeauthoryear{{Ferrari} et~al.}{{Ferrari}
  et~al.}{2008}]{2008SSRv..134...93F}
{Ferrari}, C., {Govoni}, F., {Schindler}, S., {Bykov}, A.~M.,  \& {Rephaeli},
  Y. 2008, Space Science Reviews, 134, 93

\bibitem[\protect\citeauthoryear{Field}{Field}{1958}]{4065250}
Field, G. 1958, Proceedings of the IRE, 46, 240

\bibitem[\protect\citeauthoryear{{Fletcher} \& {Shukurov}}{{Fletcher} \&
  {Shukurov}}{2007}]{2007EAS....23..109F}
{Fletcher}, A.,  \& {Shukurov}, A. 2007, in EAS Publications Series, Vol.~23,
  EAS Publications Series, ed. M.-A. {Miville-Desch{\^e}nes} \& F.~{Boulanger},
  109

\bibitem[\protect\citeauthoryear{{Foster}}{{Foster}}{1993}]{foster1993}
{Foster}, J.~C. 1993, \jgr, 98, 1675

\bibitem[\protect\citeauthoryear{{Foster} \& {Coster}}{{Foster} \&
  {Coster}}{2007}]{fosterandcoster2007}
{Foster}, J.~C.,  \& {Coster}, A.~J. 2007, Journal of Atmospheric and
  Solar-Terrestrial Physics, 69, 1241

\bibitem[\protect\citeauthoryear{{Foster} \& {Erickson}}{{Foster} \&
  {Erickson}}{2013}]{fosteranderickson2012}
{Foster}, J.~C.,  \& {Erickson}, P.~J. 2013, J. Atm. Solar-Terr. Phys.
  (submitted)

\bibitem[\protect\citeauthoryear{{Frail}, {Kulkarni}, \& {Bloom}}{{Frail}
  et~al.}{1999}]{fkb99}
{Frail}, D.~A., {Kulkarni}, S.~R.,  \& {Bloom}, J.~S. 1999, \nat, 398, 127

\bibitem[\protect\citeauthoryear{{Frail} et~al.}{{Frail}
  et~al.}{2012}]{2012ApJ...747...70F}
{Frail}, D.~A., {Kulkarni}, S.~R., {Ofek}, E.~O., {Bower}, G.~C.,  \& {Nakar},
  E. 2012, \apj, 747, 70

\bibitem[\protect\citeauthoryear{{Furlanetto}, {Oh}, \& {Briggs}}{{Furlanetto}
  et~al.}{2006}]{2006PhR...433..181F}
{Furlanetto}, S.~R., {Oh}, S.~P.,  \& {Briggs}, F.~H. 2006, \physrep, 433, 181

\bibitem[\protect\citeauthoryear{{Furlanetto}, {Zaldarriaga}, \&
  {Hernquist}}{{Furlanetto} et~al.}{2004}]{2004ApJ...613....1F}
{Furlanetto}, S.~R., {Zaldarriaga}, M.,  \& {Hernquist}, L. 2004, \apj, 613, 1

\bibitem[\protect\citeauthoryear{{Gaensler} et~al.}{{Gaensler}
  et~al.}{2005a}]{ghs05}
{Gaensler}, B., {Haverkorn}, M., {Staveley-Smith}, L., {Dickey}, J.,
  {McClure-Griffiths}, N., {Dickel}, J.,  \& {Wolleben}, M. 2005a, in The
  Magnetized Plasma in Galaxy Evolution, ed. K.~T. {Chyzy},
  K.~{Otmianowska-Mazur}, M.~{Soida}, \& R.-J. {Dettmar}, 209

\bibitem[\protect\citeauthoryear{{Gaensler} et~al.}{{Gaensler}
  et~al.}{2001}]{gdm01}
{Gaensler}, B.~M., {Dickey}, J.~M., {McClure-Griffiths}, N.~M., {Green}, A.~J.,
  {Wieringa}, M.~H.,  \& {Haynes}, R.~F. 2001, \apj, 549, 959

\bibitem[\protect\citeauthoryear{{Gaensler} et~al.}{{Gaensler}
  et~al.}{2011}]{2011Natur.478..214G}
{Gaensler}, B.~M., et~al. 2011, \nat, 478, 214

\bibitem[\protect\citeauthoryear{{Gaensler} et~al.}{{Gaensler}
  et~al.}{2005b}]{gkg+05}
{Gaensler}, B.~M., et~al. 2005b, \nat, 434, 1104

\bibitem[\protect\citeauthoryear{{Galushko} et~al.}{{Galushko}
  et~al.}{2003}]{galushkoetal2003}
{Galushko}, V.~G., {Beley}, V.~S., {Koloskov}, A.~V., {Yampolski}, Y.~M.,
  {Paznukhov}, V.~V., {Reinisch}, B.~W., {Foster}, J.~C.,  \& {Erickson}, P.
  2003, Radio Science, 38, 060000

\bibitem[\protect\citeauthoryear{{Geil}, {Gaensler}, \& {Wyithe}}{{Geil}
  et~al.}{2011}]{2011MNRAS.418..516G}
{Geil}, P.~M., {Gaensler}, B.~M.,  \& {Wyithe}, J.~S.~B. 2011, \mnras, 418, 516

\bibitem[\protect\citeauthoryear{{Geil} et~al.}{{Geil}
  et~al.}{2008}]{2008MNRAS.390.1496G}
{Geil}, P.~M., {Wyithe}, J.~S.~B., {Petrovic}, N.,  \& {Oh}, S.~P. 2008,
  \mnras, 390, 1496

\bibitem[\protect\citeauthoryear{{Ginzburg} \& {Zhelezniakov}}{{Ginzburg} \&
  {Zhelezniakov}}{1958}]{ginzburgandzheleznyakov1958}
{Ginzburg}, V.~L.,  \& {Zhelezniakov}, V.~V. 1958, Soviet Astronomy, 2, 653

\bibitem[\protect\citeauthoryear{{Green}}{{Green}}{2004}]{2004BASI...32..335G}
{Green}, D.~A. 2004, Bulletin of the Astronomical Society of India, 32, 335

\bibitem[\protect\citeauthoryear{{Grie{\ss}meier}, {Zarka}, \&
  {Spreeuw}}{{Grie{\ss}meier} et~al.}{2007}]{gzs+07}
{Grie{\ss}meier}, J.-M., {Zarka}, P.,  \& {Spreeuw}, H. 2007, \aap, 475, 359

\bibitem[\protect\citeauthoryear{{G{\"u}del}}{{G{\"u}del}}{2002}]{gudel02}
{G{\"u}del}, M. 2002, \araa, 40, 217

\bibitem[\protect\citeauthoryear{{G\"{u}del} \& {Benz}}{{G\"{u}del} \&
  {Benz}}{1993}]{gb93}
{G\"{u}del}, M.,  \& {Benz}, A.~O. 1993, \apjl, 405, L63

\bibitem[\protect\citeauthoryear{{Gupta} \& {Gangadhara}}{{Gupta} \&
  {Gangadhara}}{2003}]{gg03}
{Gupta}, Y.,  \& {Gangadhara}, R.~T. 2003, \apj, 584, 418

\bibitem[\protect\citeauthoryear{{Gurnett}}{{Gurnett}}{1974}]{gurnett74}
{Gurnett}, D.~A. 1974, \jgr, 79, 4227

\bibitem[\protect\citeauthoryear{{Hales}, {Baldwin}, \& {Warner}}{{Hales}
  et~al.}{1988}]{1988MNRAS.234..919H}
{Hales}, S.~E.~G., {Baldwin}, J.~E.,  \& {Warner}, P.~J. 1988, \mnras, 234, 919

\bibitem[\protect\citeauthoryear{{Hallinan} et~al.}{{Hallinan}
  et~al.}{2007}]{hbl+07}
{Hallinan}, G., et~al. 2007, \apjl, 663, L25

\bibitem[\protect\citeauthoryear{{Halpern} et~al.}{{Halpern}
  et~al.}{2005}]{hgb+05}
{Halpern}, J.~P., {Gotthelf}, E.~V., {Becker}, R.~H., {Helfand}, D.~J.,  \&
  {White}, R.~L. 2005, \apjl, 632, L29

\bibitem[\protect\citeauthoryear{{Hansen} \& {Lyutikov}}{{Hansen} \&
  {Lyutikov}}{2001}]{hl01}
{Hansen}, B.~M.~S.,  \& {Lyutikov}, M. 2001, \mnras, 322, 695

\bibitem[\protect\citeauthoryear{{Hassall} et~al.}{{Hassall}
  et~al.}{2012}]{hsh+12}
{Hassall}, T.~E., et~al. 2012, \aap, 543, A66

\bibitem[\protect\citeauthoryear{{Haverkorn} et~al.}{{Haverkorn}
  et~al.}{2006}]{hgm06}
{Haverkorn}, M., {Gaensler}, B.~M., {McClure-Griffiths}, N.~M., {Dickey},
  J.~M.,  \& {Green}, A.~J. 2006, \apjs, 167, 230

\bibitem[\protect\citeauthoryear{{Haverkorn} \& {Heitsch}}{{Haverkorn} \&
  {Heitsch}}{2004}]{2004A&A...421.1011H}
{Haverkorn}, M.,  \& {Heitsch}, F. 2004, \aap, 421, 1011

\bibitem[\protect\citeauthoryear{{Haverkorn}, {Katgert}, \& {de
  Bruyn}}{{Haverkorn} et~al.}{2003a}]{hav03a}
{Haverkorn}, M., {Katgert}, P.,  \& {de Bruyn}, A.~G. 2003a, \aap, 403, 1045

\bibitem[\protect\citeauthoryear{{Haverkorn}, {Katgert}, \& {de
  Bruyn}}{{Haverkorn} et~al.}{2003b}]{hkb03}
{Haverkorn}, M., {Katgert}, P.,  \& {de Bruyn}, A.~G. 2003b, \aap, 403, 1031

\bibitem[\protect\citeauthoryear{{Haverkorn}, {Katgert}, \& {de
  Bruyn}}{{Haverkorn} et~al.}{2003c}]{hav03b}
{Haverkorn}, M., {Katgert}, P.,  \& {de Bruyn}, A.~G. 2003c, \aap, 403, 1031

\bibitem[\protect\citeauthoryear{{Haverkorn}, {Katgert}, \& {de
  Bruyn}}{{Haverkorn} et~al.}{2003d}]{hav03c}
{Haverkorn}, M., {Katgert}, P.,  \& {de Bruyn}, A.~G. 2003d, \aap, 404, 233

\bibitem[\protect\citeauthoryear{{Haynes} et~al.}{{Haynes}
  et~al.}{1986}]{hmk86}
{Haynes}, R.~F., {Murray}, J.~D., {Klein}, U.,  \& {Wielebinski}, R. 1986,
  \aap, 159, 22

\bibitem[\protect\citeauthoryear{{Helmboldt}, {Lane}, \& {Cotton}}{{Helmboldt}
  et~al.}{2012}]{2012RaSc...47.5008H}
{Helmboldt}, J.~F., {Lane}, W.~M.,  \& {Cotton}, W.~D. 2012, Radio Science

\bibitem[\protect\citeauthoryear{{Helmboldt} et~al.}{{Helmboldt}
  et~al.}{2012}]{2012RaSc...47.0L02H}
{Helmboldt}, J.~F., {Lazio}, T.~J.~W., {Intema}, H.~T.,  \& {Dymond}, K.~F.
  2012, Radio Science, 47, 0

\bibitem[\protect\citeauthoryear{{Hern{\'a}ndez-Pajares}, {Juan}, \&
  {Sanz}}{{Hern{\'a}ndez-Pajares} et~al.}{2006}]{pajaresetal2006}
{Hern{\'a}ndez-Pajares}, M., {Juan}, J.~M.,  \& {Sanz}, J. 2006, Journal of
  Geophysical Research (Space Physics), 111, A07S11

\bibitem[\protect\citeauthoryear{{Hewish}, {Scott}, \& {Wills}}{{Hewish}
  et~al.}{1964}]{hewish64}
{Hewish}, A., {Scott}, P.~F.,  \& {Wills}, D. 1964, \nat, 203, 1214

\bibitem[\protect\citeauthoryear{{Hollitt} \& {Johnston-Hollitt}}{{Hollitt} \&
  {Johnston-Hollitt}}{2012}]{2012PASA...29..309H}
{Hollitt}, C.,  \& {Johnston-Hollitt}, M. 2012, \pasa, 29, 309

\bibitem[\protect\citeauthoryear{{Hollweg} et~al.}{{Hollweg}
  et~al.}{1982}]{hollweg1982}
{Hollweg}, J.~V., {Bird}, M.~K., {Volland}, H., {Edenhofer}, P., {Stelzried},
  C.~T.,  \& {Seidel}, B.~L. 1982, \jgr, 87, 1

\bibitem[\protect\citeauthoryear{{Hyman} et~al.}{{Hyman} et~al.}{2002}]{hlkb02}
{Hyman}, S.~D., {Lazio}, T.~J.~W., {Kassim}, N.~E.,  \& {Bartleson}, A.~L.
  2002, \aj, 123, 1497

\bibitem[\protect\citeauthoryear{{Hyman} et~al.}{{Hyman} et~al.}{2005}]{hlk+05}
{Hyman}, S.~D., {Lazio}, T.~J.~W., {Kassim}, N.~E., {Ray}, P.~S., {Markwardt},
  C.~B.,  \& {Yusef-Zadeh}, F. 2005, \nat, 434, 50

\bibitem[\protect\citeauthoryear{{Hyman} et~al.}{{Hyman} et~al.}{2009}]{hwl+08}
{Hyman}, S.~D., {Wijnands}, R., {Lazio}, T.~J.~W., {Pal}, S., {Starling}, R.,
  {Kassim}, N.~E.,  \& {Ray}, P.~S. 2009, \apj, 696, 280

\bibitem[\protect\citeauthoryear{{Ichikawa} et~al.}{{Ichikawa}
  et~al.}{2010}]{2010MNRAS.406.2521I}
{Ichikawa}, K., {Barkana}, R., {Iliev}, I.~T., {Mellema}, G.,  \& {Shapiro},
  P.~R. 2010, \mnras, 406, 2521

\bibitem[\protect\citeauthoryear{{Inoue}}{{Inoue}}{2004}]{inoue04}
{Inoue}, S. 2004, \mnras, 348, 999

\bibitem[\protect\citeauthoryear{{Ishwara-Chandra} et~al.}{{Ishwara-Chandra}
  et~al.}{2005}]{irp+05}
{Ishwara-Chandra}, C.~H., {Rao}, A.~P., {Pandey}, M., {Manchanda}, R.~K.,  \&
  {Durouchoux}, P. 2005, Chinese Journal of Astronomy and Astrophysics
  Supplement, 5, 87

\bibitem[\protect\citeauthoryear{{Jackson} et~al.}{{Jackson}
  et~al.}{1998}]{Jackson98}
{Jackson}, B.~V., {Hick}, P.~L., {Kojima}, M.,  \& {Yokobe}, A. 1998, \jgr,
  103, 12049

\bibitem[\protect\citeauthoryear{{Jackson} et~al.}{{Jackson}
  et~al.}{2007}]{Jackson07}
{Jackson}, B.~V., {Hick}, P.~P., {Buffington}, A., {Bisi}, M.~M., {Kojima}, M.,
   \& {Tokumaru}, M. 2007, Astronomical and Astrophysical Transactions, 26, 477

\bibitem[\protect\citeauthoryear{{Jacobson} et~al.}{{Jacobson}
  et~al.}{1995}]{1995JGR...100.1653J}
{Jacobson}, A.~R., {Carlos}, R.~C., {Massey}, R.~S.,  \& {Wu}, G. 1995, \jgr,
  100, 1653

\bibitem[\protect\citeauthoryear{{Jacobson} \& {Erickson}}{{Jacobson} \&
  {Erickson}}{1993}]{1993AnGeo..11..869J}
{Jacobson}, A.~R.,  \& {Erickson}, W.~C. 1993, Annales Geophysicae, 11, 869

\bibitem[\protect\citeauthoryear{{Jaeger} et~al.}{{Jaeger}
  et~al.}{2012}]{jhkl12}
{Jaeger}, T.~R., {Hyman}, S.~D., {Kassim}, N.~E.,  \& {Lazio}, T.~J.~W. 2012,
  \aj, 143, 96

\bibitem[\protect\citeauthoryear{{Jenet}, {Anderson}, \& {Prince}}{{Jenet}
  et~al.}{2001}]{jap01}
{Jenet}, F.~A., {Anderson}, S.~B.,  \& {Prince}, T.~A. 2001, \apj, 558, 302

\bibitem[\protect\citeauthoryear{{Jensen} et~al.}{{Jensen}
  et~al.}{2010}]{Jensen10}
{Jensen}, E.~A., {Hick}, P.~P., {Bisi}, M.~M., {Jackson}, B.~V., {Clover}, J.,
  \& {Mulligan}, T. 2010, Sol. Phys., 265, 31

\bibitem[\protect\citeauthoryear{{Johns-Krull} \& {Valenti}}{{Johns-Krull} \&
  {Valenti}}{1996}]{jv96}
{Johns-Krull}, C.~M.,  \& {Valenti}, J.~A. 1996, \apjl, 459, L95

\bibitem[\protect\citeauthoryear{{Johnston}}{{Johnston}}{2003}]{johnston03}
{Johnston}, S. 2003, \mnras, 340, L43

\bibitem[\protect\citeauthoryear{{Johnston} et~al.}{{Johnston}
  et~al.}{2005}]{jhv+05}
{Johnston}, S., {Hobbs}, G., {Vigeland}, S., {Kramer}, M., {Weisberg}, J.~M.,
  \& {Lyne}, A.~G. 2005, \mnras, 364, 1397

\bibitem[\protect\citeauthoryear{{Kaiser}}{{Kaiser}}{1987}]{1987MNRAS.227....1K}
{Kaiser}, N. 1987, \mnras, 227, 1

\bibitem[\protect\citeauthoryear{{Kaplan}}{{Kaplan}}{2008}]{kaplan08}
{Kaplan}, D.~L. 2008, in American Institute of Physics Conference Series, Vol.
  968, Astrophysics of Compact Objects, ed. Y.-F. {Yuan}, X.-D. {Li}, \&
  D.~{Lai}, 129

\bibitem[\protect\citeauthoryear{{Kaplan} et~al.}{{Kaplan}
  et~al.}{2008}]{khr+08}
{Kaplan}, D.~L., {Hyman}, S.~D., {Roy}, S., {Bandyopadh yay}, R.~M.,
  {Chakrabarty}, D., {Kassim}, N.~E., {Lazio}, T.~J.~W.,  \& {Ray}, P.~S. 2008,
  \apj, 687, 262

\bibitem[\protect\citeauthoryear{{Kaplan} et~al.}{{Kaplan}
  et~al.}{2002}]{kkfvk02}
{Kaplan}, D.~L., {Kulkarni}, S.~R., {Frail}, D.~A.,  \& {van Kerkwijk}, M.~H.
  2002, \apj, 566, 378

\bibitem[\protect\citeauthoryear{{Kaplan}, {Kulkarni}, \& {van
  Kerkwijk}}{{Kaplan} et~al.}{2003}]{kkvk03}
{Kaplan}, D.~L., {Kulkarni}, S.~R.,  \& {van Kerkwijk}, M.~H. 2003, \apjl, 588,
  L33

\bibitem[\protect\citeauthoryear{{Kaplan} \& {van Kerkwijk}}{{Kaplan} \& {van
  Kerkwijk}}{2009}]{kvk09b}
{Kaplan}, D.~L.,  \& {van Kerkwijk}, M.~H. 2009, \apj, 705, 798

\bibitem[\protect\citeauthoryear{{Karlick{\'y}}}{{Karlick{\'y}}}{2003}]{karlicky2003}
{Karlick{\'y}}, M. 2003, Space Science Review, 107, 81

\bibitem[\protect\citeauthoryear{{Katz} et~al.}{{Katz} et~al.}{2003}]{khcm03}
{Katz}, C.~A., {Hewitt}, J.~N., {Corey}, B.~E.,  \& {Moore}, C.~B. 2003, \pasp,
  115, 675

\bibitem[\protect\citeauthoryear{{Keane} \& {Kramer}}{{Keane} \&
  {Kramer}}{2008}]{kk08}
{Keane}, E.~F.,  \& {Kramer}, M. 2008, \mnras, 391, 2009

\bibitem[\protect\citeauthoryear{{Keane} et~al.}{{Keane}
  et~al.}{2011}]{2011MNRAS.415.3065K}
{Keane}, E.~F., {Kramer}, M., {Lyne}, A.~G., {Stappers}, B.~W.,  \&
  {McLaughlin}, M.~A. 2011, \mnras, 415, 3065

\bibitem[\protect\citeauthoryear{{Keane} et~al.}{{Keane} et~al.}{2010}]{kle+10}
{Keane}, E.~F., {Ludovici}, D.~A., {Eatough}, R.~P., {Kramer}, M., {Lyne},
  A.~G., {McLaughlin}, M.~A.,  \& {Stappers}, B.~W. 2010, \mnras, 401, 1057

\bibitem[\protect\citeauthoryear{{Keane} et~al.}{{Keane} et~al.}{2012}]{kskl12}
{Keane}, E.~F., {Stappers}, B.~W., {Kramer}, M.,  \& {Lyne}, A.~G. 2012,
  \mnras, 425, L71

\bibitem[\protect\citeauthoryear{{Keith} et~al.}{{Keith}
  et~al.}{2010}]{kjvs+10}
{Keith}, M.~J., et~al. 2010, \mnras, 409, 619

\bibitem[\protect\citeauthoryear{{Keller} et~al.}{{Keller}
  et~al.}{2007}]{2007PASA...24....1K}
{Keller}, S.~C., et~al. 2007, Publications of the Astronomical Society of
  Australia, 24, 1

\bibitem[\protect\citeauthoryear{{Kempner} et~al.}{{Kempner}
  et~al.}{2004}]{2004rcfg.proc..335K}
{Kempner}, J.~C., {Blanton}, E.~L., {Clarke}, T.~E., {En{\ss}lin}, T.~A.,
  {Johnston-Hollitt}, M.,  \& {Rudnick}, L. 2004, in The Riddle of Cooling
  Flows in Galaxies and Clusters of galaxies, ed. T.~{Reiprich}, J.~{Kempner},
  \& N.~{Soker}, 335

\bibitem[\protect\citeauthoryear{{Keshet}, {Waxman}, \& {Loeb}}{{Keshet}
  et~al.}{2004}]{2004ApJ...617..281K}
{Keshet}, U., {Waxman}, E.,  \& {Loeb}, A. 2004, \apj, 617, 281

\bibitem[\protect\citeauthoryear{{Kojima} et~al.}{{Kojima}
  et~al.}{1998}]{Kojima98}
{Kojima}, M., {Tokumaru}, M., {Watanabe}, H., {Yokobe}, A., {Asai}, K.,
  {Jackson}, B.~V.,  \& {Hick}, P.~L. 1998, \jgr, 103, 1981

\bibitem[\protect\citeauthoryear{{Komatsu} et~al.}{{Komatsu}
  et~al.}{2009}]{2009ApJS..180..330K}
{Komatsu}, E., et~al. 2009, \apjs, 180, 330

\bibitem[\protect\citeauthoryear{{Kondratiev}, {Stappers}, \& {the LOFAR Pulsar
  Working Group}}{{Kondratiev} et~al.}{2012}]{ks+12}
{Kondratiev}, V., {Stappers}, B.,  \& {the LOFAR Pulsar Working Group}. 2012,
  in {IAU Symposium 291: Neutron Stars and Pulsars: Challenges and
  Opportunities after 80 years}, ed. J.~{van Leeuwen}, arXiv:1210.7005

\bibitem[\protect\citeauthoryear{{Kondratiev} \& {the LOFAR Pulsar Working
  Group}}{{Kondratiev} \& {the LOFAR Pulsar Working
  Group}}{2012}]{kondratiev+12}
{Kondratiev}, V.,  \& {the LOFAR Pulsar Working Group}. 2012, in {IAU Symposium
  291: Neutron Stars and Pulsars: Challenges and Opportunities after 80 years},
  ed. J.~{van Leeuwen}, arXiv:1210.6994

\bibitem[\protect\citeauthoryear{{Kondratiev} et~al.}{{Kondratiev}
  et~al.}{2009}]{kml+09}
{Kondratiev}, V.~I., {McLaughlin}, M.~A., {Lorimer}, D.~R., {Burgay}, M.,
  {Possenti}, A., {Turolla}, R., {Popov}, S.~B.,  \& {Zane}, S. 2009, \apj,
  702, 692

\bibitem[\protect\citeauthoryear{{Kotake} et~al.}{{Kotake}
  et~al.}{2006}]{kotakeetal2006}
{Kotake}, N., {Otsuka}, Y., {Tsugawa}, T., {Ogawa}, T.,  \& {Saito}, A. 2006,
  Journal of Geophysical Research (Space Physics), 111, A04306

\bibitem[\protect\citeauthoryear{{Kramer} et~al.}{{Kramer}
  et~al.}{2006}]{klo+06}
{Kramer}, M., {Lyne}, A.~G., {O'Brien}, J.~T., {Jordan}, C.~A.,  \& {Lorimer},
  D.~R. 2006, Science, 312, 549

\bibitem[\protect\citeauthoryear{{Kulkarni} \& {Phinney}}{{Kulkarni} \&
  {Phinney}}{2005}]{kp05}
{Kulkarni}, S.~R.,  \& {Phinney}, E.~S. 2005, \nat, 434, 28

\bibitem[\protect\citeauthoryear{{Kumar}}{{Kumar}}{1962}]{kumar62}
{Kumar}, S.~S. 1962, \aj, 67, 579

\bibitem[\protect\citeauthoryear{{Law} \& {Bower}}{{Law} \&
  {Bower}}{2012}]{lb12}
{Law}, C.~J.,  \& {Bower}, G.~C. 2012, \apj, 749, 143

\bibitem[\protect\citeauthoryear{{Law} et~al.}{{Law} et~al.}{2012}]{lbp+12}
{Law}, C.~J., {Bower}, G.~C., {Pokorny}, M., {Rupen}, M.~P.,  \& {Sowinski}, K.
  2012, \apj, 760, 124

\bibitem[\protect\citeauthoryear{{Lazio} \& {Farrell}}{{Lazio} \&
  {Farrell}}{2007}]{lf07}
{Lazio}, T.~J.~W.,  \& {Farrell}, W.~M. 2007, \apj, 668, 1182

\bibitem[\protect\citeauthoryear{{Lazio} et~al.}{{Lazio} et~al.}{2004}]{lfd+04}
{Lazio}, T.~J.~W., {Farrell}, W.~M., {Dietrick}, J., {Greenlees}, E., {Hogan},
  E., {Jones}, C.,  \& {Hennig}, L.~A. 2004, \apj, 612, 511

\bibitem[\protect\citeauthoryear{{Lenc} et~al.}{{Lenc} et~al.}{2008}]{lgw+08}
{Lenc}, E., {Garrett}, M.~A., {Wucknitz}, O., {Anderson}, J.~M.,  \& {Tingay},
  S.~J. 2008, \apj, 673, 78

\bibitem[\protect\citeauthoryear{{Li} et~al.}{{Li}
  et~al.}{1991}]{1991ApJ...378...93L}
{Li}, Z., {Wheeler}, J.~C., {Bash}, F.~N.,  \& {Jefferys}, W.~H. 1991, \apj,
  378, 93

\bibitem[\protect\citeauthoryear{{Lidz} et~al.}{{Lidz}
  et~al.}{2009}]{2009ApJ...690..252L}
{Lidz}, A., {Zahn}, O., {Furlanetto}, S.~R., {McQuinn}, M., {Hernquist}, L.,
  \& {Zaldarriaga}, M. 2009, \apj, 690, 252

\bibitem[\protect\citeauthoryear{{Lidz} et~al.}{{Lidz}
  et~al.}{2008}]{2008ApJ...680..962L}
{Lidz}, A., {Zahn}, O., {McQuinn}, M., {Zaldarriaga}, M.,  \& {Hernquist}, L.
  2008, \apj, 680, 962

\bibitem[\protect\citeauthoryear{{Liu} \& {Tegmark}}{{Liu} \&
  {Tegmark}}{2011}]{2011PhRvD..83j3006L}
{Liu}, A.,  \& {Tegmark}, M. 2011, \prd, 83, 103006

\bibitem[\protect\citeauthoryear{{Liu} et~al.}{{Liu}
  et~al.}{2009}]{2009MNRAS.398..401L}
{Liu}, A., {Tegmark}, M., {Bowman}, J., {Hewitt}, J.,  \& {Zaldarriaga}, M.
  2009, \mnras, 398, 401

\bibitem[\protect\citeauthoryear{{Liu}, {Tegmark}, \& {Zaldarriaga}}{{Liu}
  et~al.}{2009}]{2009MNRAS.394.1575L}
{Liu}, A., {Tegmark}, M.,  \& {Zaldarriaga}, M. 2009, \mnras, 394, 1575

\bibitem[\protect\citeauthoryear{{Liu} et~al.}{{Liu} et~al.}{2007}]{liu2007}
{Liu}, Y., {Manchester}, W.~B., IV, {Kasper}, J.~C., {Richardson}, J.~D.,  \&
  {Belcher}, J.~W. 2007, \apj, 665, 1439

\bibitem[\protect\citeauthoryear{{Lobzin} et~al.}{{Lobzin}
  et~al.}{2010}]{lobzinetal2010}
{Lobzin}, V.~V., {Cairns}, I.~H., {Robinson}, P.~A., {Warmuth}, A., {Mann}, G.,
  {Gorgutsa}, R.~V.,  \& {Fomichev}, V.~V. 2010, \apj, 724, 1099

\bibitem[\protect\citeauthoryear{{Lonsdale} et~al.}{{Lonsdale}
  et~al.}{2009}]{2009IEEEP..97.1497L}
{Lonsdale}, C.~J., et~al. 2009, IEEE Proceedings, 97, 1497

\bibitem[\protect\citeauthoryear{{Lorimer} et~al.}{{Lorimer}
  et~al.}{2007}]{lbm+07}
{Lorimer}, D.~R., {Bailes}, M., {McLaughlin}, M.~A., {Narkevic}, D.~J.,  \&
  {Crawford}, F. 2007, Science, 318, 777

\bibitem[\protect\citeauthoryear{{Lynch} et~al.}{{Lynch} et~al.}{2013}]{lbr+12}
{Lynch}, R.~S., et~al. 2013, \apj, 763, 81

\bibitem[\protect\citeauthoryear{{Lyne} et~al.}{{Lyne} et~al.}{2009}]{lmk+09}
{Lyne}, A.~G., {McLaughlin}, M.~A., {Keane}, E.~F., {Kramer}, M., {Espinoza},
  C.~M., {Stappers}, B.~W., {Palliyaguru}, N.~T.,  \& {Miller}, J. 2009,
  \mnras, 400, 1439

\bibitem[\protect\citeauthoryear{{Maan} et~al.}{{Maan}
  et~al.}{2013}]{2013ApJS..204...12M}
{Maan}, Y., et~al. 2013, \apjs, 204, 12

\bibitem[\protect\citeauthoryear{{Macquart}}{{Macquart}}{2007}]{macquart07}
{Macquart}, J.-P. 2007, \apjl, 658, L1

\bibitem[\protect\citeauthoryear{{Maia} et~al.}{{Maia}
  et~al.}{2000}]{maiaetal2000}
{Maia}, D., {Pick}, M., {Vourlidas}, A.,  \& {Howard}, R. 2000, \apjl, 528, L49

\bibitem[\protect\citeauthoryear{{Mancuso} \& {Spangler}}{{Mancuso} \&
  {Spangler}}{2000}]{mancuso2000}
{Mancuso}, S.,  \& {Spangler}, S.~R. 2000, \apj, 539, 480

\bibitem[\protect\citeauthoryear{{Mao} et~al.}{{Mao} et~al.}{2008}]{mgs08}
{Mao}, S.~A., {Gaensler}, B.~M., {Stanimirovi{\'c}}, S., {Haverkorn}, M.,
  {McClure-Griffiths}, N.~M., {Staveley-Smith}, L.,  \& {Dickey}, J.~M. 2008,
  \apj, 688, 1029

\bibitem[\protect\citeauthoryear{{Marcy} et~al.}{{Marcy} et~al.}{2005}]{mbf+05}
{Marcy}, G., {Butler}, R.~P., {Fischer}, D., {Vogt}, S., {Wright}, J.~T.,
  {Tinney}, C.~G.,  \& {Jones}, H.~R.~A. 2005, Progress of Theoretical Physics
  Supplement, 158, 24

\bibitem[\protect\citeauthoryear{{Maron} et~al.}{{Maron} et~al.}{2000}]{mkkw00}
{Maron}, O., {Kijak}, J., {Kramer}, M.,  \& {Wielebinski}, R. 2000, \aaps, 147,
  195

\bibitem[\protect\citeauthoryear{{Matsumura} et~al.}{{Matsumura}
  et~al.}{2007}]{mdk+07}
{Matsumura}, N., et~al. 2007, \aj, 133, 1441

\bibitem[\protect\citeauthoryear{{Matsumura} et~al.}{{Matsumura}
  et~al.}{2009}]{mnk+09}
{Matsumura}, N., et~al. 2009, \aj, 138, 787

\bibitem[\protect\citeauthoryear{{McClintock} \& {Remillard}}{{McClintock} \&
  {Remillard}}{2006}]{mr06}
{McClintock}, J.~E.,  \& {Remillard}, R.~A. 2006, in Compact stellar X-ray
  sources, ed. W.~Lewin \& M.~van~der Klis (Cambridge, UK: Cambridge University
  Press), 157

\bibitem[\protect\citeauthoryear{{McLaughlin} et~al.}{{McLaughlin}
  et~al.}{2006}]{mll+06}
{McLaughlin}, M.~A., et~al. 2006, \nat, 439, 817

\bibitem[\protect\citeauthoryear{{McLaughlin} et~al.}{{McLaughlin}
  et~al.}{2007}]{mrg+07}
{McLaughlin}, M.~A., et~al. 2007, \apj, 670, 1307

\bibitem[\protect\citeauthoryear{{McLean} \& {Labrum}}{{McLean} \&
  {Labrum}}{1985}]{mcleanandlabrum1985}
{McLean}, D.~J.,  \& {Labrum}, N.~R. 1985, {Solar radiophysics: Studies of
  emission from the sun at metre wavelengths}

\bibitem[\protect\citeauthoryear{{McQuinn} et~al.}{{McQuinn}
  et~al.}{2007}]{2007MNRAS.377.1043M}
{McQuinn}, M., {Lidz}, A., {Zahn}, O., {Dutta}, S., {Hernquist}, L.,  \&
  {Zaldarriaga}, M. 2007, \mnras, 377, 1043

\bibitem[\protect\citeauthoryear{{McQuinn} et~al.}{{McQuinn}
  et~al.}{2006}]{2006ApJ...653..815M}
{McQuinn}, M., {Zahn}, O., {Zaldarriaga}, M., {Hernquist}, L.,  \&
  {Furlanetto}, S.~R. 2006, \apj, 653, 815

\bibitem[\protect\citeauthoryear{{Mejia-Ambriz} et~al.}{{Mejia-Ambriz}
  et~al.}{2010}]{Mejia10}
{Mejia-Ambriz}, J.~C., {Villanueva-Hernandez}, P., {Gonzalez-Esparza}, J.~A.,
  {Aguilar-Rodriguez}, E.,  \& {Jeyakumar}, S. 2010, \solphys, 265, 309

\bibitem[\protect\citeauthoryear{{Mellema} et~al.}{{Mellema}
  et~al.}{2006}]{2006MNRAS.372..679M}
{Mellema}, G., {Iliev}, I.~T., {Pen}, U.-L.,  \& {Shapiro}, P.~R. 2006, \mnras,
  372, 679

\bibitem[\protect\citeauthoryear{{Melrose}}{{Melrose}}{1980}]{melrose1980}
{Melrose}, D.~B. 1980, Space Science Review, 26, 3

\bibitem[\protect\citeauthoryear{{Melrose}}{{Melrose}}{2003}]{melrose03}
{Melrose}, D.~B. 2003, Plasma Physics and Controlled Fusion, 45, 523

\bibitem[\protect\citeauthoryear{{Mesa} et~al.}{{Mesa}
  et~al.}{2002}]{2002A&A...396..463M}
{Mesa}, D., {Baccigalupi}, C., {De Zotti}, G., {Gregorini}, L., {Mack}, K.-H.,
  {Vigotti}, M.,  \& {Klein}, U. 2002, \aap, 396, 463

\bibitem[\protect\citeauthoryear{{Mesinger}}{{Mesinger}}{2010}]{2010MNRAS.407.1328M}
{Mesinger}, A. 2010, \mnras, 407, 1328

\bibitem[\protect\citeauthoryear{{Mirabel} \& {Rodr{\'{\i}}guez}}{{Mirabel} \&
  {Rodr{\'{\i}}guez}}{1999}]{mr99}
{Mirabel}, I.~F.,  \& {Rodr{\'{\i}}guez}, L.~F. 1999, \araa, 37, 409

\bibitem[\protect\citeauthoryear{{Mitchell} et~al.}{{Mitchell}
  et~al.}{2008}]{2008ISTSP...2..707M}
{Mitchell}, D.~A., {Greenhill}, L.~J., {Wayth}, R.~B., {Sault}, R.~J.,
  {Lonsdale}, C.~J., {Cappallo}, R.~J., {Morales}, M.~F.,  \& {Ord}, S.~M.
  2008, IEEE Journal of Selected Topics in Signal Processing, vol.~2, issue 5,
  pp.~707-717, 2, 707

\bibitem[\protect\citeauthoryear{{Moortgat} \& {Kuijpers}}{{Moortgat} \&
  {Kuijpers}}{2005}]{mk05}
{Moortgat}, J.,  \& {Kuijpers}, J. 2005, in 22nd Texas Symposium on
  Relativistic Astrophysics, ed. P.~{Chen}, E.~{Bloom}, G.~{Madejski}, \&
  V.~{Patrosian}, 326

\bibitem[\protect\citeauthoryear{{Morales}, {Bowman}, \& {Hewitt}}{{Morales}
  et~al.}{2006}]{2006ApJ...648..767M}
{Morales}, M.~F., {Bowman}, J.~D.,  \& {Hewitt}, J.~N. 2006, \apj, 648, 767

\bibitem[\protect\citeauthoryear{{Morales} et~al.}{{Morales}
  et~al.}{2012}]{2012arXiv1202.3830M}
{Morales}, M.~F., {Hazelton}, B., {Sullivan}, I.,  \& {Beardsley}, A. 2012,
  \apj, 752, 137

\bibitem[\protect\citeauthoryear{{Morales} \& {Hewitt}}{{Morales} \&
  {Hewitt}}{2004}]{2004ApJ...615....7M}
{Morales}, M.~F.,  \& {Hewitt}, J. 2004, \apj, 615, 7

\bibitem[\protect\citeauthoryear{{Morales} et~al.}{{Morales}
  et~al.}{2005}]{mhk+05}
{Morales}, M.~F., {Hewitt}, J.~N., {Kasper}, J.~C., {Lane}, B., {Bowman}, J.,
  {Ray}, P.~S.,  \& {Cappallo}, R.~J. 2005, in ASP Conf. Ser., Vol. 345, From
  Clark Lake to the Long Wavelength Array: Bill Erickson's Radio Science, ed.
  N.~{Kassim}, M.~{Perez}, W.~{Junor}, \& P.~{Henning} (San Fransisco: ASP),
  512

\bibitem[\protect\citeauthoryear{{Nakar} \& {Piran}}{{Nakar} \&
  {Piran}}{2011}]{np11}
{Nakar}, E.,  \& {Piran}, T. 2011, \nat, 478, 82

\bibitem[\protect\citeauthoryear{{Nelson} \& {Melrose}}{{Nelson} \&
  {Melrose}}{1985}]{nelsonandmelrose1985}
{Nelson}, G.~J.,  \& {Melrose}, D.~B. 1985, {Type II bursts}, ed. {McLean,
  D.~J.~\& Labrum, N.~R.} 333

\bibitem[\protect\citeauthoryear{{Nicolls} \& {Heinselman}}{{Nicolls} \&
  {Heinselman}}{2007}]{nicollsandheinselman2007}
{Nicolls}, M.~J.,  \& {Heinselman}, C.~J. 2007, Geophysical Research Letters,
  34, L21104

\bibitem[\protect\citeauthoryear{{Nijboer}, {Pandey-Pommier}, \& {de
  Bruyn}}{{Nijboer} et~al.}{2009}]{2009Nijboer}
{Nijboer}, R.~J., {Pandey-Pommier}, M.,  \& {de Bruyn}, A.~G. 2009, SKA Memo
  \#113, http://www.skatelescope.org/publications/

\bibitem[\protect\citeauthoryear{{Nord} et~al.}{{Nord} et~al.}{2006}]{nhr+06}
{Nord}, M.~E., {Henning}, P.~A., {Rand}, R.~J., {Lazio}, T.~J.~W.,  \&
  {Kassim}, N.~E. 2006, \aj, 132, 242

\bibitem[\protect\citeauthoryear{{Noutsos} et~al.}{{Noutsos}
  et~al.}{2008}]{njkk08}
{Noutsos}, A., {Johnston}, S., {Kramer}, M.,  \& {Karastergiou}, A. 2008,
  \mnras, 386, 1881

\bibitem[\protect\citeauthoryear{{Oberoi}}{{Oberoi}}{2000}]{Oberoi00}
{Oberoi}, D. 2000, Ph.D. thesis, Indian Institue of Science

\bibitem[\protect\citeauthoryear{{Oberoi} \& {Benkevitch}}{{Oberoi} \&
  {Benkevitch}}{2010}]{OberoiandBenkevitch2010}
{Oberoi}, D.,  \& {Benkevitch}, L. 2010, \solphys, 265, 293

\bibitem[\protect\citeauthoryear{{Oberoi} \& {Kasper}}{{Oberoi} \&
  {Kasper}}{2004}]{OberoiandKasper2004}
{Oberoi}, D.,  \& {Kasper}, J.~C. 2004, Planetary and Space Science, 52, 1415

\bibitem[\protect\citeauthoryear{{Oberoi} \& {Lonsdale}}{{Oberoi} \&
  {Lonsdale}}{2013}]{Oberoi13}
{Oberoi}, D.,  \& {Lonsdale}, C.~J. 2013, Radio Science, in press

\bibitem[\protect\citeauthoryear{{Oberoi} et~al.}{{Oberoi}
  et~al.}{2011}]{oberoietal2011}
{Oberoi}, D., et~al. 2011, \apjl, 728, L27

\bibitem[\protect\citeauthoryear{{Oliver} et~al.}{{Oliver}
  et~al.}{1994}]{oliveretal1994}
{Oliver}, W.~L., {Fukao}, S., {Yamamoto}, Y., {Takami}, T., {Yamanaka}, M.~D.,
  {Yamamoto}, M., {Nakamura}, T.,  \& {Tsuda}, T. 1994, \jgr, 99, 6321

\bibitem[\protect\citeauthoryear{{Oliver} et~al.}{{Oliver}
  et~al.}{1997}]{oliveretal1997}
{Oliver}, W.~L., {Otsuka}, Y., {Sato}, M., {Takami}, T.,  \& {Fukao}, S. 1997,
  \jgr, 102, 14499

\bibitem[\protect\citeauthoryear{{Ord}, {Johnston}, \& {Sarkissian}}{{Ord}
  et~al.}{2007}]{ojs07}
{Ord}, S.~M., {Johnston}, S.,  \& {Sarkissian}, J. 2007, \solphys, 245, 109

\bibitem[\protect\citeauthoryear{{Osten} \& {Bastian}}{{Osten} \&
  {Bastian}}{2006}]{ob06}
{Osten}, R.~A.,  \& {Bastian}, T.~S. 2006, \apj, 637, 1016

\bibitem[\protect\citeauthoryear{{Osten} et~al.}{{Osten} et~al.}{2006}]{ohbr06}
{Osten}, R.~A., {Hawley}, S.~L., {Bastian}, T.~S.,  \& {Reid}, I.~N. 2006,
  \apj, 637, 518

\bibitem[\protect\citeauthoryear{{Paladini} et~al.}{{Paladini}
  et~al.}{2003}]{2003A&A...397..213P}
{Paladini}, R., {Burigana}, C., {Davies}, R.~D., {Maino}, D., {Bersanelli}, M.,
  {Cappellini}, B., {Platania}, P.,  \& {Smoot}, G. 2003, \aap, 397, 213

\bibitem[\protect\citeauthoryear{{Paladini}, {Davies}, \& {De
  Zotti}}{{Paladini} et~al.}{2004}]{2004MNRAS.347..237P}
{Paladini}, R., {Davies}, R.~D.,  \& {De Zotti}, G. 2004, \mnras, 347, 237

\bibitem[\protect\citeauthoryear{{Palenzuela}, {Lehner}, \&
  {Liebling}}{{Palenzuela} et~al.}{2010}]{pll10}
{Palenzuela}, C., {Lehner}, L.,  \& {Liebling}, S.~L. 2010, Science, 329, 927

\bibitem[\protect\citeauthoryear{{Pandey} et~al.}{{Pandey}
  et~al.}{2007}]{pri+07}
{Pandey}, M., {Rao}, A.~P., {Ishwara-Chandra}, C.~H., {Durouchoux}, P.,  \&
  {Manchanda}, R.~K. 2007, \aap, 463, 567

\bibitem[\protect\citeauthoryear{{Pandey} et~al.}{{Pandey}
  et~al.}{2006}]{prp+06}
{Pandey}, M., {Rao}, A.~P., {Pooley}, G.~G., {Durouchoux}, P., {Manchanda},
  R.~K.,  \& {Ishwara-Chandra}, C.~H. 2006, \aap, 447, 525

\bibitem[\protect\citeauthoryear{{Pandey} et~al.}{{Pandey}
  et~al.}{2005}]{prm+05}
{Pandey}, M.~D., {Rao}, A.~P., {Manchanda}, R.~K., {Durouchoux}, P.,
  {Ishwara-Chandra}, C.~H.,  \& {Kulkarni}, V.~K. 2005, Bulletin of the
  Astronomical Society of India, 33, 382

\bibitem[\protect\citeauthoryear{{Parker}}{{Parker}}{1955}]{parker55}
{Parker}, E.~N. 1955, \apj, 122, 293

\bibitem[\protect\citeauthoryear{{Patterson} et~al.}{{Patterson}
  et~al.}{2008}]{pem+08}
{Patterson}, C.~D., {Ellingson}, S.~W., {Martin}, B.~S., {Deshpande}, K.,
  {Simonetti}, J.~H., {Kavic}, M.,  \& {Cutchin}, S.~E. 2008, arXiv:0812.1255

\bibitem[\protect\citeauthoryear{{Patzold} et~al.}{{Patzold}
  et~al.}{1987}]{patzold1987}
{Patzold}, M., {Bird}, M.~K., {Volland}, H., {Levy}, G.~S., {Seidel}, B.~L.,
  \& {Stelzried}, C.~T. 1987, \solphys, 109, 91

\bibitem[\protect\citeauthoryear{{Petrosian}}{{Petrosian}}{2001}]{2001ApJ...557..560P}
{Petrosian}, V. 2001, \apj, 557, 560

\bibitem[\protect\citeauthoryear{{Petrovic} \& {Oh}}{{Petrovic} \&
  {Oh}}{2011}]{2011MNRAS.413.2103P}
{Petrovic}, N.,  \& {Oh}, S.~P. 2011, \mnras, 413, 2103

\bibitem[\protect\citeauthoryear{{Pfrommer}, {En{\ss}lin}, \&
  {Springel}}{{Pfrommer} et~al.}{2008}]{2008MNRAS.385.1211P}
{Pfrommer}, C., {En{\ss}lin}, T.~A.,  \& {Springel}, V. 2008, \mnras, 385, 1211

\bibitem[\protect\citeauthoryear{{Pfrommer} et~al.}{{Pfrommer}
  et~al.}{2006}]{2006MNRAS.367..113P}
{Pfrommer}, C., {Springel}, V., {En{\ss}lin}, T.~A.,  \& {Jubelgas}, M. 2006,
  \mnras, 367, 113

\bibitem[\protect\citeauthoryear{{Pindor} et~al.}{{Pindor}
  et~al.}{2011}]{2011PASA...28...46P}
{Pindor}, B., {Wyithe}, J.~S.~B., {Mitchell}, D.~A., {Ord}, S.~M., {Wayth},
  R.~B.,  \& {Greenhill}, L.~J. 2011, \pasa, 28, 46

\bibitem[\protect\citeauthoryear{{Popov}, {Turolla}, \& {Possenti}}{{Popov}
  et~al.}{2006}]{ptp06}
{Popov}, S.~B., {Turolla}, R.,  \& {Possenti}, A. 2006, \mnras, 369, L23

\bibitem[\protect\citeauthoryear{{Readhead}}{{Readhead}}{1994}]{readhead94}
{Readhead}, A.~C.~S. 1994, \apj, 426, 51

\bibitem[\protect\citeauthoryear{{Reich}, {Reich}, \& {Fuerst}}{{Reich}
  et~al.}{1990}]{1990A&AS...83..539R}
{Reich}, W., {Reich}, P.,  \& {Fuerst}, E. 1990, \aaps, 83, 539

\bibitem[\protect\citeauthoryear{{Remillard} \& {McClintock}}{{Remillard} \&
  {McClintock}}{2006}]{rm06}
{Remillard}, R.~A.,  \& {McClintock}, J.~E. 2006, \araa, 44, 49

\bibitem[\protect\citeauthoryear{{Rickett}}{{Rickett}}{1969}]{rickett69}
{Rickett}, B.~J. 1969, \nat, 221, 158

\bibitem[\protect\citeauthoryear{{Rickett}}{{Rickett}}{1990}]{rickett90}
{Rickett}, B.~J. 1990, \araa, 28, 561

\bibitem[\protect\citeauthoryear{{Robinson} \& {Cairns}}{{Robinson} \&
  {Cairns}}{2000}]{robinsonandcairns2000}
{Robinson}, P.~A.,  \& {Cairns}, I.~H. 2000, in Radio Astronomy at Long
  Wavelengths, 37

\bibitem[\protect\citeauthoryear{{Rogers} \& {Bowman}}{{Rogers} \&
  {Bowman}}{2008}]{2008AJ....136..641R}
{Rogers}, A.~E.~E.,  \& {Bowman}, J.~D. 2008, \aj, 136, 641

\bibitem[\protect\citeauthoryear{{Rudnick} et~al.}{{Rudnick}
  et~al.}{2009}]{2009astro2010S.253R}
{Rudnick}, L., et~al. 2009, in Astro2010: The Astronomy and Astrophysics
  Decadal Survey, Science White Papers, no. 253, arXiv: 0903.0824

\bibitem[\protect\citeauthoryear{{Ryu} et~al.}{{Ryu}
  et~al.}{2003}]{2003ApJ...593..599R}
{Ryu}, D., {Kang}, H., {Hallman}, E.,  \& {Jones}, T.~W. 2003, \apj, 593, 599

\bibitem[\protect\citeauthoryear{{Saar} \& {Linsky}}{{Saar} \&
  {Linsky}}{1985}]{sl85}
{Saar}, S.~H.,  \& {Linsky}, J.~L. 1985, \apjl, 299, L47

\bibitem[\protect\citeauthoryear{{Sagiv} \& {Waxman}}{{Sagiv} \&
  {Waxman}}{2002}]{sw02}
{Sagiv}, A.,  \& {Waxman}, E. 2002, \apj, 574, 861

\bibitem[\protect\citeauthoryear{{Salah} et~al.}{{Salah}
  et~al.}{2005}]{salahetal2005}
{Salah}, J.~E., {Lonsdale}, C.~J., {Oberoi}, D., {Cappallo}, R.~J.,  \&
  {Kasper}, J.~C. 2005, in Society of Photo-Optical Instrumentation Engineers
  (SPIE) Conference Series, Vol. 5901, Society of Photo-Optical Instrumentation
  Engineers (SPIE) Conference Series, ed. S.~{Fineschi} \& R.~A. {Viereck}, 124

\bibitem[\protect\citeauthoryear{{Santos}, {Cooray}, \& {Knox}}{{Santos}
  et~al.}{2005}]{2005ApJ...625..575S}
{Santos}, M.~G., {Cooray}, A.,  \& {Knox}, L. 2005, \apj, 625, 575

\bibitem[\protect\citeauthoryear{{Scherer} et~al.}{{Scherer}
  et~al.}{2005}]{schereretal2005}
{Scherer}, K., {Fichtner}, H., {Heber}, B.,  \& {Mall}, U. 2005, {Space
  Weather: The Physics Behind a Slogan.}

\bibitem[\protect\citeauthoryear{{Schnitzeler}, {Katgert}, \& {de
  Bruyn}}{{Schnitzeler} et~al.}{2009}]{2009A&A...494..611S}
{Schnitzeler}, D.~H.~F.~M., {Katgert}, P.,  \& {de Bruyn}, A.~G. 2009, \aap,
  494, 611

\bibitem[\protect\citeauthoryear{{Scott} \& {Rees}}{{Scott} \&
  {Rees}}{1990}]{1990MNRAS.247..510S}
{Scott}, D.,  \& {Rees}, M.~J. 1990, \mnras, 247, 510

\bibitem[\protect\citeauthoryear{{Scott}, {Rickett}, \& {Armstrong}}{{Scott}
  et~al.}{1983}]{Scott83}
{Scott}, S.~L., {Rickett}, B.~J.,  \& {Armstrong}, J.~W. 1983, \aap, 123, 191

\bibitem[\protect\citeauthoryear{{Shkolnik}, {Walker}, \&
  {Bohlender}}{{Shkolnik} et~al.}{2003}]{swb03}
{Shkolnik}, E., {Walker}, G.~A.~H.,  \& {Bohlender}, D.~A. 2003, \apj, 597,
  1092

\bibitem[\protect\citeauthoryear{{Skillman} et~al.}{{Skillman}
  et~al.}{2011}]{2011ApJ...735...96S}
{Skillman}, S.~W., {Hallman}, E.~J., {O'Shea}, B.~W., {Burns}, J.~O., {Smith},
  B.~D.,  \& {Turk}, M.~J. 2011, \apj, 735, 96

\bibitem[\protect\citeauthoryear{{Soderberg} et~al.}{{Soderberg}
  et~al.}{2010}]{scp+10}
{Soderberg}, A.~M., et~al. 2010, \nat, 463, 513

\bibitem[\protect\citeauthoryear{{Soderberg} et~al.}{{Soderberg}
  et~al.}{2007}]{snc+06}
{Soderberg}, A.~M., et~al. 2007, \apj, 661, 982

\bibitem[\protect\citeauthoryear{{Sokoloff} et~al.}{{Sokoloff}
  et~al.}{1998}]{1998MNRAS.299..189S}
{Sokoloff}, D.~D., {Bykov}, A.~A., {Shukurov}, A., {Berkhuijsen}, E.~M.,
  {Beck}, R.,  \& {Poezd}, A.~D. 1998, \mnras, 299, 189

\bibitem[\protect\citeauthoryear{{Spangler} \& {Whiting}}{{Spangler} \&
  {Whiting}}{2009}]{Spangler09}
{Spangler}, S.~R.,  \& {Whiting}, C.~A. 2009, in IAU Symposium, Vol. 257, IAU
  Symposium, ed. N.~{Gopalswamy} \& D.~F. {Webb}, 529

\bibitem[\protect\citeauthoryear{{Stelzried} et~al.}{{Stelzried}
  et~al.}{1970}]{stelzried1970}
{Stelzried}, C.~T., {Levy}, G.~S., {Sato}, T., {Rusch}, W.~V.~T., {Ohlson},
  J.~E., {Schatten}, K.~H.,  \& {Wilcox}, J.~M. 1970, \solphys, 14, 440

\bibitem[\protect\citeauthoryear{{Subrahmanyan} \& {Ekers}}{{Subrahmanyan} \&
  {Ekers}}{2002}]{2002astro.ph..9569S}
{Subrahmanyan}, R.,  \& {Ekers}, R.~D. 2002, ArXiv Astrophysics e-prints

\bibitem[\protect\citeauthoryear{{Sun} et~al.}{{Sun}
  et~al.}{2008}]{2008A&A...477..573S}
{Sun}, X.~H., {Reich}, W., {Waelkens}, A.,  \& {En{\ss}lin}, T.~A. 2008, \aap,
  477, 573

\bibitem[\protect\citeauthoryear{{Suzuki} \& {Dulk}}{{Suzuki} \&
  {Dulk}}{1985}]{suzukianddulk1985}
{Suzuki}, S.,  \& {Dulk}, G.~A. 1985, {Bursts of Type III and Type V}, ed.
  {McLean, D.~J.~\& Labrum, N.~R.} 289

\bibitem[\protect\citeauthoryear{{Tammann}, {Loeffler}, \&
  {Schroeder}}{{Tammann} et~al.}{1994}]{1994ApJS...92..487T}
{Tammann}, G.~A., {Loeffler}, W.,  \& {Schroeder}, A. 1994, \apjs, 92, 487

\bibitem[\protect\citeauthoryear{{Taylor}, {Stil}, \& {Sunstrum}}{{Taylor}
  et~al.}{2009}]{2009ApJ...702.1230T}
{Taylor}, A.~R., {Stil}, J.~M.,  \& {Sunstrum}, C. 2009, \apj, 702, 1230

\bibitem[\protect\citeauthoryear{{Taylor} \& {Cordes}}{{Taylor} \&
  {Cordes}}{1993}]{tc93}
{Taylor}, J.~H.,  \& {Cordes}, J.~M. 1993, \apj, 411, 674

\bibitem[\protect\citeauthoryear{{Tegmark}}{{Tegmark}}{1997}]{1997PhRvD..55.5895T}
{Tegmark}, M. 1997, \prd, 55, 5895

\bibitem[\protect\citeauthoryear{{Thompson} \& {Duncan}}{{Thompson} \&
  {Duncan}}{1995}]{td95}
{Thompson}, C.,  \& {Duncan}, R.~C. 1995, \mnras, 275, 255

\bibitem[\protect\citeauthoryear{Tingay et~al.}{Tingay et~al.}{2013}]{mwa_tech}
Tingay, S.~J., et~al. 2013, Publications of the Astronomical Society of
  Australia, 30

\bibitem[\protect\citeauthoryear{{Tokumaru} et~al.}{{Tokumaru}
  et~al.}{2011}]{tokumaru11}
{Tokumaru}, M., {Kojima}, M., {Fujiki}, K., {Maruyama}, K., {Maruyama}, Y.,
  {Ito}, H.,  \& {Iju}, T. 2011, Radio Science, 46, 0

\bibitem[\protect\citeauthoryear{{Tomsick} et~al.}{{Tomsick}
  et~al.}{2003}]{tcf+03}
{Tomsick}, J.~A., {Corbel}, S., {Fender}, R., {Miller}, J.~M., {Orosz}, J.~A.,
  {Tzioumis}, T., {Wijnands}, R.,  \& {Kaaret}, P. 2003, \apj, 582, 933

\bibitem[\protect\citeauthoryear{{Usov} \& {Katz}}{{Usov} \&
  {Katz}}{2000}]{uk00}
{Usov}, V.~V.,  \& {Katz}, J.~I. 2000, \aap, 364, 655

\bibitem[\protect\citeauthoryear{{Uyaniker} et~al.}{{Uyaniker}
  et~al.}{2003}]{2003ApJ...585..785U}
{Uyaniker}, B., {Landecker}, T.~L., {Gray}, A.~D.,  \& {Kothes}, R. 2003, \apj,
  585, 785

\bibitem[\protect\citeauthoryear{{Vr{\v s}nak} \& {Cliver}}{{Vr{\v s}nak} \&
  {Cliver}}{2008}]{vrsnakandcliver2008}
{Vr{\v s}nak}, B.,  \& {Cliver}, E.~W. 2008, \solphys, 253, 215

\bibitem[\protect\citeauthoryear{{Warmuth} \& {Mann}}{{Warmuth} \&
  {Mann}}{2004}]{warmuthandmann2005}
{Warmuth}, A.,  \& {Mann}, G. 2004, in Lecture Notes in Physics, Berlin
  Springer Verlag, Vol. 656, Lecture Notes in Physics, Berlin Springer Verlag,
  ed. {K.~Scherer, H.~Fichter, \& B.~Herber}, 49

\bibitem[\protect\citeauthoryear{{Wieringa} et~al.}{{Wieringa}
  et~al.}{1993}]{wie93}
{Wieringa}, M.~H., {de Bruyn}, A.~G., {Jansen}, D., {Brouw}, W.~N.,  \&
  {Katgert}, P. 1993, \aap, 268, 215

\bibitem[\protect\citeauthoryear{{Wild}, {Smerd}, \& {Weiss}}{{Wild}
  et~al.}{1963}]{wildetal1963}
{Wild}, J.~P., {Smerd}, S.~F.,  \& {Weiss}, A.~A. 1963, \araa, 1, 291

\bibitem[\protect\citeauthoryear{{Wilms} et~al.}{{Wilms} et~al.}{2007}]{wpp+07}
{Wilms}, J., {Pottschmidt}, K., {Pooley}, G.~G., {Markoff}, S., {Nowak}, M.~A.,
  {Kreykenbohm}, I.,  \& {Rothschild}, R.~E. 2007, \apjl, 663, L97

\bibitem[\protect\citeauthoryear{{Windhorst} et~al.}{{Windhorst}
  et~al.}{1985}]{1985ApJ...289..494W}
{Windhorst}, R.~A., {Miley}, G.~K., {Owen}, F.~N., {Kron}, R.~G.,  \& {Koo},
  D.~C. 1985, \apj, 289, 494

\bibitem[\protect\citeauthoryear{{Wolleben} et~al.}{{Wolleben}
  et~al.}{2006}]{wlr06}
{Wolleben}, M., {Landecker}, T.~L., {Reich}, W.,  \& {Wielebinski}, R. 2006,
  \aap, 448, 411

\bibitem[\protect\citeauthoryear{{Woods} \& {Thompson}}{{Woods} \&
  {Thompson}}{2006}]{wt06}
{Woods}, P.~M.,  \& {Thompson}, C. 2006, in Compact stellar X-ray sources, ed.
  W.~{Lewin} \& M.~{van der Klis} (Cambridge, UK: Cambridge University Press),
  547

\bibitem[\protect\citeauthoryear{{Wouthuysen}}{{Wouthuysen}}{1952}]{1952AJ.....57R..31W}
{Wouthuysen}, S.~A. 1952, \aj, 57, 31

\bibitem[\protect\citeauthoryear{{Wu} \& {Lee}}{{Wu} \& {Lee}}{1979}]{wl79}
{Wu}, C.~S.,  \& {Lee}, L.~C. 1979, \apj, 230, 621

\bibitem[\protect\citeauthoryear{{Wyithe}}{{Wyithe}}{2008}]{2008MNRAS.387..469W}
{Wyithe}, J.~S.~B. 2008, \mnras, 387, 469

\bibitem[\protect\citeauthoryear{{Wyithe} \& {Loeb}}{{Wyithe} \&
  {Loeb}}{2008}]{2008MNRAS.383..606W}
{Wyithe}, J.~S.~B.,  \& {Loeb}, A. 2008, \mnras, 383, 606

\bibitem[\protect\citeauthoryear{{Wyithe}, {Loeb}, \& {Barnes}}{{Wyithe}
  et~al.}{2005}]{2005ApJ...634..715W}
{Wyithe}, J.~S.~B., {Loeb}, A.,  \& {Barnes}, D.~G. 2005, \apj, 634, 715

\bibitem[\protect\citeauthoryear{{Wyithe}, {Loeb}, \& {Schmidt}}{{Wyithe}
  et~al.}{2007}]{2007MNRAS.380.1087W}
{Wyithe}, J.~S.~B., {Loeb}, A.,  \& {Schmidt}, B.~P. 2007, \mnras, 380, 1087

\bibitem[\protect\citeauthoryear{{Wyithe} \& {Morales}}{{Wyithe} \&
  {Morales}}{2007}]{2007MNRAS.379.1647W}
{Wyithe}, J.~S.~B.,  \& {Morales}, M.~F. 2007, \mnras, 379, 1647

\bibitem[\protect\citeauthoryear{{Xiong} et~al.}{{Xiong}
  et~al.}{2011}]{Xiong11}
{Xiong}, M., {Breen}, A.~R., {Bisi}, M.~M., {Owens}, M.~J., {Fallows}, R.~A.,
  {Dorrian}, G.~D., {Davies}, J.~A.,  \& {Thomasson}, P. 2011, Journal of
  Atmospheric and Solar-Terrestrial Physics, 73, 1270

\bibitem[\protect\citeauthoryear{{Zarka} et~al.}{{Zarka} et~al.}{2001}]{ztrr01}
{Zarka}, P., {Treumann}, R.~A., {Ryabov}, B.~P.,  \& {Ryabov}, V.~B. 2001,
  \apss, 277, 293

\bibitem[\protect\citeauthoryear{{Zaroubi} \& {Silk}}{{Zaroubi} \&
  {Silk}}{2005}]{2005MNRAS.360L..64Z}
{Zaroubi}, S.,  \& {Silk}, J. 2005, \mnras, 360, L64

\bibitem[\protect\citeauthoryear{{Zhang}}{{Zhang}}{2007}]{Zhang07}
{Zhang}, X. 2007, Chinese Journal of Astronomy and Astrophysics, 7, 712

\end{thebibliography}

\end{document}